\documentclass[longauth]{aa}

\usepackage{graphicx}
\usepackage[hidelinks]{hyperref}
\usepackage{multirow}
\usepackage{tabularx}
\usepackage{longtable,tabu}
\usepackage{lscape}
\usepackage{subfigure}
\usepackage{natbib}
\usepackage{placeins}
\usepackage[dvipsnames]{xcolor}

\bibpunct{(}{)}{;}{a}{}{,} 

\newcommand{\planet}{KELT-20b/MASCARA-2b}
\newcommand{\system}{KELT-20/MASCARA-2}

\begin{document}

\title{The GAPS programme at TNG. XLV. H{\sc i} Balmer lines transmission spectroscopy and NLTE atmospheric modelling of the ultra-hot Jupiter KELT-20b/MASCARA-2b}

\titlerunning{Transmission spectroscopy and NLTE atmospheric modelling of KELT-20b/MASCARA-2b}

\authorrunning{Fossati et al.}

\author{L. Fossati\inst{1} \and        
        F. Biassoni\inst{2,3} \and
        G. M. Cappello\inst{1,4,5} \and
        F. Borsa\inst{3} \and
        D. Shulyak\inst{6} \and
        A. S. Bonomo\inst{7} \and
        D. Gandolfi\inst{4} \and
        F. Haardt\inst{2,3} \and
        T. Koskinen\inst{8} \and
        A. F. Lanza\inst{9} \and
        V. Nascimbeni\inst{10} \and
        D. Sicilia\inst{9} \and
        M. Young\inst{11} \and
        G. Aresu\inst{12} \and
        A. Bignamini\inst{13} \and
        M. Brogi\inst{4,7} \and
        I. Carleo\inst{14,7} \and
        R. Claudi\inst{10} \and
        R. Cosentino\inst{15} \and
        G. Guilluy\inst{7} \and
        C. Knapic\inst{13} \and
        L. Malavolta\inst{16,10} \and
        L. Mancini\inst{17,18,7} \and
        D. Nardiello\inst{10} \and
        M. Pinamonti\inst{7} \and
        L. Pino\inst{19} \and
        E. Poretti\inst{14,3} \and
        M. Rainer\inst{3} \and
        F. Rigamonti\inst{2,3,20} \and
        A. Sozzetti\inst{7}        
        }




\institute{Space Research Institute, Austrian Academy of Sciences, Schmiedlstrasse 6, 8042, Graz, Austria\\
\email{Luca.Fossati@oeaw.ac.at}
\and
DISAT, Universit\`a degli Studi dell'Insubria, via Valleggio 11, 22100, Como, Italy
\and
INAF -- Osservatorio Astronomico di Brera, via E. Bianchi 46, 23807, Merate, Italy
\and
Dipartimento di Fisica, Universit\`a degli Studi di Torino, via Pietro Giuria 1, 10125, Torino, Italy
\and
Institute of Physics, University of Graz, Universit\"atsplatz 5, 8010, Graz, Austria
\and
Instituto de Astrof\'isica de Andaluc\'ia (CSIC), Glorieta de la Astronom\'ia s/n, 18008, Granada, Spain
\and
INAF -- Osservatorio Astrofisico di Torino, Via Osservatorio 20, 10025, Pino Torinese, Italy
\and
Lunar and Planetary Laboratory, University of Arizona, 1629 East University Boulevard, Tucson, AZ, 85721-0092, USA
\and
INAF -- Osservatorio Astrofisico di Catania, Via S. Sofia 78, 95123, Catania, Italy
\and
INAF -- Osservatorio Astronomico di Padova, Vicolo dell'Osservatorio 5, 35122, Padova, Italy
\and
Department of Physics, University of Oxford, Denys Wilkinson Building, Keble Road, Oxford, OX1 3RH, UK
\and
INAF -- Osservatorio Astronomico di Cagliari, Via della Scienza 5, 09047, Selargius (CA), Italy 
\and
INAF -- Osservatorio Astronomico di Trieste, via Tiepolo 11, 34143, Trieste, Italy
\and
Instituto de Astrof\'{i}sica de Canarias (IAC), 38205 La Laguna, Tenerife, Spain
\and
Fundaci\'on Galileo Galilei - INAF, Rambla Jos\'e Ana Fernandez P\'erez 7, 38712, Bre\~na Baja, TF, Spain
\and
Dipartimento di Fisica e Astronomia ``Galileo Galilei'' - Universit\`a di Padova, Vicolo dell'Osservatorio 3, 35122, Padova, Italy
\and
Department of Physics, University of Rome Tor Vergata, Via della Ricerca Scientifica 1, 00133, Roma, Italy
\and
Max Planck Institute for Astronomy, K\"onigstuhl 17, 69117, Heidelberg, Germany
\and
INAF -- Osservatorio Astrofisico di Arcetri, Largo Enrico Fermi 5, I-50125 Firenze, Italy
\and
INFN, Sezione di Milano-Bicocca, Piazza della Scienza 3, I-20126 Milano, Italy
}
\date{Received date ; Accepted date }
\abstract
{}
{We aim at extracting the transmission spectrum of the H{\sc i} Balmer lines of the ultra-hot Jupiter (UHJ) \planet\ from observations and to further compare the results with what obtained through forward modelling accounting for non-local thermodynamic equilibrium (NLTE) effects.}
{We extract the line profiles from six transits obtained with the HARPS-N high-resolution spectrograph attached to the Telescopio Nazionale Galileo telescope. We compute the temperature-pressure (TP) profile employing the {\sc helios} code in the lower atmosphere and the {\sc Cloudy} NLTE code in the middle and upper atmosphere. We further use {\sc Cloudy} to compute the theoretical planetary transmission spectrum in LTE and NLTE for comparison with observations.}
{We detected the H$\alpha$ (0.79$\pm$0.03\%; 1.25\,R$_{\rm p}$), H$\beta$ (0.52$\pm$0.03\%; 1.17\,R$_{\rm p}$), and H$\gamma$ (0.39$\pm$0.06\%; 1.13\,R$_{\rm p}$) lines, while we detected the H$\delta$ line at almost 4$\sigma$ (0.27$\pm$0.07\%; 1.09\,R$_{\rm p}$). The models predict an isothermal temperature of $\approx$2200\,K at pressures $>$10$^{-2}$\,bar and of $\approx$7700\,K at pressures $<$10$^{-8}$\,bar, with a roughly linear temperature rise in between. In the middle and upper atmosphere, the NLTE TP profile is up to $\sim$3000\,K hotter than in LTE. The synthetic transmission spectrum derived from the NLTE TP profile is in good agreement with the observed H{\sc i} Balmer line profiles, validating our obtained atmospheric structure. Instead, the synthetic transmission spectrum derived from the LTE TP profile leads to significantly weaker absorption compared to the observations.}
{Metals appear to be the primary agents leading to the temperature inversion in UHJs and the impact of NLTE effects on them increases the magnitude of the inversion. We find that the impact of NLTE effects on the TP profile of \planet\ is larger than for the hotter UHJ KELT-9b, and thus NLTE effects might be relevant also for planets cooler than \planet.}
\maketitle
\section{Introduction}
Transmission and emission spectroscopy, both from ground and space, have led to significant advances in our understanding of exoplanetary atmospheres. In this context, ultra-hot Jupiters (UHJs), that is planets with an equilibrium temperature ($T_{\rm eq}$) greater than $\approx$2000\,K and for which H$^-$ opacity and thermal dissociation are significant, play a key role. This is because the high atmospheric temperatures of these planets lead to large pressure scale heights and thus more easily detectable spectral features. Furthermore, these planets are typically detected orbiting rather bright intermediate-mass stars (i.e. F- and A-type), which eases gathering high-quality observations, particularly at optical wavelengths.

With an equilibrium temperature of nearly 4000\,K, KELT-9b \citep{gaudi2017} is the most extreme of the UHJs and so far also the most studied both observationally and theoretically. Because of the brightness of the host star ($V$\,$\approx$\,7.6\,mag), KELT-20b \citep{lund2017}, also known as MASCARA-2b \citep{talens2018}, is one of the most studied UHJs in terms of atmospheric characterisation observations, mainly through ground-based high-resolution spectroscopy.

\planet\ orbits the A2V host star (HD\,185603) with a period of about 3.47\,days, which implies an equilibrium temperature of about 2200\,K, assuming zero albedo and complete heat redistribution \citep{lund2017}. Photometric transit observations provided a rather precise measurement of the planetary radius of about 1.83\,$R_{\rm J}$ \citep{talens2018}, but the broad spectral lines due to the rapid rotation of the host star \citep[$\nu\sin{i}$\,=\,116.23\,km\,s$^{-1}$; $P_{\rm rot}$\,=\,0.695\,days;][]{rainer2021} hinder the precise measurement of the planetary mass for which just a 3$\sigma$ upper limit of 3.51\,$M_{\rm J}$ has been obtained \citep{lund2017}. 

Multiple ground-based transmission spectroscopy observations enabled the clear detection of the H$\alpha$ and H$\beta$ hydrogen Balmer lines, as well as of the Na{\sc i}\,D doublet, the Ca{\sc ii} infrared triplet, the Ca{\sc ii}\,H\&K lines, and of multiple Fe{\sc ii} lines \citep{casasayas2018_kelt20b,Casasayas_2019,nugroho2020}. The cross-correlation technique led to the further detection of Fe{\sc i}, Mg{\sc i}, and Cr{\sc ii} in the planetary atmosphere, as well as to the confirmation of some of the previously detected species \citep{stangret2020,hoeijmakers2020,nugroho2020,rainer2021,cont2022,belloArufe2022,johnson2022}.

Thanks to its high equilibrium temperature, \planet\ has been also observed during and around secondary eclipse both from the ground and from space. Applications of the cross-correlation technique on high-resolution ground-based dayside observations enabled the detection of Fe{\sc i}, Fe{\sc ii}, Si{\sc i}, and Cr{\sc i} on the atmospheric day-side \citep{borsa2022_mascara2b,cont2022,johnson2022,yan2022_mascara2b,kasper2023}. Space-based secondary eclipse observations led to the further detection of water in the lower atmosphere, measurement of the brightness temperature in different bands, and constraints on the atmospheric metallicity \citep{fu2022}. These observations have also been used to constrain the atmospheric temperature-pressure (TP) profile at pressures higher than 10$^{-5}$\,bar, which is probed by the cores of metal lines in the optical band \citep{borsa2022_mascara2b,fu2022,yan2022_mascara2b}. The detection of emission features close to secondary eclipse clearly indicates the presence of a temperature inversion, that is an atmospheric temperature increasing with decreasing pressure. Retrievals of the TP profile performed on the observations concur on setting the temperature increase at the $\sim$0.1\,bar level with a rather isothermal profile of about 2200\,K at higher pressures, which is supported by forward modelling \citep{borsa2022_mascara2b,fu2022,yan2022_mascara2b}.

Several previous studies sought to identify the radiatively active species responsible for the temperature inversion, particularly focusing on the search for molecules such as TiO, VO, and FeH. These attempts have not been successful \citep{nugroho2020,johnson2022}, except for FeH that has been tentatively detected by \citet{kesseli2020}, but not confirmed by follow-up observations \citep{johnson2022}. These non-detections and the simultaneous detection of a number of neutral and singly ionised atomic species support the idea that metal absorption of the stellar radiation, particularly of ultraviolet (UV) photons, is at the origin of the temperature inversion, as predicted by models of UHJs \citep{lothringer2018,lothringer2019,fossati2021}. This was particularly evident in the case of KELT-9b, for which a broad range of atomic species has been detected \citep[e.g.][]{hoeijmakers2018,hoeijmakers2019,yan2019_kelt9b,turner2020_kelt9b,borsa2022_kelt9b_OI,pino2020}. Furthermore, for KELT-9b \citet{fossati2021} showed that non-local thermodynamic equilibrium (NLTE) effects lead to a significant overpopulation of Fe{\sc ii} and underpopulation of Mg{\sc ii} that are the key agents, respectively, driving heating and cooling in the planetary atmosphere. In particular, the NLTE-driven overpopulation of excited Fe{\sc ii} significantly increases the absorption of stellar near-UV radiation, further increasing the heating rate. This stems from the fact that the near-UV spectral range, which is also where the host star's spectral energy distribution peaks, contains several Fe{\sc ii} lines rising from low energy levels. This conclusion was supported by the fact that the forward model presented by \citet{fossati2021} accounting for NLTE effects was able not only to fit the observed hydrogen Balmer line profiles, but also to predict the presence of the O{\sc i} infrared triplet and its strength in the planetary transmission spectrum, which has been then confirmed by observations \citep{borsa2022_kelt9b_OI}.    

Here, we present the results of transmission spectroscopy of \planet\ based on six transit observations that aim at detecting hydrogen Balmer line absorption and constraining the shape of the absorption line profiles. This work follows the observations of \citet{rainer2021} and \citet{borsa2022_mascara2b} on this same planet that focused respectively on measuring the Rossiter-McLaughlin (RM) effect during transit and detecting species through dayside measurements. We interpret the hydrogen Balmer line observations by using a forward model that spans from the lower atmosphere (10\,bar) to the upper atmosphere ($\sim$10$^{-12}$\,bar) and includes NLTE effects, comparable to the model used by \citet{fossati2021} to explain the observations of KELT-9b.

This paper is organised as follows. Section~\ref{sec:observations} presents the observations and the methodology employed to analyse the data. Section~\ref{sec:modelling} describes the atmospheric modelling. Section~\ref{sec:results} presents the modelling results (Section~\ref{sec:TP_chemistry}) and the hydrogen Balmer line profiles obtained from the observations (Section~\ref{sec:balmer}). In Section~\ref{sec:discussion}, we compare the observational and modelling results, present the synthetic UV-to-infrared transmission spectrum, and extract more information about the properties of the planetary atmosphere. Finally, we gather the conclusions in Section~\ref{sec:conclusions}.
\section{Observations and data analysis}\label{sec:observations}
\planet\ was observed in the optical band with the high resolution ($R$\,$\sim$\,115\,000) HARPS-N spectrograph \citep{Cosentino2012} located at Telescopio Nazionale Galileo (TNG) during six different transits between the years 2017 and 2022. The first three transits that we analyse are taken from the TNG archive and have been previously analysed by \citet{Casasayas_2019} to look for the hydrogen Balmer lines in the planetary atmosphere, as well as for other chemical species. We add here further three transits taken in the context of the atmospheric characterisation part of the GAPS programme \citep[e.g.][]{borsa2019,Giacobbe2021,guilluy2022}, for which the H Balmer lines were not analysed before.
\begin{table*}
\centering
\caption{Log of the transit observations of KELT-20b/MASCARA-2b  used in this work.}
\label{log_observation}
\resizebox{18.8cm}{!}{
\begin{tabular}{c|c|c|c|c|c|c|c}

\hline
\hline
Night \#&
Night date      &
Program   &
PI &
$T_{\rm exp}$ [sec] &
\# of spectra (Out/In) &
 <S/N>@ 550nm&
 airmass (min/max)\\
\hline
1 &
2017-08-16          &
CAT17A-38  &
Rebolo&
200                 &
90 (33/57)          &
61               &
1/2.13 \\
2 &
2018-07-12          &
CAT18A-34  &
Casasayas-Barris&
200                 &
105 (53/52)         &
96               &
1/1.58    \\
3 &
2018-07-19          &
CAT18A-34  &
Casasayas-Barris&
300                 &
78 (39/39)          &
105              &
1/1.42    \\
4 &
2019-08-26          &
GAPS       &
Micela&
600                 &
30 (10/20)          &
164              &
1/2.09    \\
5 &
2019-09-02          &
GAPS       &
Micela&
 600                 &
29 (8/21)           &
176              &
1/1.55    \\
6 &
2022-07-31          &
GAPS       &
Micela&
600                 &
25 (10/15)           &
125               &
1/1.44    \\

\hline
    \end{tabular}
    }
\end{table*}

Table~\ref{log_observation} summarises the observations considered in this work. Nights 2 and 6 present an unstable signal-to-noise ratio (S/N), thereby for these nights we decided to discard all spectra with S/N\,$<$\,50. Furthermore, nights 2 and 3 suffered a problem that affected the telescope's Atmospheric Dispersion Corrector \citep[ADC; see][for more details]{Casasayas_2019}, which causes wavelength dependent flux losses in particular in the blue part of the spectrum, and thus called for particular attention in the normalisation process. All spectra have been reduced with the standard data reduction software (DRS) v3.7 and we analysed the one-dimensional pipeline products (i.e. s1d), which cover the 3800--6900\,\AA\ wavelength range and have a constant wavelength step of 0.01\,\AA.
 
\begin{figure}[t!]
		\centering
		\includegraphics[width=9cm]{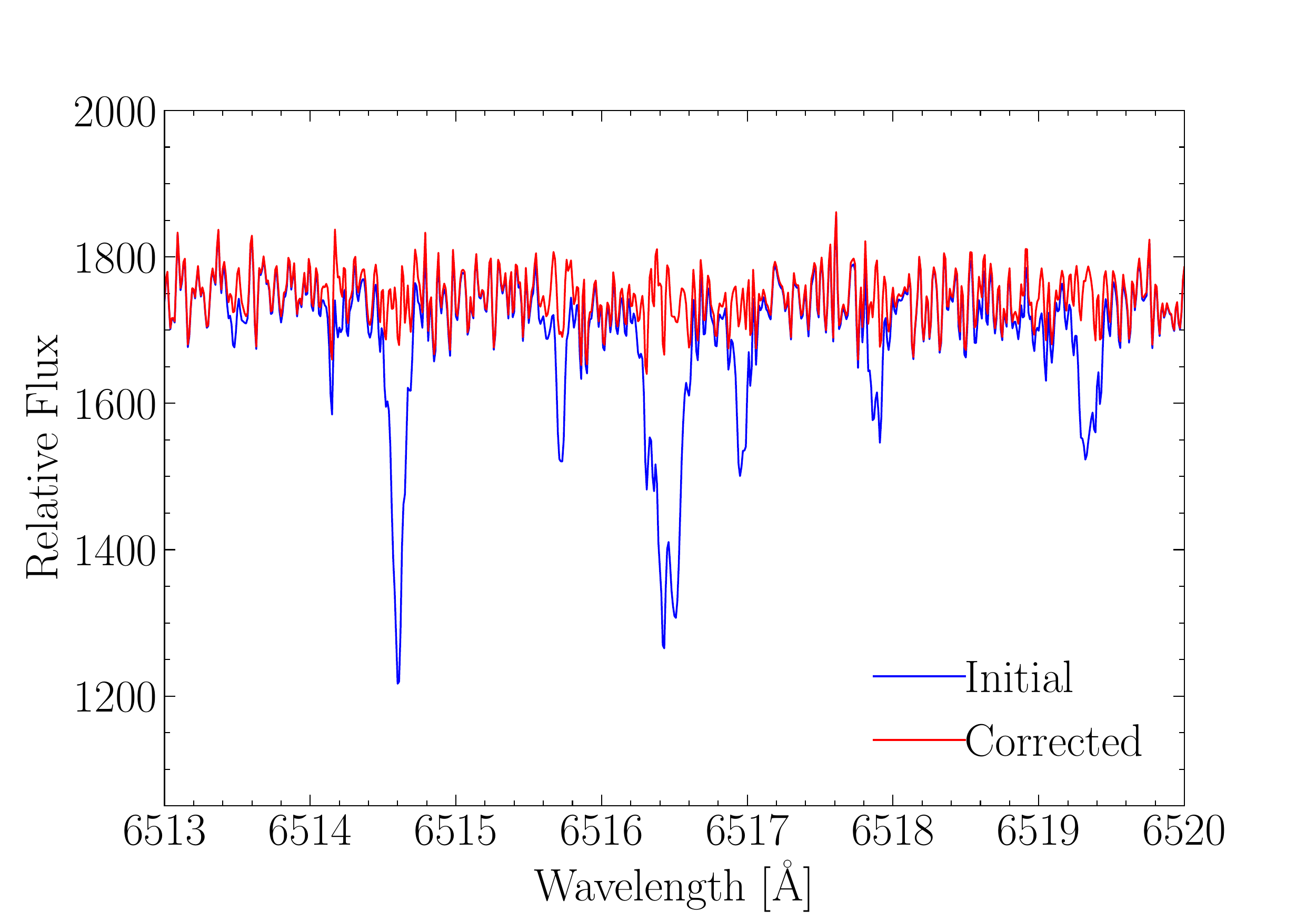}
		\caption{Example of {\sc Molecfit} tellurics correction in proximity of the H$\alpha$ line.}
  		\label{fig:molecfit}
\end{figure}

We focused our attention on the Balmer lines H$\alpha$, H$\beta$, H$\gamma$, and H$\delta$. For each transit, we calculated the transmission spectra following a procedure similar to that described by \citet{Wyttenbach2015}. We started by correcting telluric lines across the whole wavelength range, employing {\sc Molecfit} v4.2.3 \citep{Smette_2015, Kausch_2015} and following the prescriptions of \citet{Allart_2017}. In particular, we corrected for absorption caused by telluric O$_2$ and H$_2$O. An example of the correction performed is presented in Figure~\ref{fig:molecfit}. We then Doppler-shifted all spectra in the stellar reference frame by subtracting the theoretical stellar radial velocity at each orbital phase, and normalised them to the same continuum level in a narrow range around each considered line (i.e. H$\alpha$ 6525--6595\,\AA; H$\beta$ 4810--4910\,\AA; H$\gamma$ 4310--4370\,\AA; H$\delta$ 4080--4120\,\AA). We then calculated an average out-of-transit spectrum (master-out, $M_{\rm out}$) employing a weighted average on the stellar flux, divided each spectrum by $M_{\rm out}$, and normalised again to remove any remaining small linear trend that has propagated throughout the analysis procedure. This last step was particularly important for the analysis of nights 2 and 3, because of the faulty ADC affecting the continuum flux.
 
At this point, we corrected the residual spectra for the contamination given by stellar rotation (i.e. the Rossiter-McLaughlin effect) and center-to-limb variations (CLV), which are known to possibly affect the transmission spectrum and eventually cause false detections \citep[e.g.][]{Borsa2018,casasayas2020}. To this end, we created a model of the contamination as in \citet{Borsa_2021}, that follows the approach of \citet{Yan_2017}, by inserting ATLAS9 stellar atmosphere models and a VALD linelist \citep{Ryabchikova_2015} into Spectroscopy Made Easy \citep[SME;][]{Piskunov_2017}, based on the system parameters listed in Table~\ref{tab:systemParameters}. The modelled contamination has been calculated for each observed orbital phase and removed from the data by dividing for it. The extension of the planetary radius R$_{p,\lambda}$ used to model the CLV and RM effects on each of the Balmer lines has been calculated iteratively, until reaching a value for which the final amplitude of the Gaussian fit to the line in the transmission spectrum (see Section~\ref{sec:balmer}) and the value of R$_{p,\lambda}$ used in the CLV+RM model coincide within 0.5$\sigma$, with $\sigma$ being the error-bar on the line depth calculated as described in Section~\ref{sec:balmer}.
 
We then calculated the final transmission spectrum for each night by shifting all residual spectra in the planetary reference frame and performing a weighted average of all residual spectra taken during the full part of the transit (i.e. between the second and third contact points). We finally obtained the average transmission spectrum of each Balmer line by performing a weighted average of the transmission spectra obtained from each of the six nights of observation. 
\begin{table}[]
\caption{Adopted parameters of the \system\ system.}
\begin{tabular}{l|c|l}
\hline
\hline
Parameter & Value & Source \\
\hline
$T_{\rm eff}$ [K] & 8980 & \citet{talens2018} \\
$M_{\rm s}$ [$M_{\odot}$] & 1.89 & \citet{talens2018} \\
$R_{\rm s}$ [$R_{\odot}$] & 1.60 & \citet{talens2018} \\
Sp.T. & A2 & \citet{talens2018} \\
\hline
$M_{\rm p}$ [$M_{\rm J}$] & 3.51 & \citet{borsa2022_mascara2b} \\
$R_{\rm p}$ [$R_{\rm J}$] & 1.83 & \citet{talens2018} \\
\hline
$a$ [AU] & 0.0542 & \citet{talens2018} \\
$P$ [days] & 3.474119 & \citet{talens2018} \\
$T_0$ [BJD] & 2457909.5875 & \citet{hoeijmakers2020} \\
$T_{\rm ingress}$ [days] & 0.01996 & \citet{cont2022} \\ 
$T_{14}$ [days] & 0.14882 & \citet{rainer2021} \\
$K_{\rm s}$ [km\,s$^{-1}$] & 0.32251 & \citet{borsa2022_mascara2b} \\
$K_{\rm p}$ [km\,s$^{-1}$] & 173 & \citet{borsa2022_mascara2b} \\ 
$V_{\rm sys}$ [km\,s$^{-1}$] & $-$24.48 & \citet{borsa2022_mascara2b} \\
$b$ & 0.503 & \citet{lund2017} \\
$\lambda$ [deg] & 3.4 & \citet{lund2017} \\
$e$ & 0 & \citet{lund2017} \\
$i$ [deg] & 86.12 & \citet{lund2017} \\
\hline
\end{tabular}
\label{tab:systemParameters}
\end{table}
%
\section{Atmospheric modelling}\label{sec:modelling}
We computed the theoretical atmospheric TP profile of \planet\ at the sub-stellar point employing the scheme described in detail by \citet{fossati2021}. It consists of the separate computation of TP profiles of the lower ($P\gtrsim10^{-4}$\,bar) atmosphere with the {\sc helios} code \citep{malik2017,malik2019} and of the upper ($P\lesssim10^{-4}$\,bar) atmosphere with the {\sc Cloudy} NLTE radiative transfer code \citep{ferland2017}, the latter through the {\sc Cloudy} for Exoplanets interface \citep[CfE;][]{fossati2021}. The {\sc helios} and {\sc Cloudy} TP profiles are then joined together to obtain a single TP profile that is used as starting point to derive the atmospheric chemical composition and transmission spectra. We adopt this scheme, because {\sc helios} does not account for NLTE effects that are relevant in the middle and upper atmosphere, while {\sc Cloudy}, which considers NLTE effects, is unreliable in the lower atmosphere at densities greater than 10$^{15}$\,cm$^{-3}$ (see \citealt{ferland2017} and \citealt{fossati2021} for more details).

{\sc helios}\footnote{\tt https://github.com/exoclime/HELIOS} is a radiative-convective equilibrium code taking as input planetary mass, radius, and atmospheric abundances, orbital semi-major axis, and stellar radius and effective temperature, further assuming equilibrium abundances, which we computed using the \textsc{FastChem}\footnote{\tt https://github.com/exoclime/FastChem} code \citep{stock2018}. For the simulation, we divided the atmosphere into 100 layers logarithmically distributed in the 100--10$^{-9}$\,bar pressure range. We computed the {\sc helios} model with a heat redistribution parameter, $f$, which accounts for the day-to-night side heat redistribution efficiency, equal to 0.25. This leads to a TP profile in the lower atmosphere comparable to that obtained from retrievals of day-side observations \citep{borsa2022_mascara2b,fu2022}. Given the high planetary atmospheric temperature, we considered additional opacities not present in the public version of the {\sc helios} code as described by \citet{fossati2021}.

{\sc Cloudy} is a general-purpose plane-parallel microphysics radiative transfer code that accounts for (photo)chemistry and NLTE effects \citep{ferland1998,ferland2013,ferland2017}. For the calculations presented here, we employed {\sc Cloudy} version 17.03. {\sc Cloudy} computations include a wide range of atomic (i.e. all elements up to Zn) and molecular species, and are valid across a wide interval of plasma temperatures (3--10$^{10}$\,K) and densities ($<$10$^{15}$\,cm$^{-3}$). This covers the parameter space of what is expected in upper planetary atmospheres. {\sc Cloudy} is a hydrostatic code and we come back to the validity of this assumption in the case of \planet\ in Section~\ref{sec:discussion}. Details of the code relevant to exoplanet atmospheric calculations are given in Section~2.2.1 of \citet{fossati2021}.

To set up the {\sc Cloudy} runs, we employed the CfE interface, which writes {\sc Cloudy} input files on the basis of input parameters given by the user, runs {\sc Cloudy}, and reads {\sc Cloudy} output files. Then, CfE uses the information contained in the output files to set up a new {\sc Cloudy} calculation in an iterative process until the temperature profile has converged. The details of how CfE sets up {\sc Cloudy} input files and the iteration procedure are described in Section~2.2.2 of \citet{fossati2021}.

For the atmospheric modelling, we considered the system parameters listed in Table~\ref{tab:systemParameters}. For the calculations, we employed a synthetic spectral energy distribution computed with the {\sc phoenix} code\footnote{{\tt https://phoenix.astro.physik.uni-goettingen.de/}} \citep{husser2013}. Because the host star is earlier than spectral type A3--A4, we did not add any X-ray and extreme ultraviolet (EUV) emission to the photospheric fluxes provided by the {\sc phoenix} model \citep{fossati2018_AstarPlanets}. However, rapidly rotating early-type stars could present magnetic activity close to the equator as a result of the low local effective temperature caused by gravity darkening, but this is not the case for \system, because its rotational velocity is just about 45\% of the critical break-up velocity. This further justifies our assumption of no X-ray and EUV emission in addition to the photospheric one.

The 10$^{15}$\,cm$^{-3}$ density limit above which {\sc Cloudy}'s computation of the heating and cooling rates becomes unreliable lies at a pressure of about 0.3\,mbar, while the continuum lies at a pressure of about 6\,mbar. This is the pressure at which the {\sc helios} model gives an optical depth of 2/3 around 5000\,\AA, which is the wavelength corresponding to the peak efficiency of the MASCARA optical system used to discover the planet \citep{talens2017}. For this reason, we run CfE setting the planetary transit radius of 1.83\,$R_{\rm J}$ \citep{talens2018} at the reference pressure ($p_0$) of 6\,mbar. The top panel of Figure~\ref{fig:test_p0_molecules} shows a comparison between {\sc Cloudy} TP profiles obtained setting the planetary transit radius at different $p_0$ values of 100, 6, and 1\,mbar, indicating that uncertainties on the location of the reference pressure do not impact the results. To save on computational time, for the {\sc Cloudy} calculations we considered all elements up to Zn and only hydrogen molecules (i.e. H$_2$, H$_2^+$, H$_3^+$), because the inclusion of all molecules present in the {\sc Cloudy} database did not affect the resulting TP profile (see the middle panel of Figure~\ref{fig:test_p0_molecules}). Finally, in agreement with \citet{fossati2021}, we find that the number of layers considered for the computation of the TP profile (i.e. 180) does not impact the results (see the bottom panel of Figure~\ref{fig:test_p0_molecules}).

To mimic atmospheric heat redistribution in the computation of the {\sc Cloudy} TP profiles, we scaled the spectral energy distribution multiplying it by a factor $f_1$\,$\leq$\,1. Following \citet{fossati2021}, we computed TP profiles with varying $f_1$ values looking for the one leading to the TP profile that best matches the {\sc helios} one around the 10$^{-4}$\,bar level. We finally adopted $f_1$\,=\,1.0, but we remark that the value employed for $f_1$ has no significant impact on the TP profile at pressures lower than about 10$^{-5}$\,bar \citep{fossati2021}. For all calculations, we assumed solar atmospheric composition \citep{lodders2003}. 
\section{Results}\label{sec:results}
\subsection{Atmospheric TP and abundance profiles}\label{sec:TP_chemistry}
We resampled the final theoretical TP profile, obtained by joining the {\sc helios} and {\sc Cloudy} profiles, over 200 layers equally spaced in $\log{p}$ ranging from 10\,bar to 4$\times$10$^{-12}$\,bar. The considered pressure range is wide enough to cover the formation region of ultraviolet, optical, and infrared lines in the transmission spectrum. 
\begin{figure}[ht!]
		\centering
		\includegraphics[width=9cm]{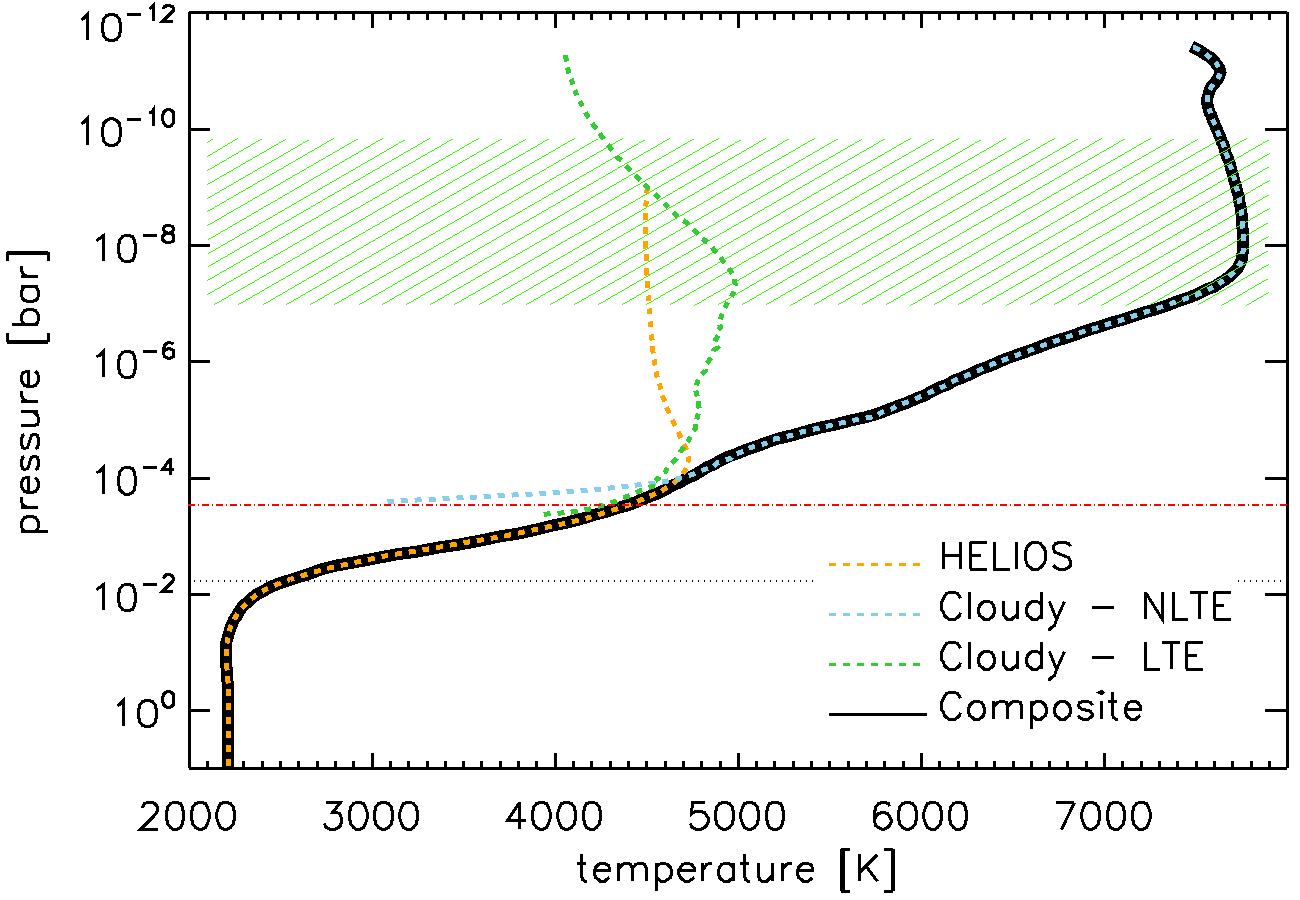}
		\includegraphics[width=9cm]{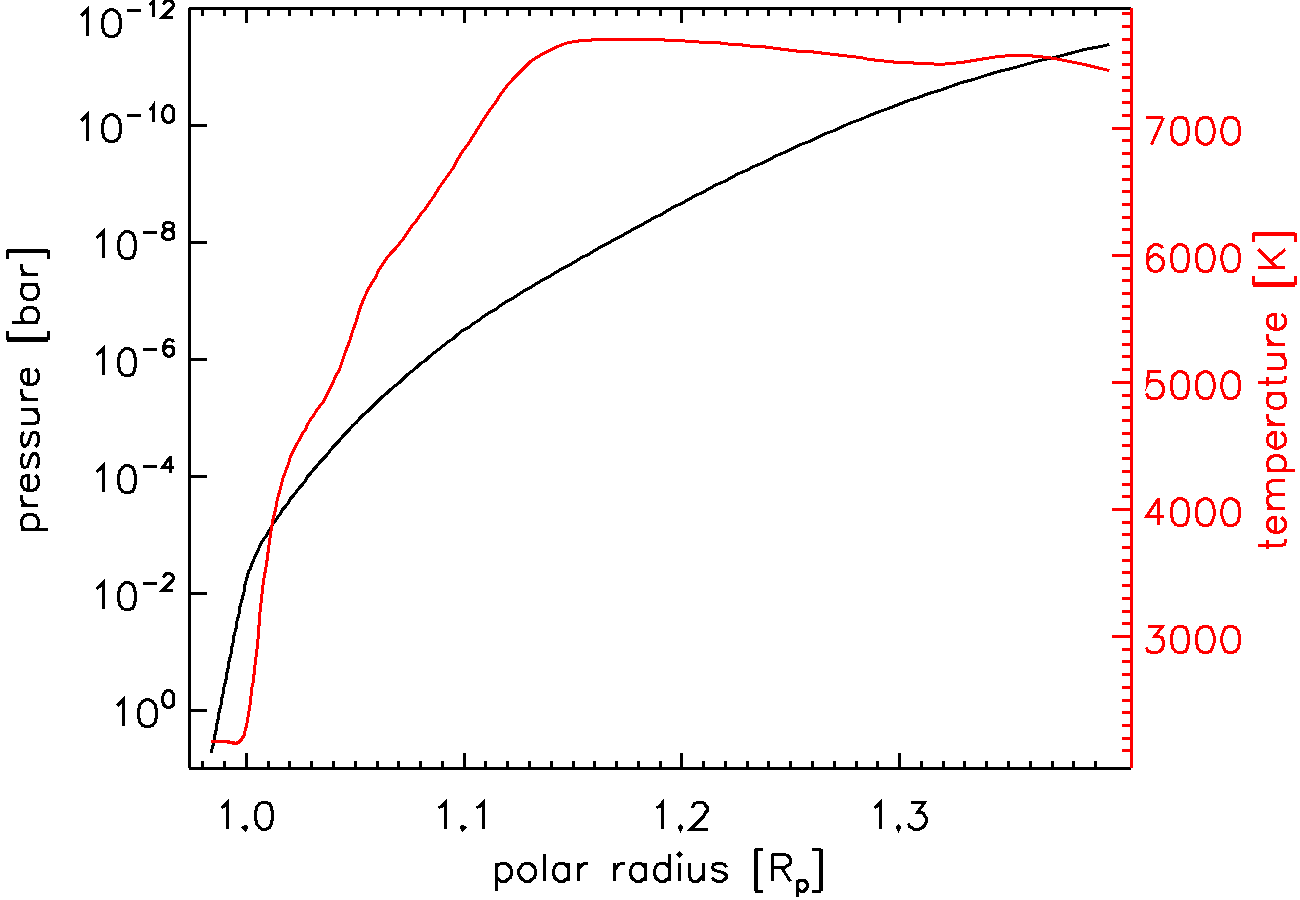}
		\caption{Theoretical atmospheric structure obtained for \planet. Top: {\sc helios} (orange dashed line) and {\sc Cloudy} (in NLTE; cyan dashed line) TP profiles. The black solid line shows the composite TP profile. The horizontal black dotted line indicates the location of the continuum according to the {\sc helios} model. The horizontal red dash-dotted line gives the location of {\sc Cloudy}’s upper-density limit of 10$^{15}$\,cm$^{-3}$. The dark green dashed line shows the {\sc Cloudy} TP profile computed assuming LTE. The hatched area indicates the H$\alpha$ line formation region. Bottom: Pressure (black; left y-axis) and temperature (red; right y-axis) composite theoretical profiles as a function of the planetary polar radius.} 
  		\label{fig:TPfrankenstein}
\end{figure}

We combined the {\sc helios} and {\sc Cloudy} theoretical TP profiles at a pressure of 10$^{-4}$\,bar and then fitted the entire profile with a polynomial to smooth the edge at the joining point, further interpolating on the final pressure scale. Figure~\ref{fig:TPfrankenstein} shows the composite theoretical TP profile in comparison to the original {\sc helios} and {\sc Cloudy} theoretical profiles. The final profile is roughly isothermal at pressures higher than about 10\,mbar and lower than about 10$^{-8}$\,bar, while the section in between those pressures is characterised by a linear temperature rise from about 2200\,K up to about 7700\,K. Therefore, the temperature varies by about 5500\,K from the lower to the upper atmosphere, which implies that an isothermal approximation for the entire atmosphere would be inappropriate for this planet. 

We identified the H$\alpha$ line formation region by computing {\sc Cloudy} models starting from the top of the atmosphere (at 10$^{-12}$\,bar) and subsequently increasing the pressure of the lower considered layer, further looking at the transmitted spectrum around the H$\alpha$ line. The top boundary of the line formation region (i.e. at low pressure) was then set where, with increasing pressure of the lower considered layer, the transmitted spectrum started to show H$\alpha$ absorption, while the bottom boundary of the line formation region (i.e. at high pressure) was set where the H$\alpha$ absorption stopped increasing with increasing pressure of the lower considered layer. In this way, we obtained that the H$\alpha$ line formation region is confined in a rather narrow pressure range between about 10$^{-10}$ and 10$^{-7}$\,bar, where the temperature is higher than about 7000\,K. This range is similar to that found for KELT-9b using the same modeling scheme employed here \citep{turner2019,fossati2021}. Both host stars (i.e. KELT-9 and \system) are not supposed to have a chromosphere \citep{fossati2018_AstarPlanets}, and thus their Ly$\alpha$ lines are totally absorbed, which prevents hydrogen photoexcitation. Therefore, the excitation of the hydrogen atoms to the $n$\,=\,2 level, which then leads to the formation of the Balmer lines, has to occur mostly thermally. This is further confirmed by the fact that photoionisation and subsequent recombination to the $n$\,=\,2 level have a low probability in comparison to thermal excitation, because of the very low EUV emission of the host star.

The derived temperature inversion supports the results of secondary eclipse observations \citep{borsa2022_mascara2b,fu2022,yan2022_mascara2b} as well as previous general atmospheric modelling of UHJs \citep{lothringer2018}, though the inclusion of NLTE effects strongly increases the magnitude of the inversion compared to previous LTE modelling. The retrieval approach followed by \citet{borsa2022_mascara2b} and \citet{yan2022_mascara2b} to constrain the TP profile from their secondary eclipse observations assumes a two-point TP profile where the temperature below the higher pressure point and above the lower pressure point was considered to be isothermal, with a linear gradient in between. At the two nodes of the two-point TP profile, they obtained a lower temperature value lying around 2200\,K with the turning point located in the 0.1--1\,bar range and a higher temperature value of about 5300\,K with the turning point at a pressure of about 10$^{-5}$\,bar. Instead, \citet{fu2022} did not make any assumption on the shape of the TP profile obtaining an isothermal profile at $\approx$2200\,K at pressures higher than about 10$^{-2}$\,bar, with a roughly linearly decreasing temperature at lower pressures up to 10$^{-4}$\,bar, where they stopped their calculation. The assumed two-point shape of the TP profile resembles well the temperature profile obtained by combining {\sc helios} and {\sc Cloudy}. As expected by the choice of parameters used to compute the {\sc helios} TP profile, in the lower atmosphere ($>$10$^{-5}$\,bar) the retrieved profiles match ours well (i.e. within 1$\sigma$). Instead, in the upper atmosphere ($<$10$^{-5}$\,bar) the retrieved temperatures are more than 1500\,K cooler and the turning points located at a 1000 times higher pressure compared to what is predicted by {\sc Cloudy} (see Figure~\ref{fig:TPliteratureCompare}). This difference can be ascribed to the fact that emission observations do not probe high enough in the atmosphere to cover the pressures below 10$^{-5}$\,bar \citep{borsa2022_mascara2b,fu2022,yan2022_mascara2b}.

\begin{figure}[ht!]
		\centering
		\includegraphics[width=9cm]{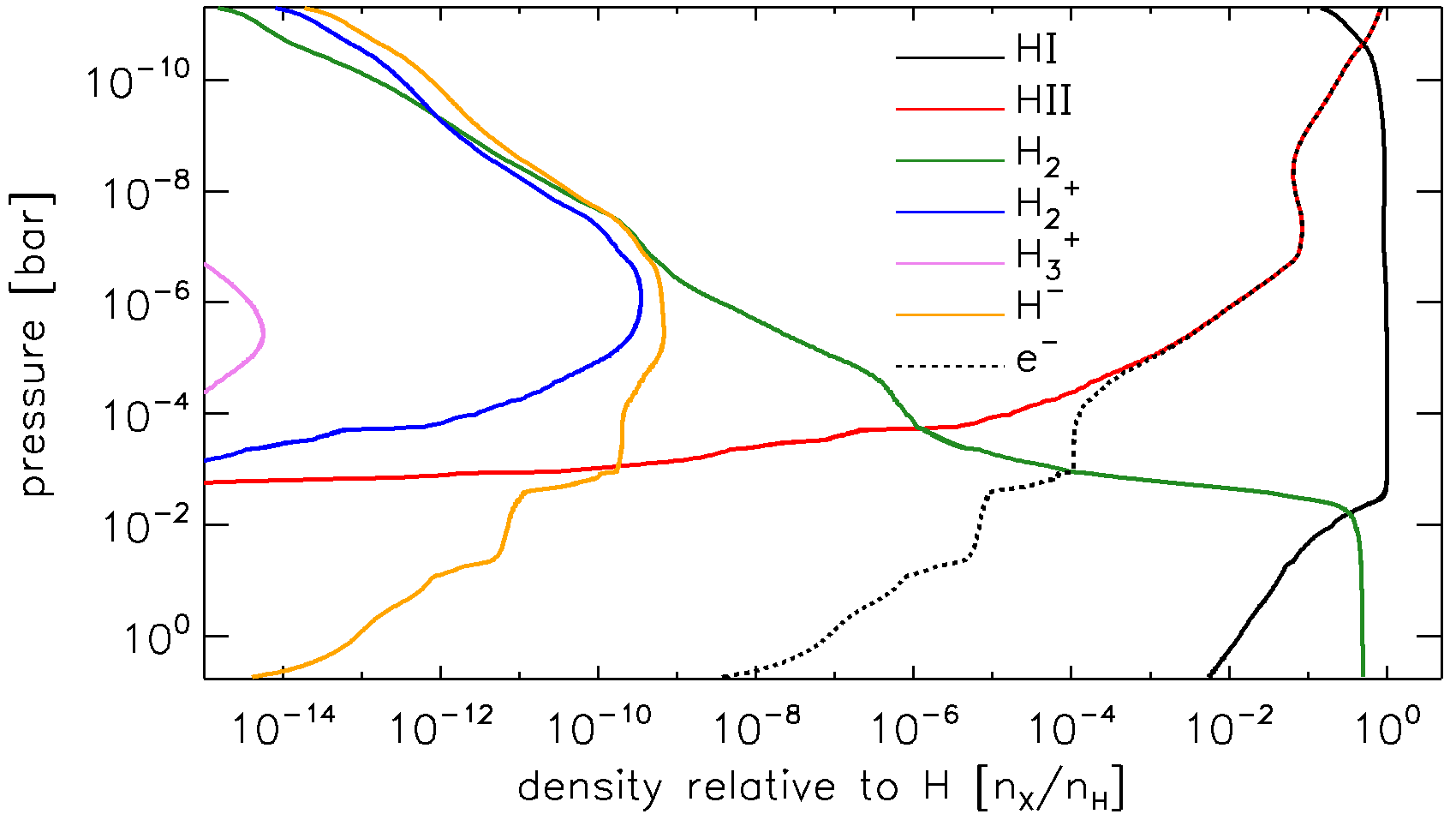}
		\caption{Density relative to the total density of hydrogen for neutral hydrogen (H{\sc i}; black solid), protons (H{\sc ii}; red), molecular hydrogen (H$_2$; dark green), H$_2^+$ (blue), H$_3^+$ (violet), H$^-$ (orange), and electrons (e$^-$; black dashed).} 
  		\label{fig:NLTEchemistryHydrogen}
\end{figure}

To obtain a homogeneous synthetic chemical atmospheric structure, we passed the composite TP profile to {\sc Cloudy} \citep[see][for more details]{fossati2021}. Figure~\ref{fig:NLTEchemistryHydrogen} shows the theoretical density profiles with respect to the total hydrogen density of the main hydrogen-bearing species (neutral hydrogen H{\sc i}, protons H{\sc ii}, H$^-$, molecular hydrogen H$_2$, H$_2^+$, H$_3^+$), plus electrons (e$^-$). The middle and upper atmosphere ($<$10$^{-3}$\,bar) are largely dominated by neutral hydrogen, which is significantly ionised only at the very top, around 5$\times$10$^{-11}$\,bar, namely a thousand times lower pressure than what the same model predicted for KELT-9b \citep{fossati2021}. This is ultimately caused by the fact that KELT-9b is more irradiated by EUV photons as a result of the hotter host star \citep[both stars have only photospheric emission;][]{fossati2018_AstarPlanets} and closer orbital separation. Instead, the lower atmosphere is dominated by H$_2$ and neutral hydrogen, with H$_2$ rapidly decreasing with decreasing pressure below the 10\,mbar level. As expected given that \planet\ is an ultra-hot Jupiter \citep[e.g.][]{arcangeli2018}, H$^-$ is relatively abundant with its density first increasing with decreasing pressure up to the 1\,$\mu$bar level and then decreasing at lower pressures.

\begin{figure}[ht!]
		\centering
		\includegraphics[width=9cm]{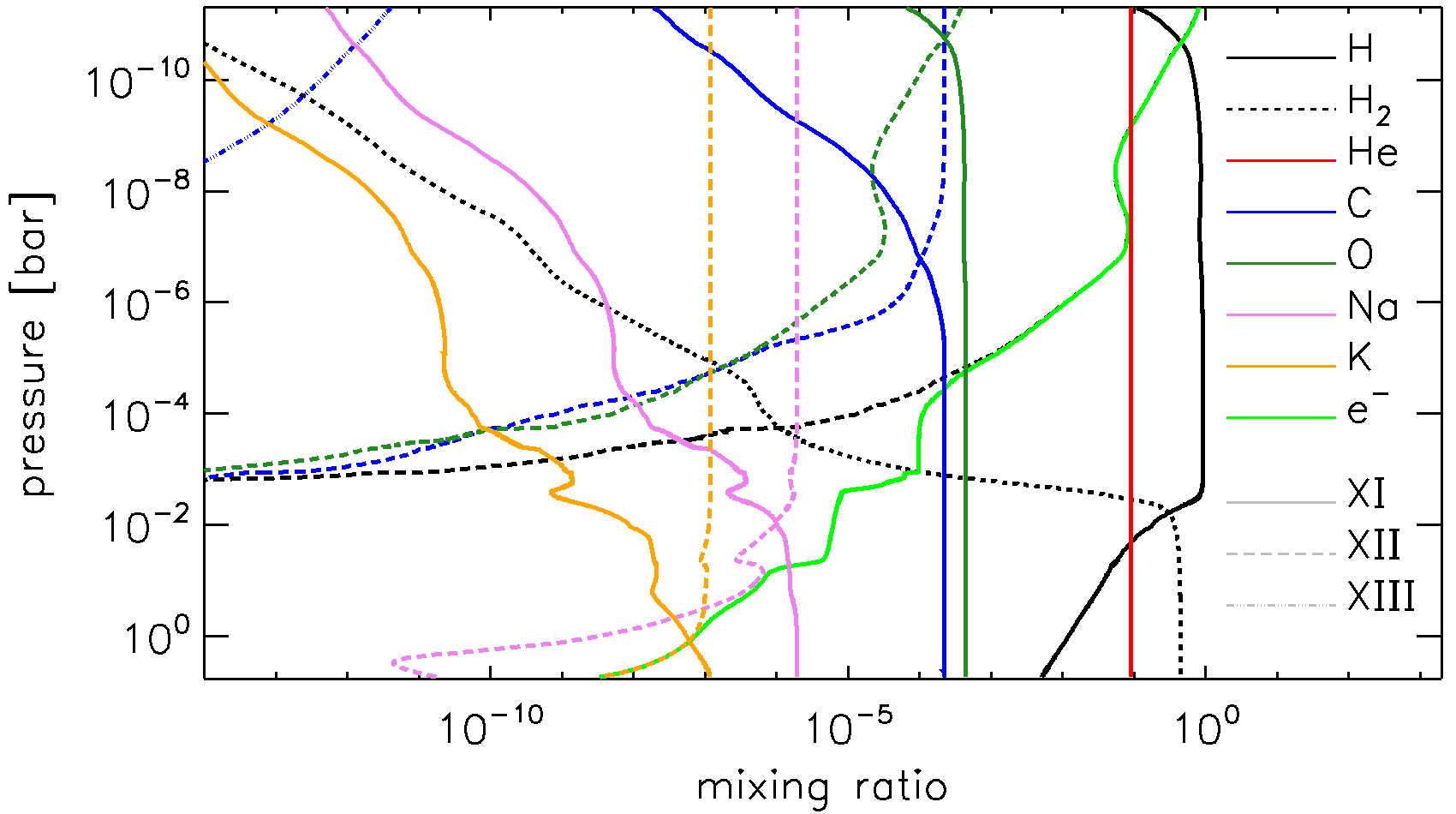}
		\includegraphics[width=9cm]{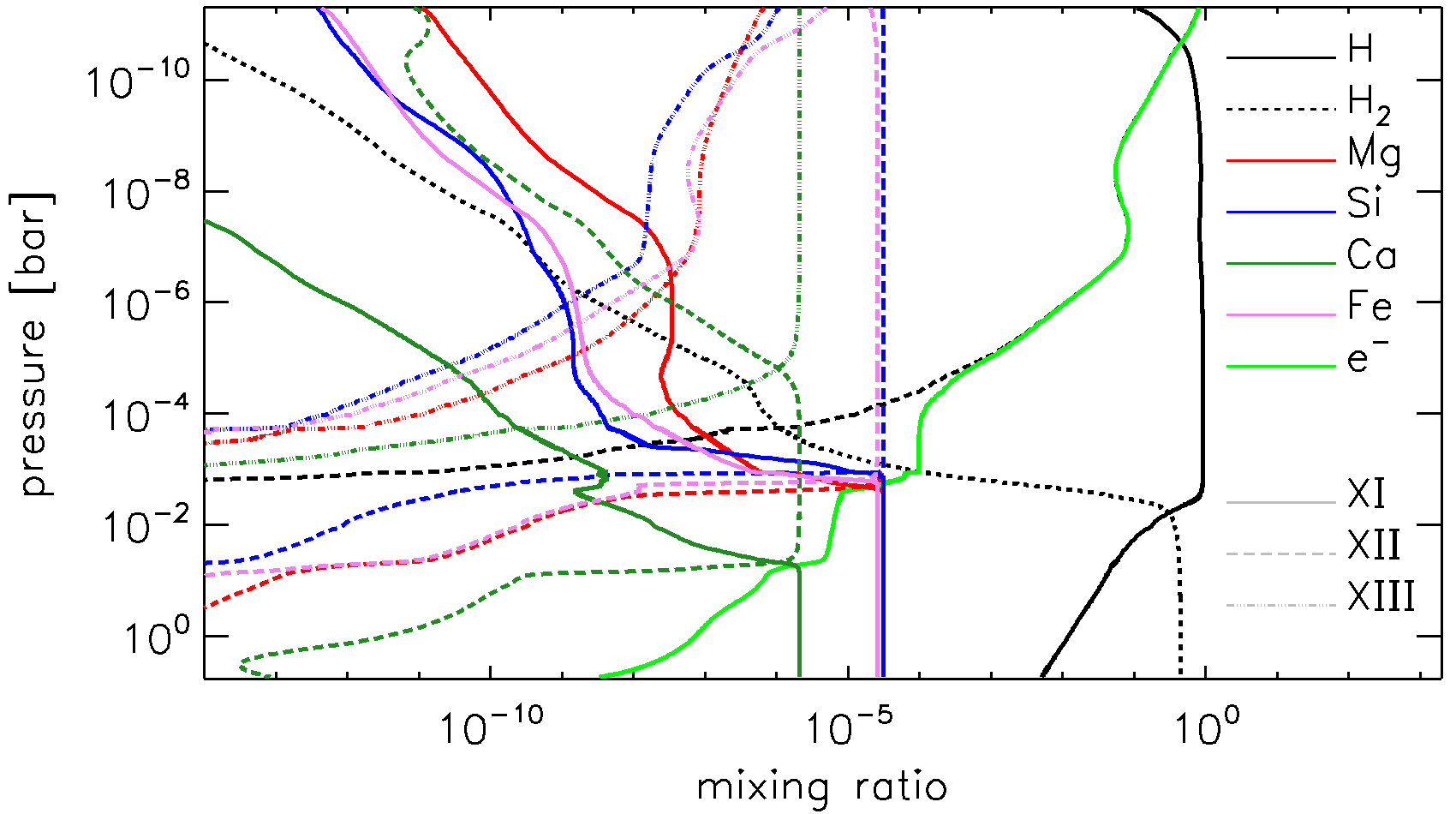}
		\caption{Atmospheric mixing ratios obtained for metals. Top: Mixing ratios for hydrogen (black), H$_2$ (black dotted), He (red), C (blue), O (dark green), Na (violet), K (orange), and electrons (bright green) as a function of atmospheric pressure. Neutral (XI), singly ionised (XII), and doubly ionised (XIII) species are shown as solid, dashed, and dash-dotted lines, respectively. Bottom: same as the top panel, but for Mg (red), Si (blue), Ca (dark green), and Fe (violet). The hydrogen, H$_2$, and e$^-$ mixing ratios are shown in both panels for reference.} 
  		\label{fig:NLTEchemistryMetals}
\end{figure}

Figure~\ref{fig:NLTEchemistryMetals} shows the theoretical mixing ratio as a function of pressure for some of the most relevant elements in terms of abundance and observability. As a consequence of the assumption of solar composition, helium is the second most abundant element throughout. Thanks to its high ionisation energy, oxygen remains in its neutral state almost up to the top of the considered pressure range (similarly to hydrogen), while carbon, which has a slightly lower ionisation energy, starts to ionise significantly at the 1\,nbar level. Sodium and potassium have similar ionisation energies and indeed behave similarly, with the ionisation occurring in the 0.01--1\,bar pressure range. Also as a result of their similar ionisation energies, magnesium, silicon, and iron have comparable behaviours with the singly ionised atoms becoming dominant at the mbar level. This result confirms that Fe{\sc i} and Fe{\sc ii} lines are likely to form at different altitudes in the planetary atmosphere, as suggested by the different velocities and widths of these features detected through the cross-correlation technique \citep{stangret2020,nugroho2020,hoeijmakers2020}. Among those shown in Figure~\ref{fig:NLTEchemistryMetals}, calcium is the only element for which the second ionised species become dominant within the simulated pressure range.

Figure~\ref{fig:TPfrankenstein} shows that in the upper atmosphere the {\sc Cloudy} TP profile is about 3000\,K hotter than predicted by the {\sc helios} model. Remarkably, this difference is about 1000\,K larger than that found for KELT-9b. Following \citet{fossati2021}, we tested whether this difference could be ascribed to NLTE effects by computing an additional TP profile with {\sc Cloudy}, but assuming LTE. We remark that even when enforcing the LTE assumption, {\sc Cloudy} computes the populations of the first two energy levels of hydrogen in NLTE. In the middle and upper atmosphere, the {\sc Cloudy} LTE TP profile, shown in Figure~\ref{fig:TPfrankenstein}, is significantly cooler than the NLTE one and lies close to that computed with {\sc helios}. Furthermore, the similarity between the {\sc Cloudy} LTE and {\sc helios} TP profiles in the middle atmosphere suggests that molecules do not have a significant impact on the shape of the TP profile at pressures lower than 0.1\,mbar.

\begin{figure}[ht!]
		\centering
		\includegraphics[width=9cm]{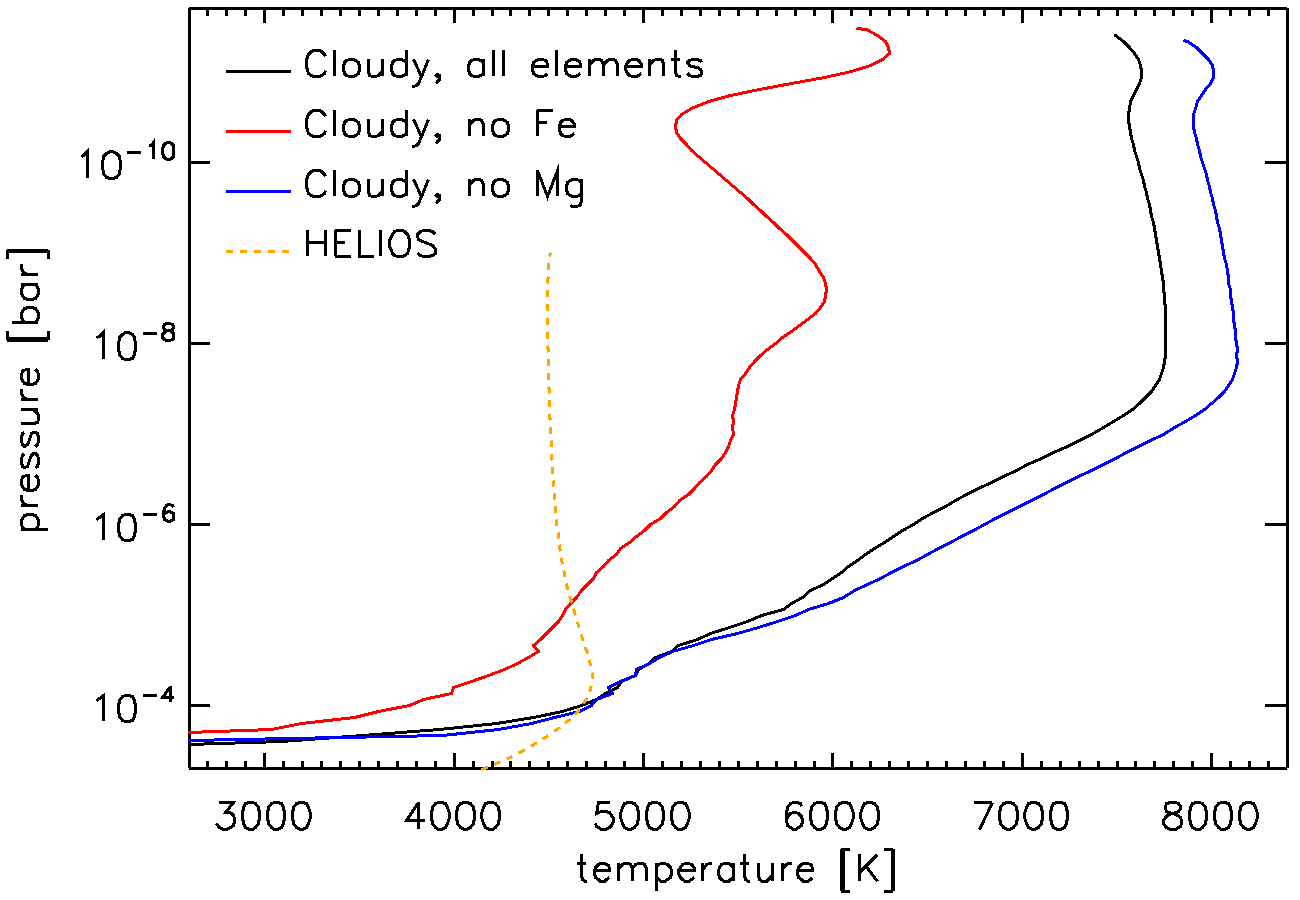}
		\caption{Comparison between the NLTE {\sc Cloudy} TP profiles obtained considering all elements up to Zn (black solid line) and excluding Fe (red solid line) or Mg (blue solid line). The orange dashed line shows the {\sc helios} (i.e. LTE) TP profile for reference.} 
  		\label{fig:NLTEnoFeMg}
\end{figure}

To identify the elements primarily responsible for the difference between the LTE and NLTE TP profiles, we computed {\sc Cloudy} NLTE TP profiles excluding one of the elements at a time, except for H and He that we always kept in each model. Similarly to the case of KELT-9b, we found that Fe and Mg are the elements that most shape the TP profile. In particular, Fe dominates the heating and Mg the cooling in the middle and upper atmosphere (Figure~\ref{fig:NLTEnoFeMg}). Removing Fe leads to a TP profile that in the middle and upper atmosphere is between 1000 and 2000\,K cooler than that obtained including Fe. Instead, excluding Mg leads to an about 300\,K hotter TP profile. The other elements, instead, contribute for less than 50\,K to the heating or cooling in the planetary atmosphere.

\begin{figure}[ht!]
		\centering
		\includegraphics[width=9cm]{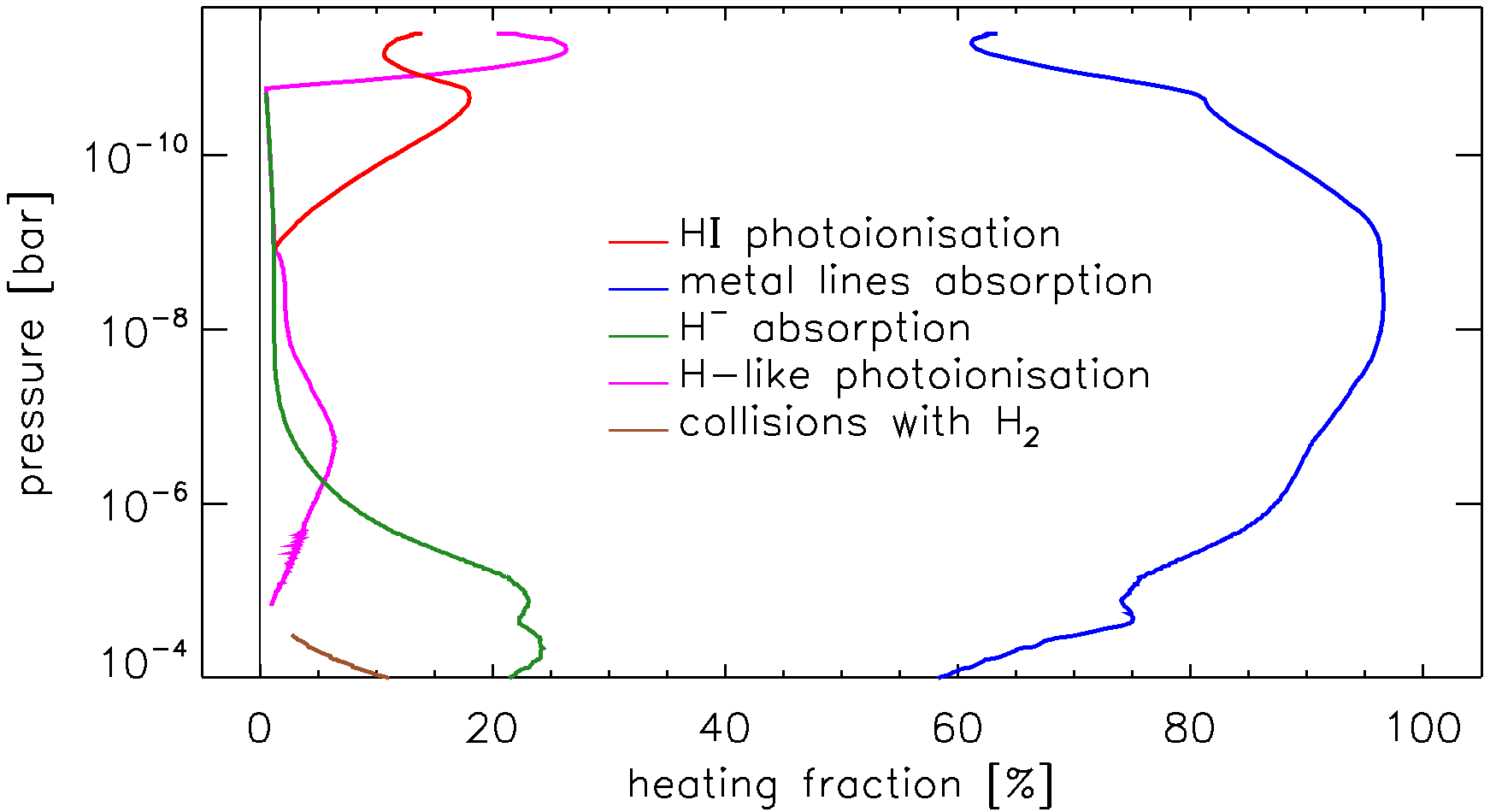}
		\includegraphics[width=9cm]{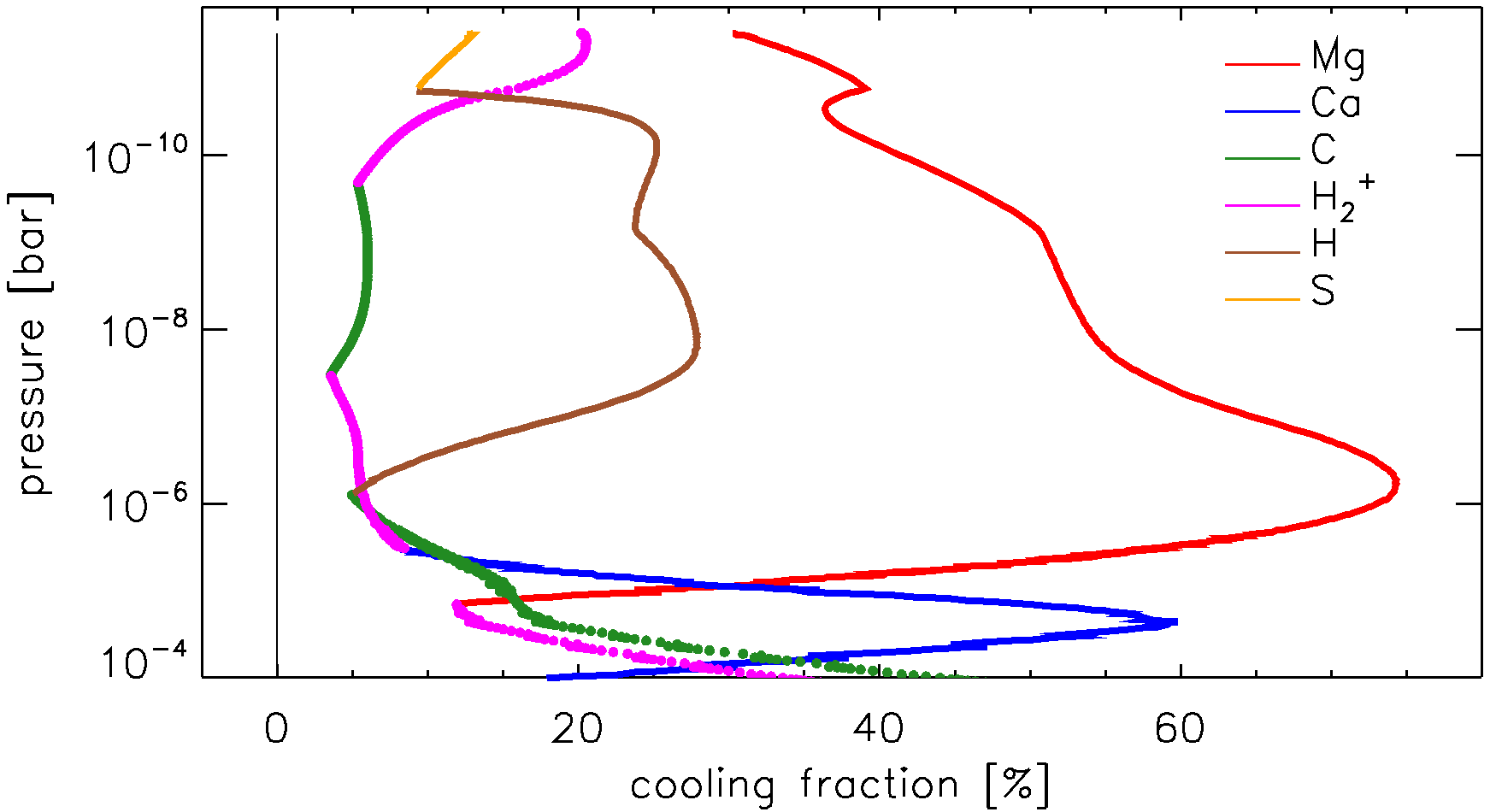}
		\caption{Heating and cooling contributions in the atmosphere of \planet. Top: Contribution to the total heating as a function of pressure. At each pressure bin, the plot shows the three most important heating processes. The main heating processes occurring in the middle and upper atmosphere are hydrogen photoionisation (red; photoionisation of H{\sc i} lying in the ground state), metal line absorption (blue), H$^-$ absorption (green), photoionisation of hydrogenic species (photoionisation of excited H{\sc i}; magenta), and collisions with H$_2$ (brown). Bottom: Contribution to the total cooling as a function of pressure. At each pressure bin, the plot shows the three most important cooling agents.} 
  		\label{fig:heatingCooling}
\end{figure}

Compared to the case of KELT-9b \citep[see Figure~7 of][]{fossati2021}, the TP profile computed excluding Fe shows a significantly smaller temperature increase in the upper atmosphere ($<$10$^{-10}$\,bar), which is most likely due to the weaker EUV emission of \system\ compared to KELT-9. This suggests that metals are primarily responsible for the heating and cooling in the middle and upper atmosphere. To confirm this, we extracted from the {\sc Cloudy} run the three main heating and cooling agents and display them in Figure~\ref{fig:heatingCooling}. We find that metal line absorption (particularly of Fe; Figure~\ref{fig:NLTEnoFeMg}) is the main heating mechanism throughout the entire middle and upper atmosphere, with photoionisation playing a secondary role, while numerous species contribute to the cooling of the middle and upper atmosphere, but Mg is by far the most important atmospheric coolant. 

\begin{figure}[ht!]
		\centering
		\includegraphics[width=9cm]{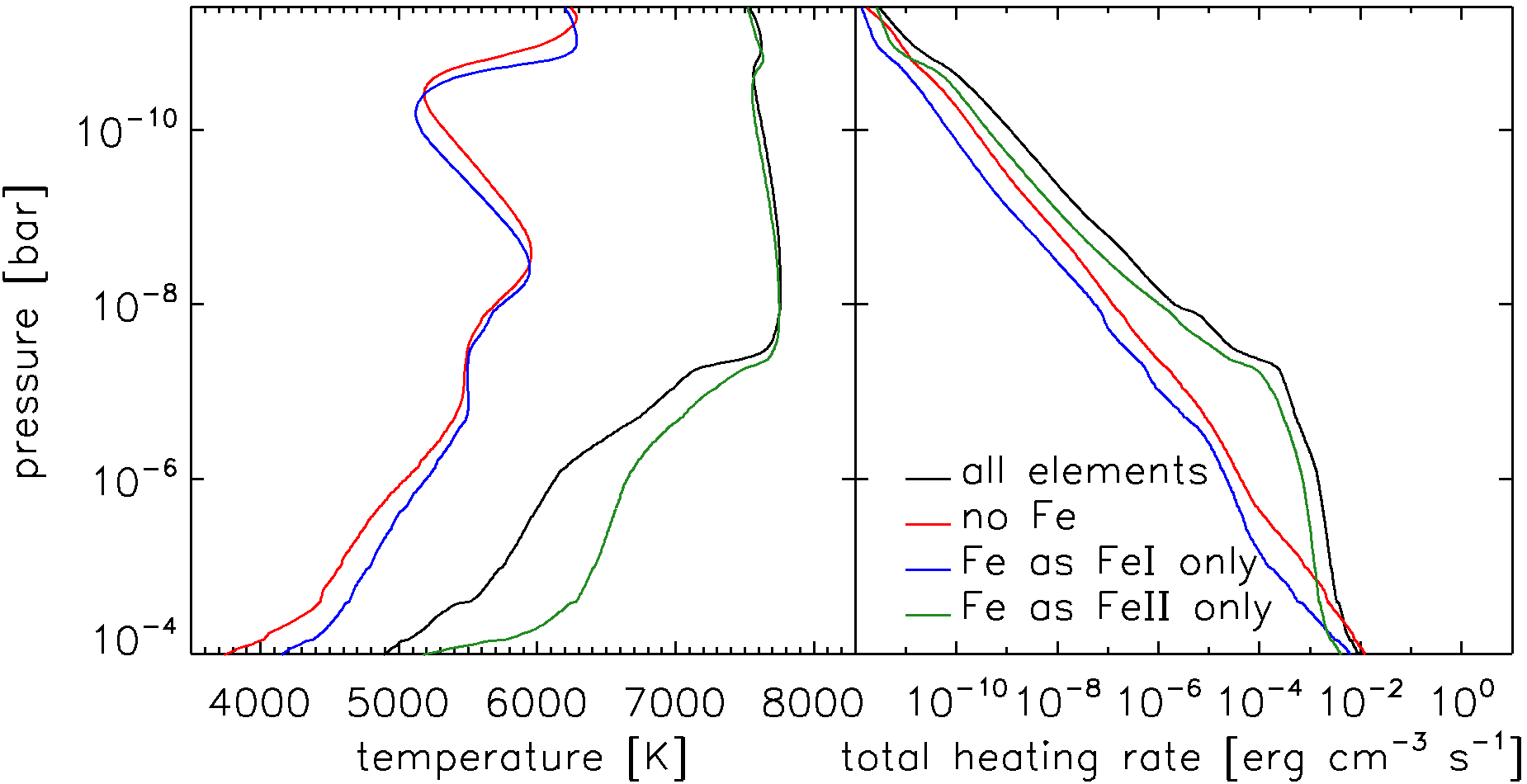}
		\includegraphics[width=9cm]{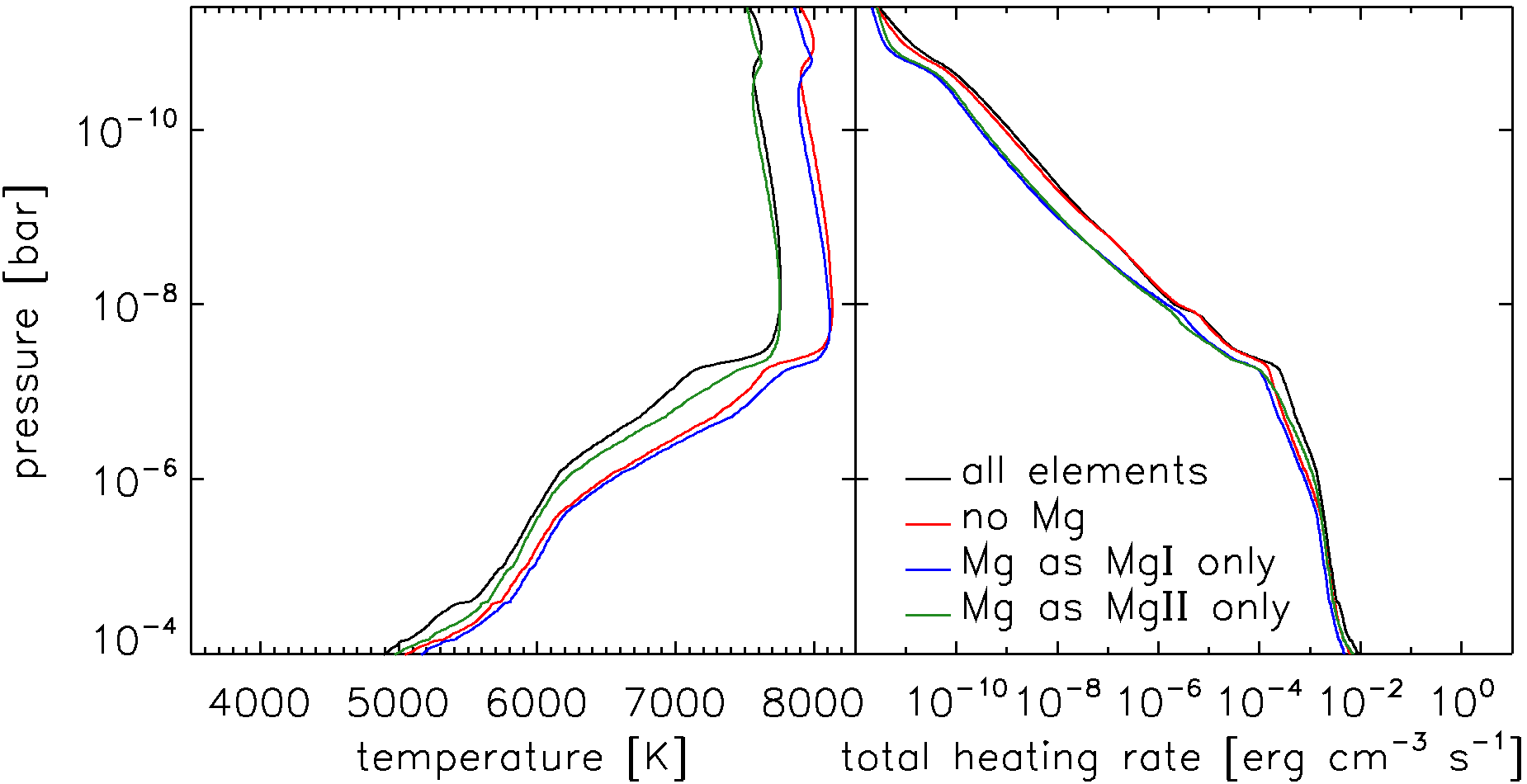}
		\caption{Atmospheric structure and heating rate obtained from isolating the impacts of Fe and Mg ions. Top: Temperature (left) and total heating rate (right) as a function of pressure obtained accounting for all elements (black; same as in Figures~\ref{fig:TPfrankenstein} and \ref{fig:NLTEnoFeMg}), considering that in the planetary atmosphere Fe exists only in the form of Fe{\sc i} (blue), only in the form of Fe{\sc ii} (green), or is absent (i.e. no Fe; same as Figure~\ref{fig:NLTEnoFeMg}; red). Bottom: Same as top, but for Mg.} 
  		\label{fig:fe1fe2mg1mg2}
\end{figure}

To deepen our understanding of the roles played by Mg and Fe in shaping the TP profile, we took the output obtained after the last NLTE CfE iteration and used it as input for a further {\sc Cloudy} run, but considering only Fe{\sc i}/Mg{\sc i} (i.e. Fe{\sc i}/Mg{\sc i} is not allowed to ionise) or Fe{\sc ii}/Mg{\sc ii} (i.e. Fe{\sc ii}/Mg{\sc ii} is not allowed to ionise or recombine), fixing the Fe{\sc i}/Mg{\sc i} or Fe{\sc ii}/Mg{\sc ii} density profile to that obtained accounting for all elements and ions. Figure~\ref{fig:fe1fe2mg1mg2} shows the temperature and the total heating rate as a function of pressure obtained in all these cases and it clearly indicates that most of the heating is caused by Fe{\sc ii}, while most of the cooling is caused by Mg{\sc ii}. Therefore, the combined impact of NLTE effects on the level populations of Fe and Mg (see Figures~\ref{fig:dep.coeff.Fe1}, \ref{fig:dep.coeff.Fe2}, \ref{fig:dep.coeff.Mg1}, \ref{fig:dep.coeff.Mg2}) leads to a general temperature increase in the middle and upper atmosphere when compared to the LTE profile.

Interestingly, in the case of \planet, the inclusion of NLTE effects leads to a larger temperature increase in the middle and upper atmosphere compared to KELT-9b, although the latter is hotter and orbits a hotter star. This can be understood by looking at the departure coefficients that are defined as
\begin{equation}
    b = \frac{n_{\rm NLTE}}{n_{\rm LTE}}\,,
\end{equation}
where $n_{\rm NLTE}$ and $n_{\rm LTE}$ are the densities of a given atom lying in a certain level in NLTE and LTE, respectively, with the $n_{\rm LTE}$ profiles obtained through the Boltzmann equation. A comparison of the Mg and Fe departure coefficients computed by {\sc Cloudy} for the two planets in the middle and upper atmosphere (see Figures~\ref{fig:dep.coeff.Fe1} to \ref{fig:dep.coeff.Mg2} for \planet\ and Figures~C.1 to C.5 of \citealt{fossati2021}) indicates that the $b$ profiles obtained for Fe{\sc i} and Mg for \planet\ are on average significantly smaller (larger in modulus) than those obtained for KELT-9b. Therefore, the middle and upper atmosphere of \planet\ has less Mg{\sc ii} available for driving the cooling, and thus the larger difference between the LTE and NLTE TP profiles obtained for \planet\ compared to KELT-9b might be ascribed to a lack of cooling rather than to an increase in heating. 
\subsection{Hydrogen Balmer lines}\label{sec:balmer}
%
\begin{figure*}[ht!]
		\centering
		\includegraphics[width=18cm]{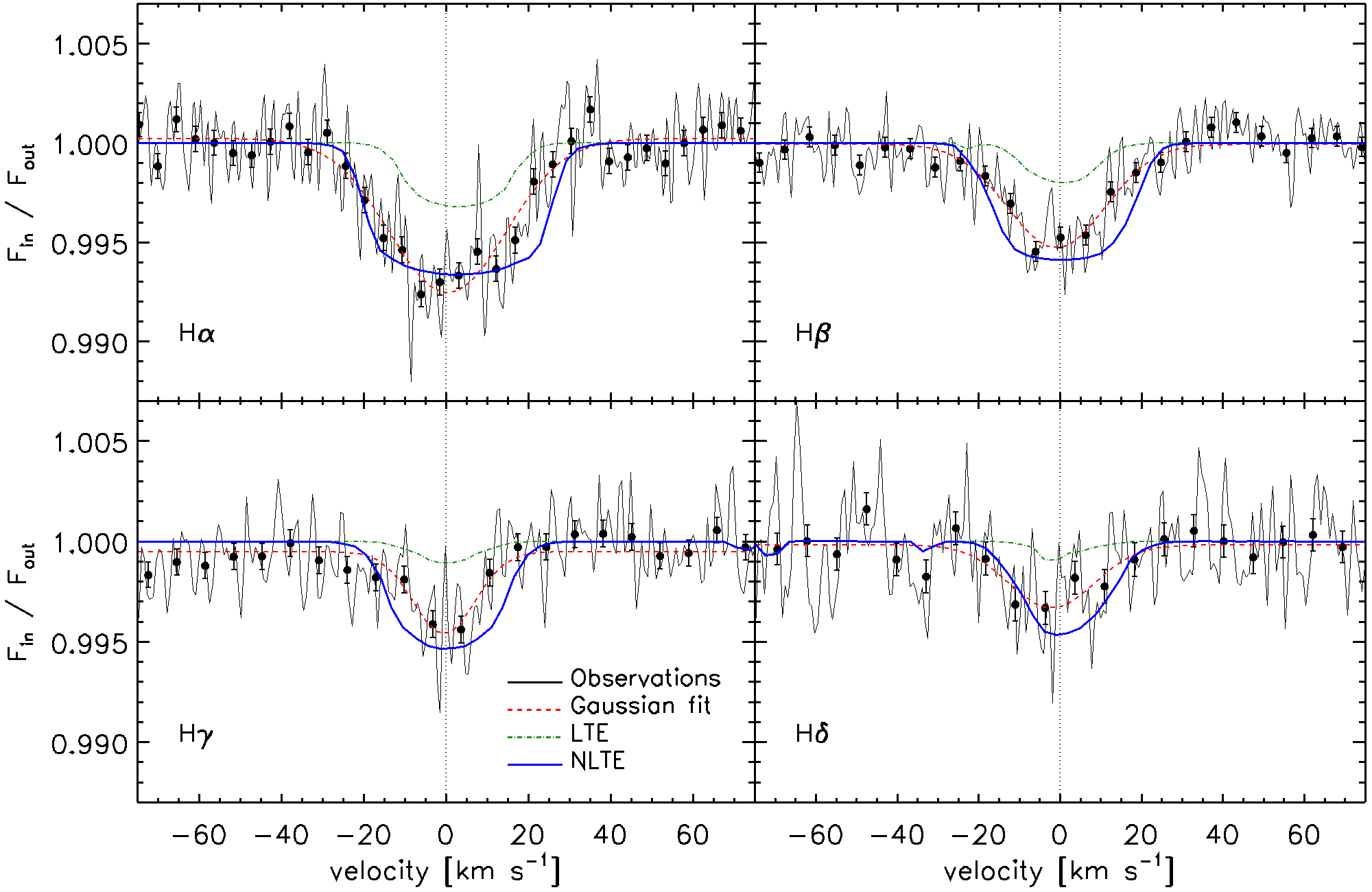}
		\caption{Comparison between observed and synthetic H$\alpha$ (top-left), H$\beta$ (top-right), H$\gamma$ (bottom-left), and H$\delta$ (bottom-right) line profiles. The black solid line shows the observations, while to guide the eye the black dots show the observed profiles rebinned to about 4.5\,km\,s$^{-1}$. The red dashed line indicates the Gaussian fit to the observations. The solid blue and dash-dotted green lines show the {\sc Cloudy} synthetic line profiles computed in NLTE and LTE, respectively. The central wavelengths of the Balmer lines (in vacuum) used to convert the wavelengths into velocities are 6564.60\,\AA\ for H$\alpha$, 4862.71\,\AA\ for H$\beta$, 4341.69\,\AA\ for H$\gamma$, and 4102.892\,\AA\ for H$\delta$.}
  		\label{fig:Balmer_lines_comparison}
\end{figure*}

We detect H$\alpha$, H$\beta$, and H$\gamma$ with an absorption depth of 0.789$\pm$0.034\,\% ($\approx$23$\sigma$), 0.517$\pm$0.034\,\% ($\approx$15$\sigma$), and 0.394$\pm$0.057\,\% ($\approx$7$\sigma$), respectively. We also report the detection of H$\delta$ at almost 4$\sigma$ with an absorption depth of 0.272$\pm$0.070\,\%. We note that \planet\ is only the second planet, after KELT-9b, for which H$\delta$ has been detected \citep{Wyttenbach_2020}. The obtained line depths translate into planetary radii of about 1.25\,R$_{\rm p}$, 1.17\,R$_{\rm p}$, 1.13\,R$_{\rm p}$, and 1.09\,R$_{\rm p}$ for H$\alpha$, H$\beta$, H$\gamma$, and H$\delta$, respectively, under the assumption of a symmetrically distributed atmosphere. The line profiles obtained from the HARPS-N observations are shown in Figure~\ref{fig:Balmer_lines_comparison}.

H$\alpha$, H$\beta$, and marginally H$\gamma$ have already been identified in the atmosphere of \planet\ by \citet{Casasayas_2019} from the analysis of the transits collected during nights 1 to 3, plus an additional transit taken with CARMENES. We note that in the present work we do not use the CARMENES data to avoid any possible instrumental systematics, since one more transit would not add much to our HARPS-N six transits dataset. Our results are in agreement within 1$\sigma$ with theirs (see Figure~\ref{fig:comparisonWcasasayas}), despite the fact that \citet{Casasayas_2019} optimised the radial velocity semi-amplitude of the planet K$_{\rm p}$ for each transit with a Markov Chain Monte Carlo (MCMC) analysis, while we used the same theoretical K$_{\rm p}$ (Table~\ref{tab:systemParameters}) for all six datasets. Planetary atmospheric variations during transit could mimic K$_{\rm p}$ variations. When the part of the atmosphere in view during transit changes (because of the orbital movement and planetary rotation), regions of the atmosphere with different dynamics could become visible, which could lead to discrepancies between the actual planetary velocity and the system of velocity of the atmosphere itself. This phenomenon has been clearly observed in the case of WASP-76b \citep{ehrenreich2020,kesseli2021}. The possibility that this could happen also for \planet\ has been mentioned by \citet{rainer2021}, but their analysis of Fe cross-correlation functions did not lead to a statistically significant detection of this phenomenon.

Our approach of using the same K$_{\rm p}$ for all the transits is justified by the fact that we aim to obtain an average planetary line profile to compare with theoretical simulations, which do not account for any possible transit-by-transit variability. To support this approach, we looked for any variation of the H$\alpha$ and H$\beta$ absorption depths, center, and full width half maximum (FWHM) among the six transits when using the same K$_{\rm p}$ value. The absorption depth, center, and FWHM values were derived by performing a Gaussian fit of average transmission spectrum given by each transit with a linear regression\footnote{We specified the use of an intrinsic scatter in the fit model to take into account the noise present in the data.} using {\fontfamily{pcr}\selectfont CPNest}\footnote{\url{https://github.com/johnveitch/cpnest}} \citep{Del_Pozzo_2022}, which is a \textsc{python} implementation of the nested sampling algorithm \citep{Skilling_2006}. By looking at the transmission spectra individually, we could find neither significant variations beyond the 2.1$\sigma$ level nor correspondences in the pattern shown by the absorption depths, center, and FWHM between the two lines (see Figure~\ref{fig:Ha_Hb_variability} and Table~\ref{tab:Ha_Hb_variability}). A further check of our results has been performed independently by using the Sloppy pipeline \citep{Sicilia2022}, obtaining fully compatible results.

\begin{figure}[h]
    \centering
    \includegraphics[width=0.50\textwidth]{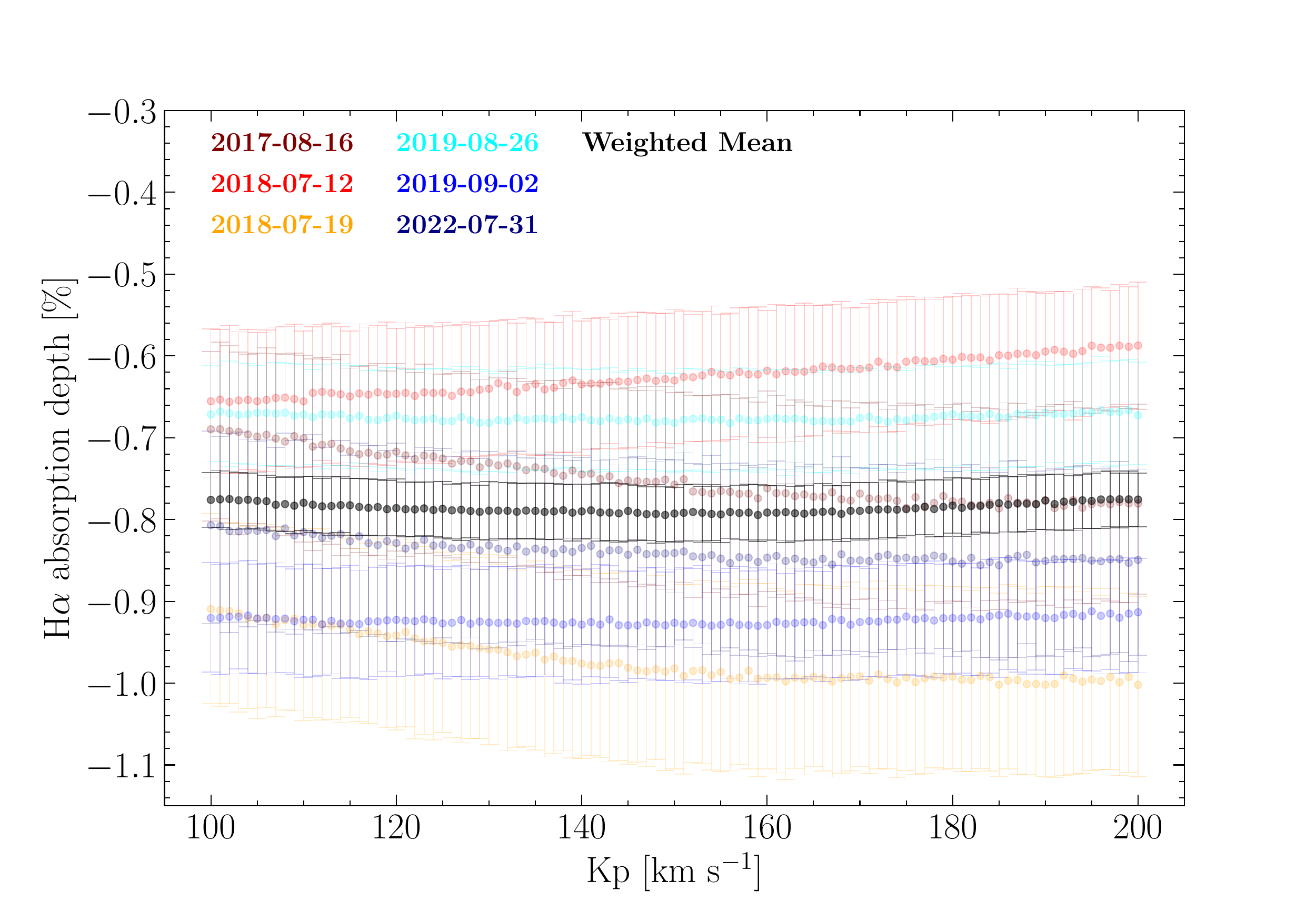}
    \caption{Variation of the H$\alpha$ absorption depth as a function of K$_{\rm p}$.}
    \label{fig:Kp_fit}
\end{figure}

\begin{figure}[h]
    \centering
    \includegraphics[width=0.50\textwidth]{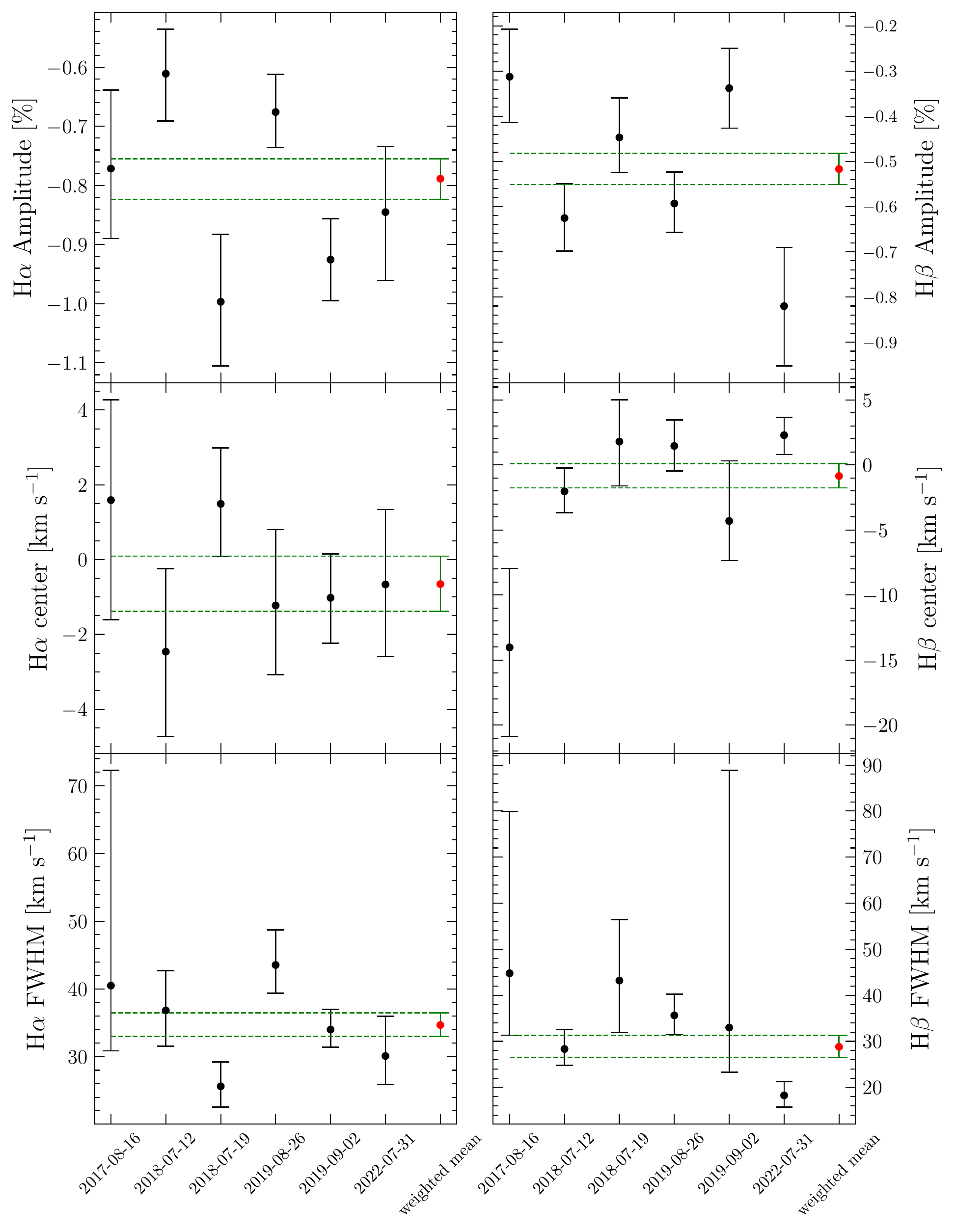}
    \caption{Amplitude, center, and FWHM values of the Gaussian fit performed on the Balmer H$\alpha$ and H$\beta$ lines for the average transmission spectrum obtained within each night. No significant variations are present among the six transits, with the largest discrepancies in the absorption depth shown during night 2 (H$\alpha$, 1.8$\sigma$) and night 6 (H$\beta$, 2.1$\sigma$). 
    The full set of values is presented in Table~\ref{tab:Ha_Hb_variability}.}
    \label{fig:Ha_Hb_variability}
\end{figure}

To further validate the use of the same K$_{\rm p}$, we compared the depths of the H$\alpha$ line with varying K$_{\rm p}$. Within a reasonable range of K$_{\rm p}$ values (Figure~\ref{fig:Kp_fit}), the depth of the average line is compatible within error bars. This fact is a consequence that the width of the line profile is quite large, and the deviation of the average line depth value becomes significant only for K$_{\rm p}$ values $\lesssim$50\,km\,s$^{-1}$ or $\gtrsim$280\,km\,s$^{-1}$.
%
\section{Discussion}\label{sec:discussion}
\subsection{Comparison with the observations}\label{sec:discussion:comparison}
To compare the observations with the modelling results and further explore the impact of NLTE effects, we used {\sc Cloudy} to compute synthetic transmission spectra in both LTE and NLTE. To this end, we followed the procedure described by \citet{young2020} and \citet{fossati2020}, dividing the entire atmosphere into 100 layers equally spaced in $\log{p}$ and considering a spectral resolution of 100\,000. The theoretical NLTE transmission spectrum has been computed employing the composite TP profile shown in Figure~\ref{fig:TPfrankenstein} and enabling NLTE throughout the entire transmission spectrum calculation. Instead, the theoretical LTE transmission spectrum has been computed employing the LTE TP profile shown in Figure~\ref{fig:TPfrankenstein}, further joined together with the {\sc helios} profile as done to obtain the composite NLTE TP profile (see Section~\ref{sec:TP_chemistry}), and imposing the LTE assumption also for the {\sc Cloudy} transmission spectrum calculation. 

The {\sc Cloudy} calculations presented above have considered the sub-stellar point, but the computation of transmission spectra implies a different geometry (i.e. from emission geometry to transmission geometry), which then calls for a new calibration of the reference pressure $p_0$ corresponding to the reference radius $R_0$, that is the measured transit radius $R_{\rm p}$. To obtain a more robust calibration, particularly when comparing the model with the observations, we employed a procedure different from that followed by \citet{fossati2021}. We computed several transmission spectra by setting $p_0$ at different pressure levels in the 0.001--0.1\,bar pressure range. We then convolved each synthetic transmission spectrum with the MASCARA instrument bandpass \citep{talens2017} to obtain $R_{\rm p}/R_{\rm s}$ as a function of $p_0$. Finally, for the comparison with the observations we considered the synthetic transmission spectrum computed with the $p_0$ value leading to the transmission spectrum best matching the observed $R_{\rm p}/R_{\rm s}$ of 0.1175 \citep{talens2018}. For the LTE and NLTE theoretical transmission spectra, we find best-fitting reference pressure values of 0.001\,bar and 0.002\,bar, respectively.

To explore the possible error introduced by an inaccurate $p_0$--$R_{\rm p}$ calibration, Figure~\ref{fig:comparisonTSdifferentP0} shows theoretical NLTE transmission spectra in the region of the H$\alpha$ line computed by setting the transit radius at the reference pressure of 0.1, 0.01, and 0.001\,bar, both before and after normalisation. As expected, the non-normalised synthetic transmission spectra have different continua, with the difference decreasing with increasing $p_0$, because the atmosphere becomes more and more compact with increasing pressure. Following normalisation to the continuum, we find that the actual choice of reference pressure value has no significant impact on the line absorption depth, particularly when compared with the noise level of the observations. Together with the top panel of Figure~\ref{fig:test_p0_molecules}, this shows that the choice of reference pressure, in the computation of both the theoretical TP profile and transmission spectrum, has no significant impact on the modelling results.

Figure~\ref{fig:Balmer_lines_comparison} shows the comparison between the observed and the synthetic H$\alpha$, H$\beta$, H$\gamma$, and H$\delta$ line profiles (for context, Figure~\ref{fig:dep.coeff.H1} shows the profiles of the H{\sc i} departure coefficients). As also shown by the $\chi^2$ and reduced $\chi^2$ ($\chi^2_{\rm red}$) values listed in Table~\ref{tab:chi2}, the NLTE synthetic spectra are a good match to the data, while the synthetic LTE line profiles are systematically weaker than the observations. This is mostly due to the difference in underlying TP profiles employed to compute the theoretical transmission spectra, particularly because, even in LTE, {\sc Cloudy} computes the hydrogen population of the first two energy levels accounting for NLTE effects \citep[see][]{fossati2021}. 

The NLTE synthetic spectra slightly overestimate the line absorption depth compared to the observations. Furthermore, the predicted line profiles are significantly broader and rounder than the observed ones, which are instead more triangular. These differences are probably caused by the fact that we consider only the day-side temperature profile, and thus assume that the TP profile computed at the sub-stellar point is valid across the entire planet (i.e. both day and night sides). This leads us to overestimate the gas temperature at the terminator region and thus the line absorption depth and broadening, because the night side has a lower temperature. Without this assumption the lines forming on the night side of the terminator region would not be as strong and broad, and would mostly contribute to the shape of the line core, making the synthetic lines weaker and more triangular, similar to the observed ones. Furthermore, a lower temperature would also lead to a less extended atmosphere, and thus decreased line absorption depth. The shape of the theoretical H$\delta$ line appears to be asymmetric, which is probably due to contamination by a nearby Fe{\sc i} line.
\begin{table}[]
\renewcommand{\arraystretch}{1.3}
\caption{$\chi^2$ and reduced $\chi^2$ ($\chi^2_{\rm red}$) values obtained from the comparison between the synthetic LTE and NLTE transmission spectra with the non-binned observations. The last column lists the number of degrees of freedom (DOF).}
\begin{tabular}{l|cc|cc|r}
\hline
\hline
Line & \multicolumn{2}{c|}{NLTE} & \multicolumn{2}{c|}{LTE} & DOF \\
     & $\chi^2$ & $\chi^2_{\rm red}$ & $\chi^2$ & $\chi^2_{\rm red}$ \\
\hline
H$\alpha$ & 159.80 & 1.23 & 412.35 & 3.17 & 130 \\
H$\beta$  & 121.70 & 1.27 & 246.41 & 2.57 &  96 \\
H$\gamma$ &  92.34 & 1.07 & 123.65 & 1.44 &  86 \\
H$\delta$ &  51.19 & 0.63 &  73.46 & 0.91 &  81 \\
\hline
\end{tabular}
\label{tab:chi2}
\end{table}
%
\subsection{Ultraviolet to infrared transmission spectrum}\label{sec:transmission_spectrum}
\begin{figure}[ht!]
		\centering
		\includegraphics[width=9cm]{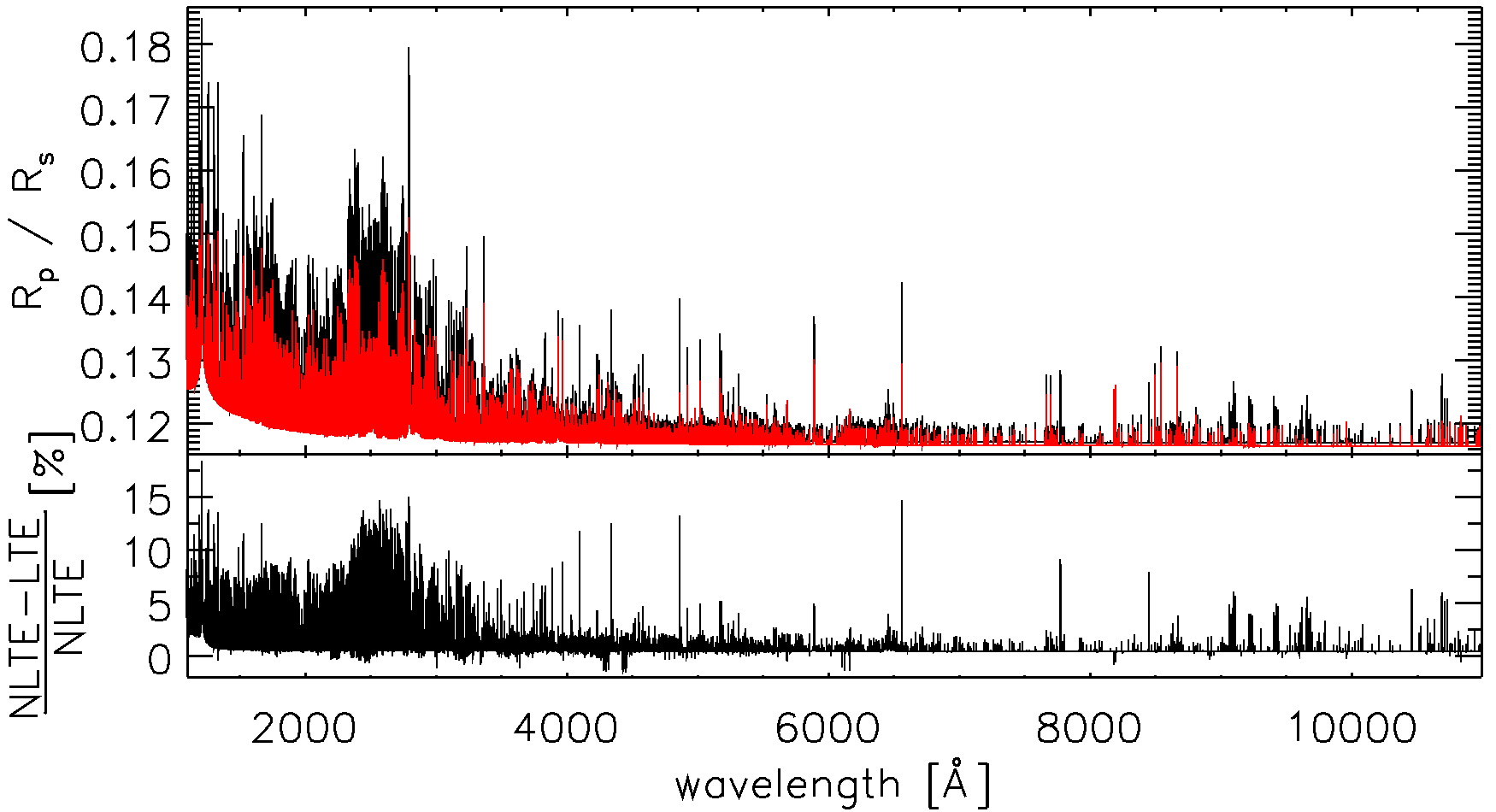}
		\includegraphics[width=9cm]{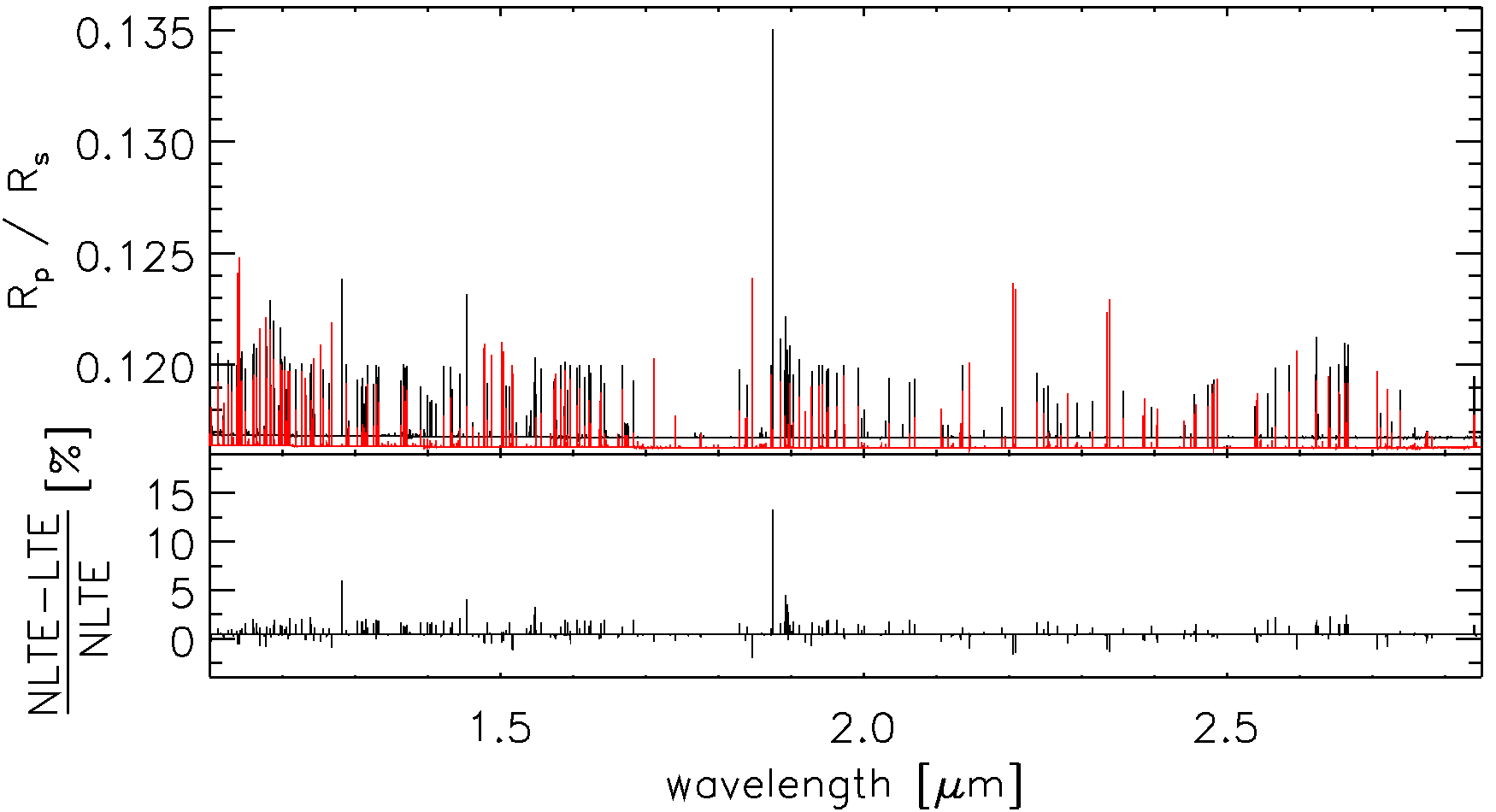}
		\caption{Comparison between the theoretical LTE (red) and NLTE (black) transmission spectra. The top plot covers the UV and optical range, while the bottom plot covers the infrared band. The transmission spectra have been computed by applying a spectral resolution of 100\,000. Within each plot, the bottom panel shows the deviation from LTE (in \%).}
  		\label{fig:transmission_spectra}
\end{figure}

To enable future comparisons with observations obtained in a broad range of wavelengths and by different facilities, we computed the synthetic NLTE and LTE transmission spectra ranging from the far-UV to the mid-infrared (i.e. from 912\,\AA\ to 2.85\,$\mu$m). Figure~\ref{fig:transmission_spectra} shows the LTE and NLTE synthetic transmission spectra across the whole considered wavelength range. Similar plots, but zooming into specific wavelength ranges for better visibility can be found in Appendix (Figures~\ref{fig:transmission_spectra_1100-1550}, \ref{fig:transmission_spectra_1500-2350}, \ref{fig:transmission_spectra_2300-3050}, \ref{fig:transmission_spectra_3000-4050}, \ref{fig:transmission_spectra_4000-6100}, \ref{fig:transmission_spectra_6000-11000}). For completeness and easier comparison with observations, we show in Figure~\ref{fig:NLTEvsLTE_transit_depth_difference} the transit depth difference between the theoretical LTE and NLTE transmission spectra.

Across the simulated wavelength range, the synthetic LTE transmission spectrum tends to underestimate the absorption, although there is a small sample of lines, mostly belonging to Ca{\sc i} and Fe-peak elements, where the LTE assumption overestimates the absorption. As for KELT-9b \citep{fossati2021}, we find the strongest deviation from LTE in the UV band, with the deviation from LTE lying mostly between 5 and 15\%, with a peak of about 17.5\% for the Ly$\alpha$ line. On average, in the optical band the deviation from LTE is smaller and lies below 5\%, but there are exceptions, such as the hydrogen Balmer lines where the deviation from LTE reaches between 10 and 15\%. As in the case of KELT-9b \citep{fossati2021,borsa2022_kelt9b_OI}, the O{\sc i} triplet at about 7780\,\AA\ shows a prominent deviation from LTE. However, in relation to the H$\alpha$ line, the absorption depth of the O{\sc i} triplet is smaller in this case compared to KELT-9b, which suggests that it might be harder to detect these lines in the atmosphere of \planet. The infrared band is dominated by the Paschen\,$\alpha$ line, for which we find a deviation from LTE of about 13\%. The predicted absorption depth of this line is comparable to that of the H$\gamma$ and H$\delta$ lines and might thus be detectable.

As described by \citet{fossati2021}, the strong difference between the theoretical LTE and NLTE transmission spectra, particularly in the UV wavelength range, is due to the difference in underlying TP profiles. Most of the spectral lines lying in the UV belong to ionised species, which are more abundant in the NLTE model as a consequence of the higher temperature of the NLTE TP profile compared to the LTE TP profile, particularly in the line forming region. Furthermore, the hotter temperature of the NLTE model leads to a higher pressure scale height compared to the LTE model, which also affects the absorption depth of the lines in the transmission spectrum.
\subsection{Upper atmosphere}\label{sec:discussion:escape}
\begin{figure}[ht!]
		\centering
		\includegraphics[width=9cm]{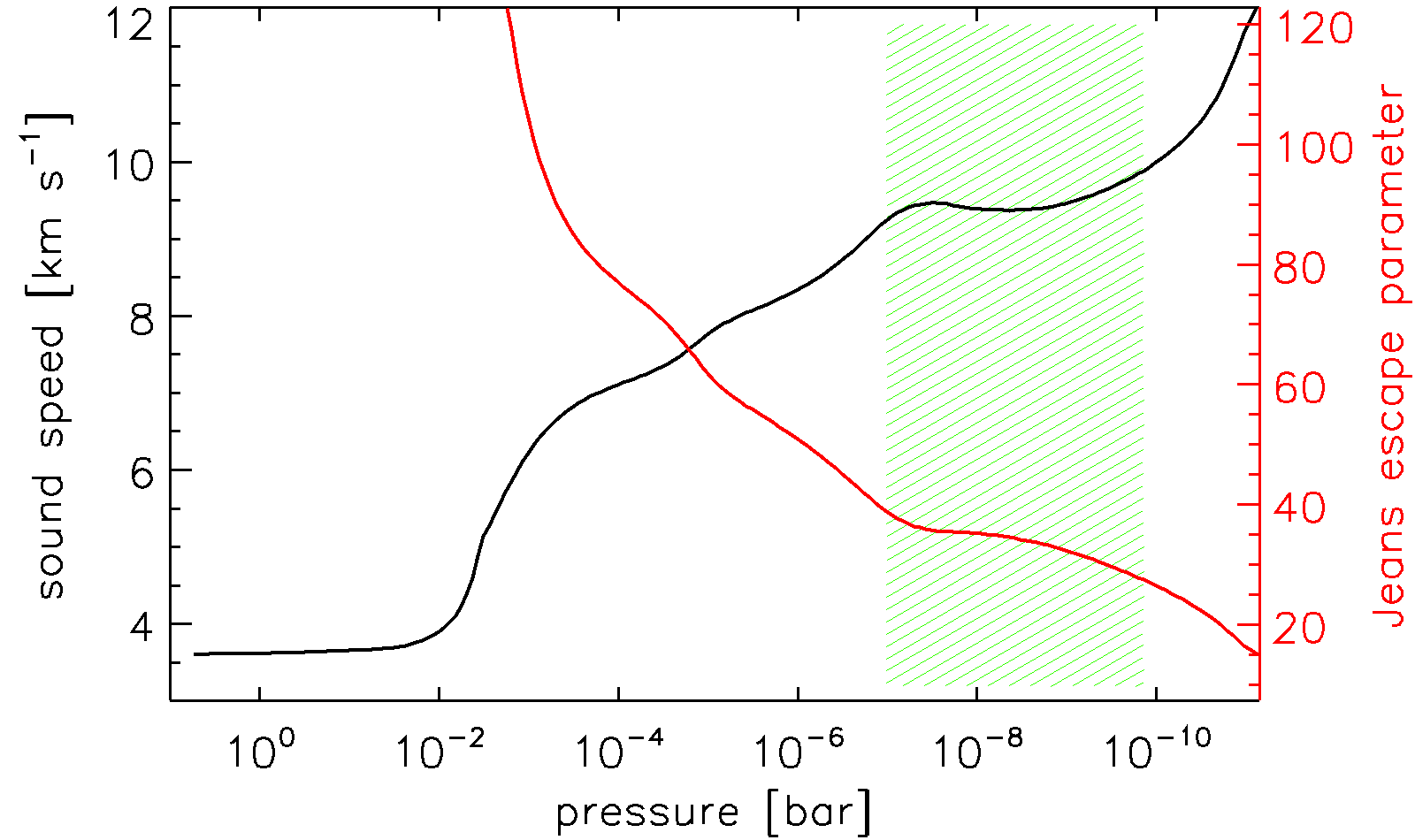}
		\caption{Atmospheric sound speed (black; left y-axis) and Jeans escape parameter (red; right y-axis) profiles as a function of pressure computed on the basis of the composite TP profile. The hatched area indicates the H$\alpha$ line formation region.}
  		\label{fig:Cs_lambda}
\end{figure}

{\sc Cloudy} is a hydrostatic code and thus the fact that the composite theoretical TP profile produces a good fit to the observed hydrogen Balmer lines (see Figure~\ref{fig:Balmer_lines_comparison}) implies that the atmosphere, at least in the line formation region and underneath it, is not in the hydrodynamic escape regime. In Figure~\ref{fig:Cs_lambda}, we plot the theoretical sound speed ($C_{\rm s}$) and Jeans escape parameter as a function of pressure, where the latter has been calculated by using Equation~(7) of \citet[][see also \citealt{volkov2011}]{fossati2017} for zero bulk flow velocity. This version of the Jeans escape parameter is based on the gravitational potential difference between a point in the atmosphere and the Roche lobe, which lies at about 9.5\,$R_{\rm J}$ (i.e. about 5.2\,$R_{\rm p}$), and thus the Jeans escape parameter goes to zero at the Roche lobe. The Jeans escape parameter is well above 20 in the line formation region and thus any outflow in this region of the atmosphere is subsonic, explaining why the density profiles are close to hydrostatic \citep[e.g.][]{koskinen2013b}. In agreement with \citet{Casasayas_2019}, we conclude that the Balmer lines do not directly constrain atmospheric escape and the planetary mass-loss rate.
%
\subsection{Impact of planetary mass}\label{sec:discussion:Mp}
\begin{figure}[ht!]
		\centering
		\includegraphics[width=9cm]{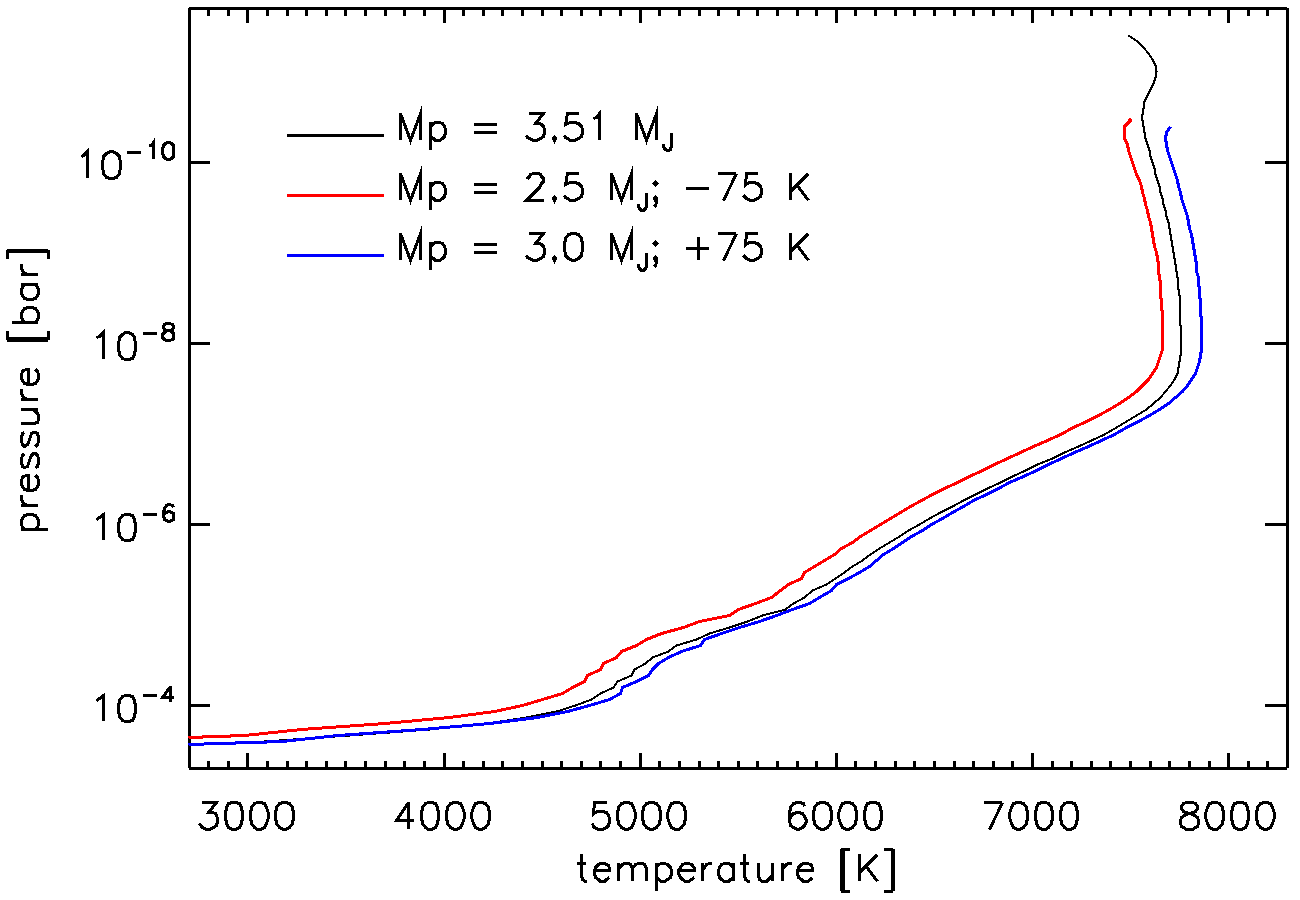}
		\includegraphics[width=9cm]{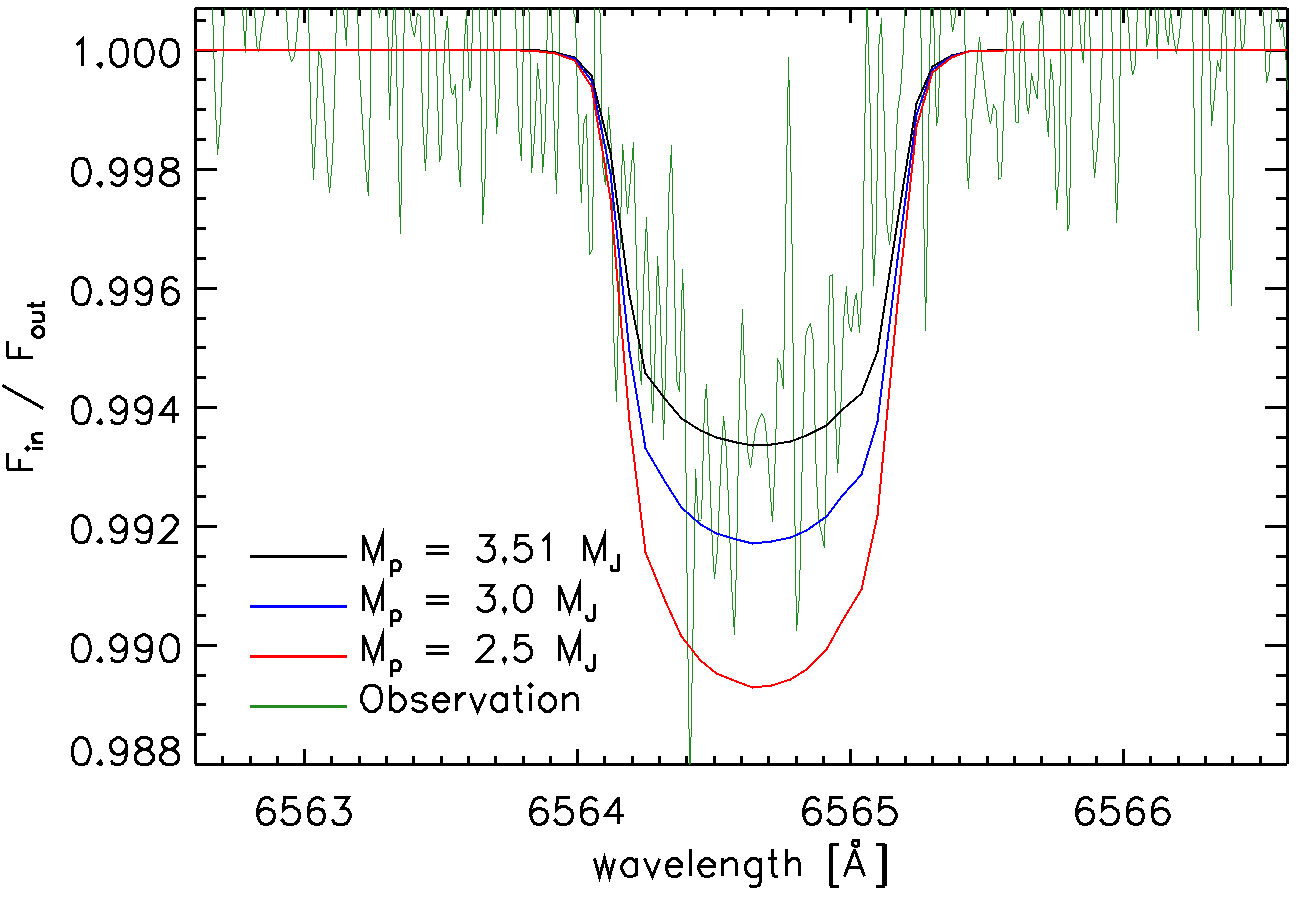}
		\includegraphics[width=9cm]{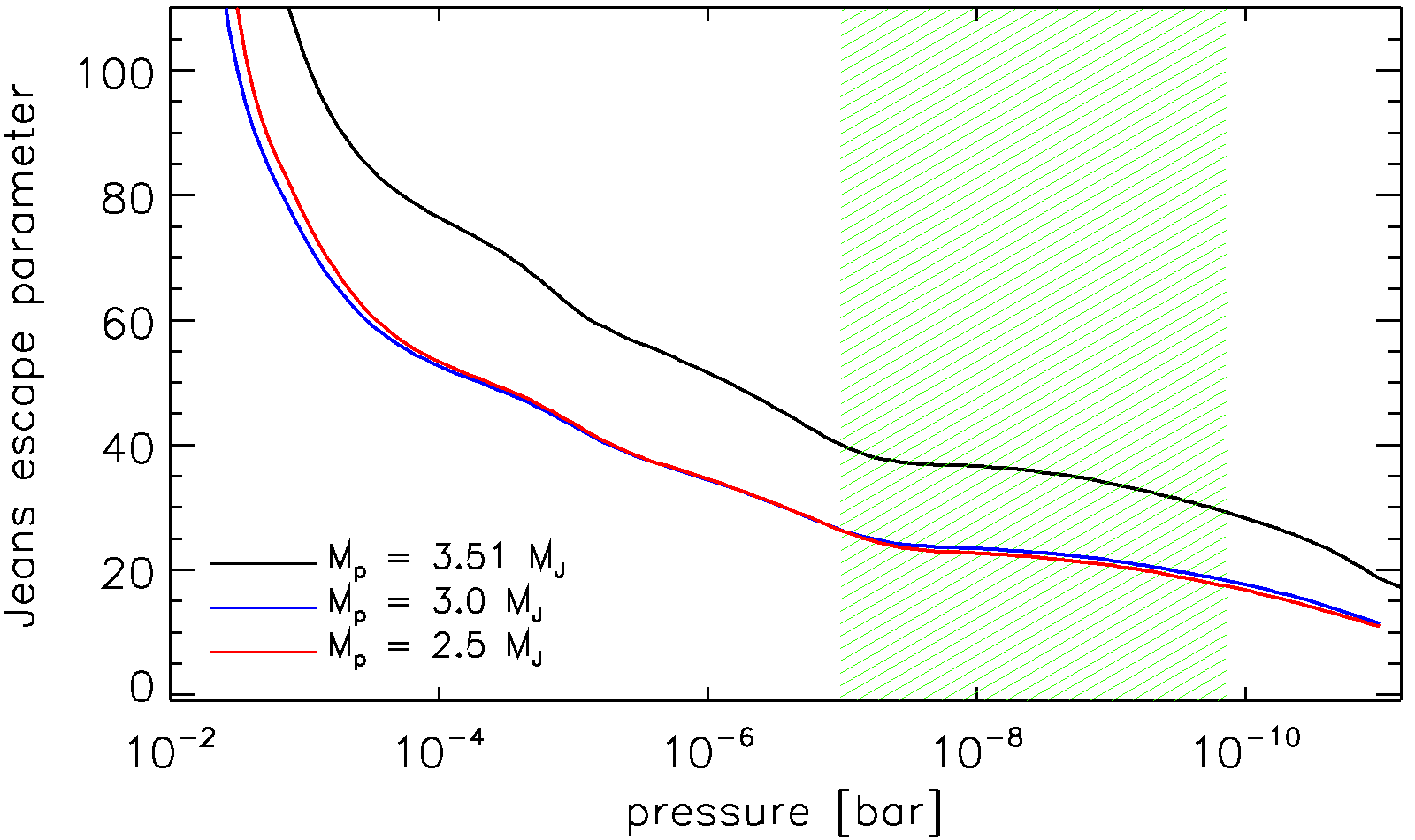}
        \caption{Comparison among the {\sc Cloudy} TP profiles (top), normalised H$\alpha$ transmission spectra (middle), and Jeans escape parameter (bottom) computed by setting the planetary mass equal to 3.51\,$M_{\rm J}$ \citep[black; ][]{borsa2022_mascara2b}, 2.5\,$M_{\rm J}$ (red), and 3.0\,$M_{\rm J}$ (blue). In the top panel, the TP profiles are rigidly shifted horizontally by the value indicated in the legend for visualisation purposes. In the middle panel, the green line shows the observed H$\alpha$ transmission spectrum.}
  		\label{fig:Mplanet}
\end{figure}

The radial velocity measurements enabled one to set just an upper limit on the planetary mass. Therefore, we tested the impact of the choice of a planetary mass of 3.51\,$M_{\rm J}$ \citep{borsa2022_mascara2b} on the main results. The top panel of Figure~\ref{fig:Mplanet} shows a comparison between the {\sc Cloudy} TP profiles computed considering a planetary mass equal to 2.5, 3.0, and 3.51\,$M_{\rm J}$. The three theoretical TP profiles are almost identical, which indicates that planetary mass has no impact on the obtained temperature structure. The profiles computed for 2.5 and 3.0\,$M_{\rm J}$ do not extend in the upper atmosphere as much as that computed for 3.51\,$M_{\rm J}$, because of {\sc Cloudy} convergence problems at low pressures for the low-mass models.

We also computed {\sc helios} TP profiles assuming planetary masses of 2.5 and 3.0\,$M_{\rm J}$ obtaining results essentially identical to those obtained for a planetary mass of 3.51\,$M_{\rm J}$. For each of the two additional values of the planetary mass, following the procedure described in Section~\ref{sec:TP_chemistry} we combined the {\sc helios} and {\sc Cloudy} TP profiles to obtain the composite theoretical TP profile, and then we followed the procedure described in Section~\ref{sec:discussion:comparison} to compute the synthetic transmission spectra. From the calibration of the transmission spectra to the observed planet-to-star radius ratio, we obtained a reference pressure value of 0.002\,bar, which is the same as that we derived for a planetary mass of 3.51\,$M_{\rm J}$. 

The middle panel of Figure~\ref{fig:Mplanet} shows the normalised H$\alpha$ synthetic transmission spectra as a function of planetary mass, further comparing them to the observed profile. As expected, the line absorption depth increases with decreasing planetary mass and the higher considered mass of 3.51\,$M_{\rm J}$ leads to a better fit to the data compared to what is obtained with lower masses. Under the assumption that the obtained theoretical TP profiles are representative of the planetary atmosphere, we conclude that the mass of \planet\ should lie between 3.0 and 3.51\,$M_{\rm J}$.
\section{Conclusions}\label{sec:conclusions}
We presented the transmission spectrum of the UHJ \planet\ covering the H$\alpha$, H$\beta$, H$\gamma$, and H$\delta$ line profiles extracted from six transit observations obtained with the HARPS-N high-resolution spectrograph. The H$\alpha$, H$\beta$, and H$\gamma$ are definitely detected with absorption depths of 0.79$\pm$0.03\% (1.25\,R$_{\rm p}$), 0.52$\pm$0.03\% (1.17\,R$_{\rm p}$), and 0.39$\pm$0.06\% (1.13\,R$_{\rm p}$), respectively, while the H$\delta$ line is detected at the $\sim$4$\sigma$ level with an absorption depth of 0.27$\pm$0.07\% (1.09\,R$_{\rm p}$). Our H$\alpha$ and H$\beta$ observed line profiles are in agreement with those obtained by \citet{Casasayas_2019} based on three transit observations, while ours is the first detection of the H$\gamma$ and H$\delta$ lines in the atmosphere of \planet, where the latter might be contaminated by a nearby Fe{\sc i} line.

We further presented the results of forward modelling of the TP profile of the planetary atmosphere at the sub-stellar point covering from the 10\,bar to the 10$^{-12}$\,bar level computed combining the {\sc helios} (LTE) and {\sc Cloudy} (NLTE) codes, thus accounting for NLTE effects in the middle and upper atmosphere. We found an isothermal temperature at $\approx$2200\,K at pressures higher than $\approx$10$^{-2}$\,bar and at $\approx$7700\,K at pressures lower than $\approx$10$^{-8}$\,bar, while in between those pressure values the temperature increases roughly linearly. From comparing the LTE and NLTE theoretical TP profiles, we found that accounting for NLTE effects leads to a temperature increase in the middle and upper atmosphere of up to about 3000\,K. As it has been found for KELT-9b, the temperature inversion is caused by metal-line absorption of the stellar photons. Furthermore, the higher temperature of the NLTE TP profile, in comparison to the LTE one, is caused by the impact of NLTE effects on the level populations of Fe and Mg, which play a significant role in heating and cooling, respectively. This result supports the idea that accounting for Fe, Mg, and their level population is of critical importance for adequately modelling the atmospheric energy balance of UHJs. This agrees with the result of \citet{nugroho2020} and \citet{johnson2022}, who looked for the signature of potential molecular species that might cause the temperature inversion, without finding it. Metals cause the temperature inversion and the impact of NLTE effects on those metals increases the magnitude of the temperature rise.

In the model of KELT-9b, the inclusion of NLTE effects leads to a temperature increase in the upper atmosphere of about 2000\,K compared to the LTE profile \citep{fossati2021}, while in the model of \planet, which is cooler and orbits a cooler host compared to KELT-9b, we find a temperature increase of about 3000\,K. Comparisons of the departure coefficients computed for the two planets in the middle and upper atmosphere led us to conclude that this difference might be ascribed to a lack of cooling in the atmosphere of \planet, rather than to an increase in heating. Therefore, NLTE effects might significantly impact the TP profile also of planets cooler than UHJs. This calls for a dedicated parameter study to gain insight into the impact that the system parameters have on the deviation from LTE on the atmospheric TP profile.

We remark that the shape of the stellar spectral energy distribution is likely to play a fundamental role in the properties of the upper atmosphere of UHJs, and in particular on whether a planet hosts a hydrostatic or hydrodynamic atmosphere. Both KELT-9b and \planet\ host hydrostatic atmospheres and the spectral lines lying in the optical range (including H$\alpha$) do not probe the exosphere and do not directly constrain mass loss. This is because both host stars are earlier than spectral type A3--A4 and thus do not have a strong X-ray and EUV emission \citep{fossati2018_AstarPlanets}. This prevents the atmospheric hydrogen, which is by far the most abundant element, to heat up significantly and thus the upper atmosphere to reach temperatures high enough to become hydrodynamic. Instead, for planets orbiting stars later than A3--A4, the X-ray and EUV irradiation is expected to be significant, which in turn leads to strong hydrogen heating and thus most likely to a hydrodynamic atmosphere. For this reason, the observations of planets orbiting stars earlier and later than A3--A4 should not be directly compared to infer general properties of planets orbiting early-type stars.

We employed the LTE and NLTE theoretical TP profiles to compute high-resolution synthetic transmission spectra covering from the UV to the infrared wavelength range, and thus including the hydrogen Balmer lines. We find that the NLTE synthetic transmission spectrum is a good fit to the observed Balmer lines, while, as a result of the cooler temperature structure, the LTE synthetic line profiles are systematically weaker than the observations. The synthetic NLTE profiles are slightly stronger and systematically broader than the observed ones, which is probably caused by the use of the sub-stellar TP profile as the underlying structure for computing transmission spectra that probe instead the terminator region. From comparing the theoretical LTE and NLTE transmission spectra over the entire considered wavelength range, we find the strongest deviation from LTE in the UV spectral band, while in the optical and infrared the hydrogen Balmer and Paschen lines are those presenting the strongest deviation from LTE.

Finally, we considered the NLTE atmospheric structure to compute the sound speed and the Jeans escape parameter across the planetary atmosphere. We found that the sound speed increases with decreasing pressure, up to a value of about 12\,km\,s$^{-1}$ at top of our simulation domain. Instead, the Jeans escape parameter suggests that the planetary atmosphere remains hydrostatic within the entire simulated pressure range.
\begin{acknowledgements}
This paper is based on observations collected with the 3.58m Telescopio Nazionale Galileo (TNG), operated on the island of La Palma (Spain) by the Fundaci\'{o}n Galileo Galilei of the INAF (Istituto Nazionale di Astrofisica) at the Spanish Observatorio del Roque de los Muchachos, in the frame of the programme Global Architecture of Planetary Systems (GAPS). This research used the facilities of the Italian Center for Astronomical Archive (IA2) operated by INAF at the Astronomical Observatory of Trieste. We acknowledge support from PRIN INAF 2019. D.S. acknowledges funding from project PID2021-126365NB-C21(MCI/AEI/FEDER, UE) and financial support from the grant CEX2021-001131-S funded by MCIN/AEI/10.13039/501100011033. We thank the anonymous referee for the useful comments that led to improve the paper.
\end{acknowledgements}

\bibliography{ref}	
\bibliographystyle{aa}
\begin{appendix}
%
%
%
\section{Additional figures}
%
\begin{figure}[ht!]
		\centering
		\includegraphics[width=9cm]{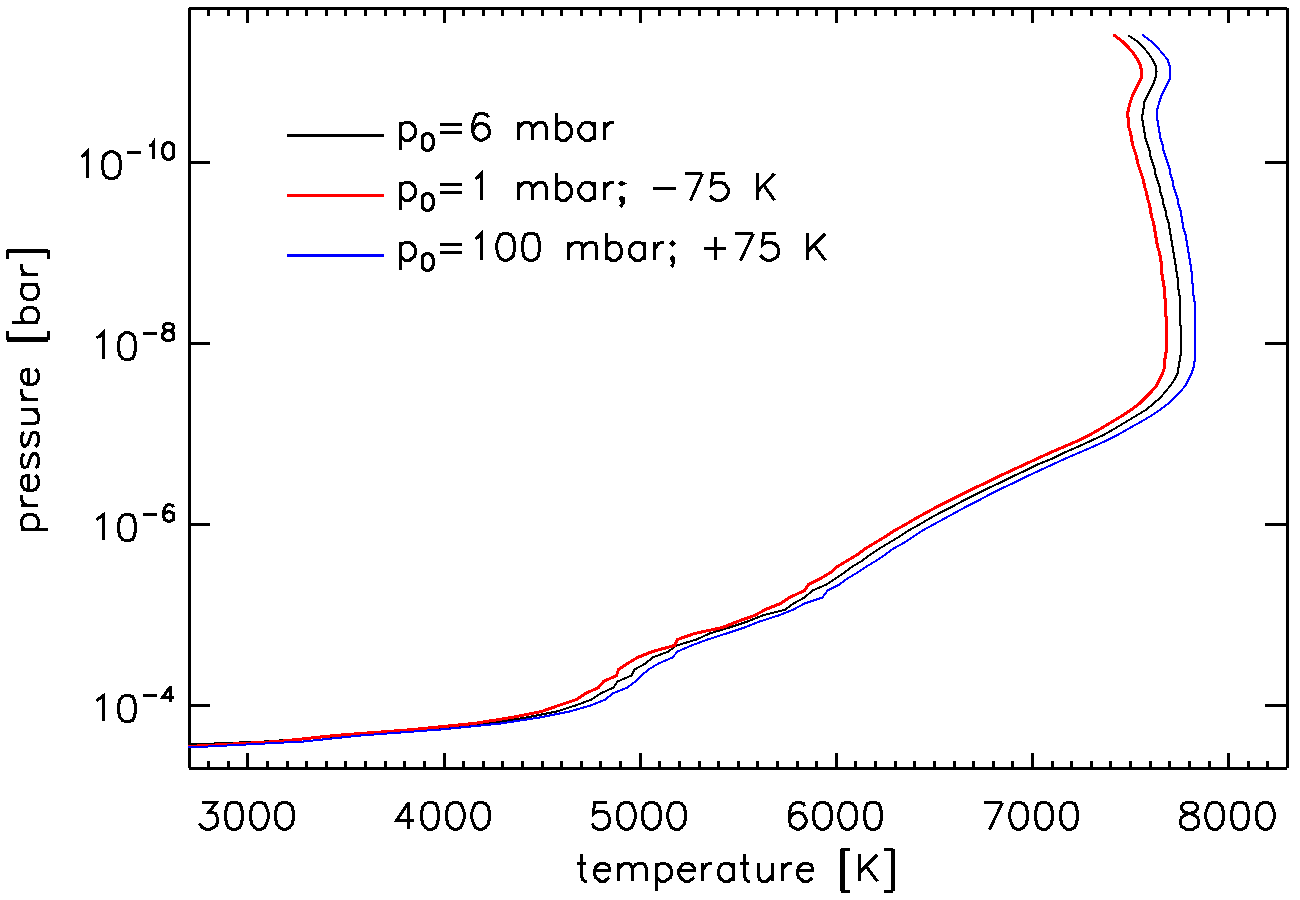}
		\includegraphics[width=9cm]{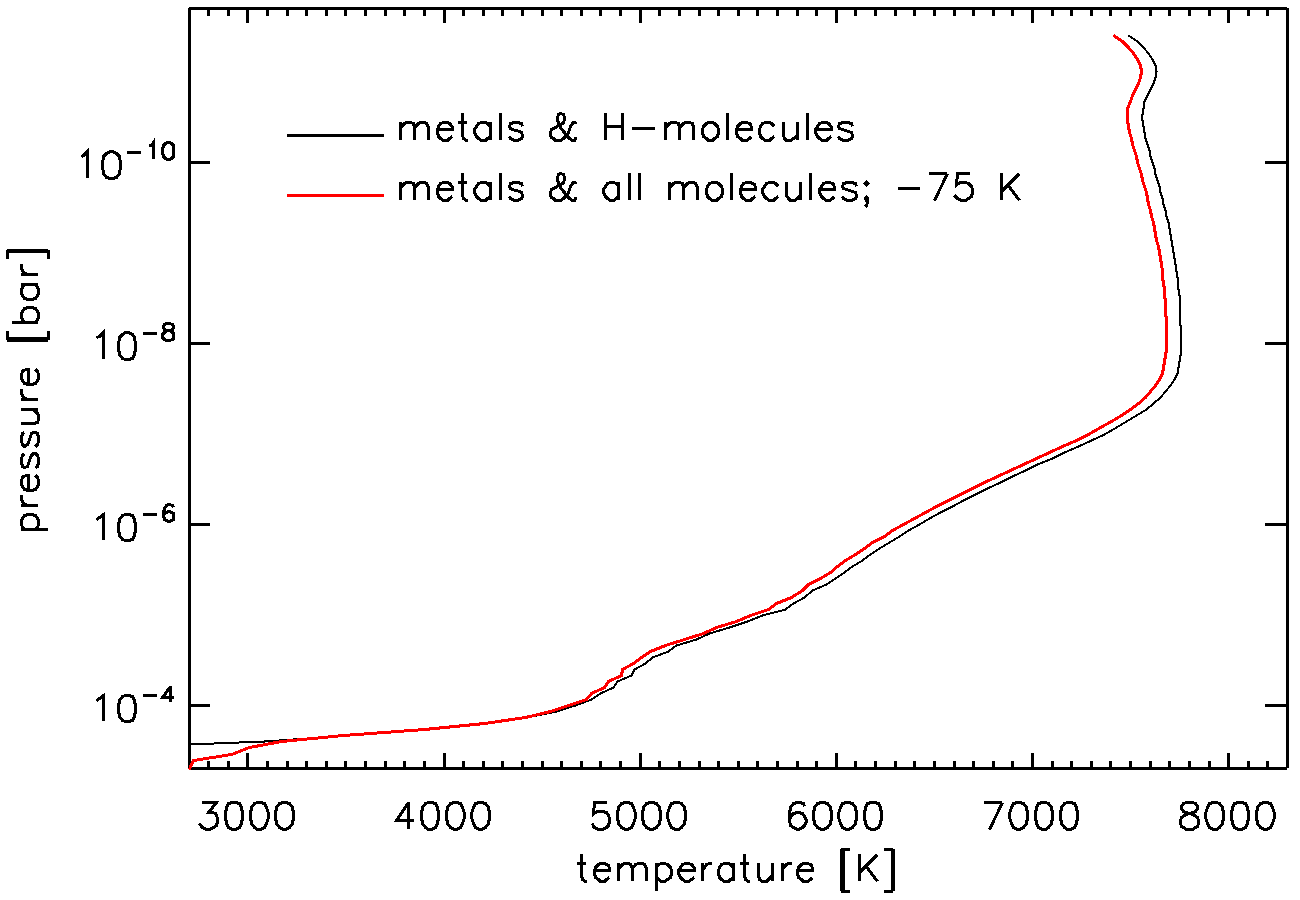}
		\includegraphics[width=9cm]{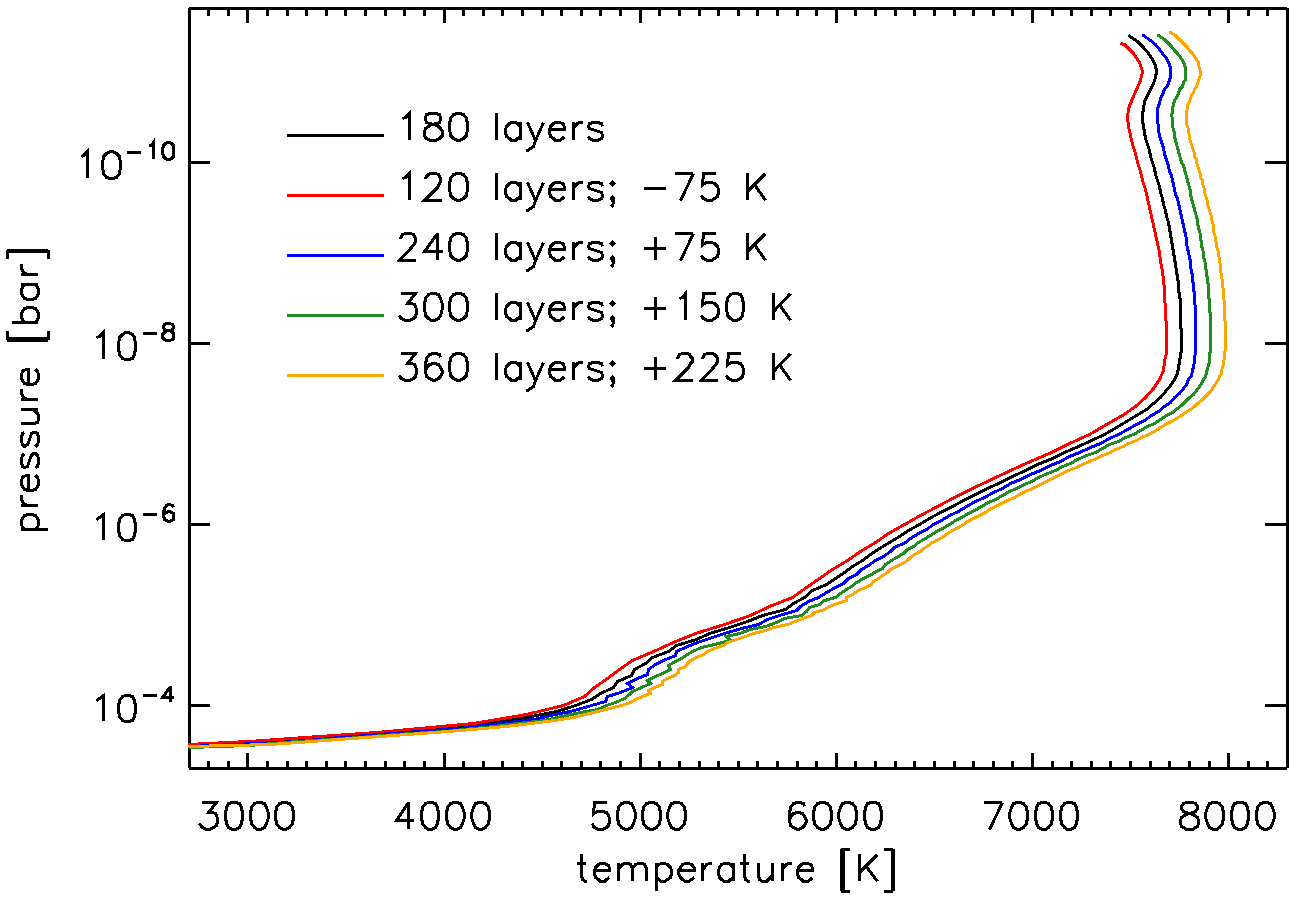}
	\caption{Comparison among TP profiles computed considering different assumptions for reference pressure, atmospheric composition, and number of layers. Top: Comparison among {\sc Cloudy} TP profiles obtained fixing the planetary transit radius at reference pressure ($p_0$) values of 100\,mbar (blue), 6\,mbar (black), and 1\,mbar (red). Middle: Comparison between {\sc Cloudy} TP profiles computed accounting for metals plus only hydrogen molecules (black) and for metals plus all molecules present in the {\sc Cloudy} database (red). Bottom: Comparison among {\sc Cloudy} TP profiles obtained considering different number of layers indicated in the legend. For visualisation purposes, the TP profiles are rigidly shifted horizontally by the value indicated in the legend.}
  		\label{fig:test_p0_molecules}
\end{figure}
\begin{figure}[ht!]
		\centering
		\includegraphics[width=9cm]{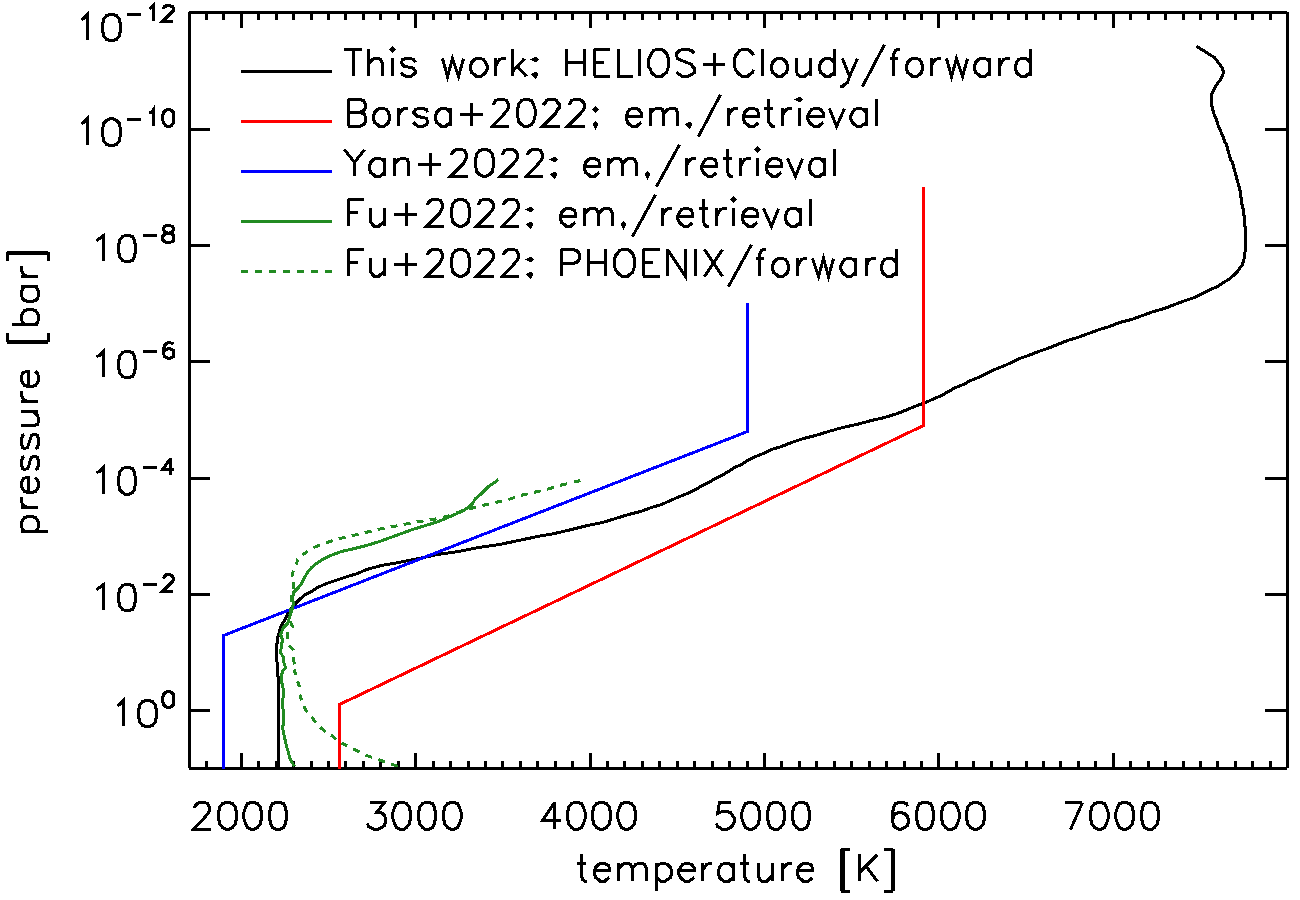}
		\caption{Comparison between our composite TP profile (black) with those published in the literature retrieved from secondary eclipse observations by \citet[][red]{borsa2022_mascara2b}, \citet[][blue]{yan2022_mascara2b}, and \citet[][green]{fu2022}. As a further comparison, the dashed green line shows the result of the forward model presented by \citet{fu2022}. We recall that the retrieved profiles are an average over the illuminated planet disk, while the black line is computed for the sub-stellar point.} 
  		\label{fig:TPliteratureCompare}
\end{figure}
\begin{figure*}[h!]
\includegraphics[width=18cm]{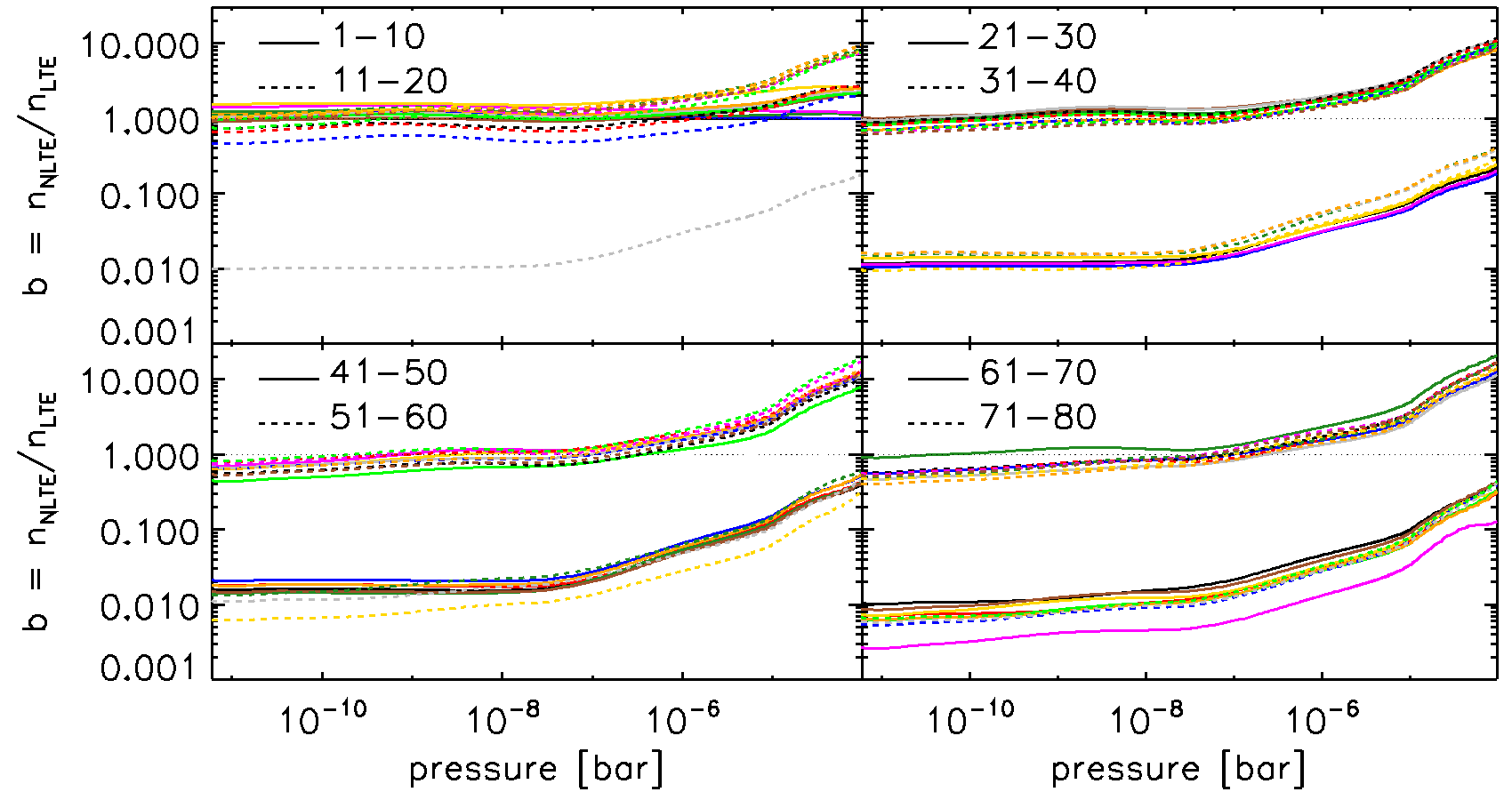}
\caption{Departure coefficients as a function of pressure for the first 80 energy levels of Fe{\sc i}. The energy levels are numbered from 1 to 80, and are separated into groups of ten levels and two line styles (solid and dashed) for each panel. Within each group of ten energy levels, the order of the line colors corresponding to increasing energy is black, red, blue, dark green, magenta, yellow, brown, grey, bright green, and orange. The dotted line (at 1.0) indicating $n_{\rm NLTE}$\,=\,$n_{\rm LTE}$ is for reference.}
\label{fig:dep.coeff.Fe1}
\end{figure*}
\begin{figure*}[h!]
\includegraphics[width=18cm]{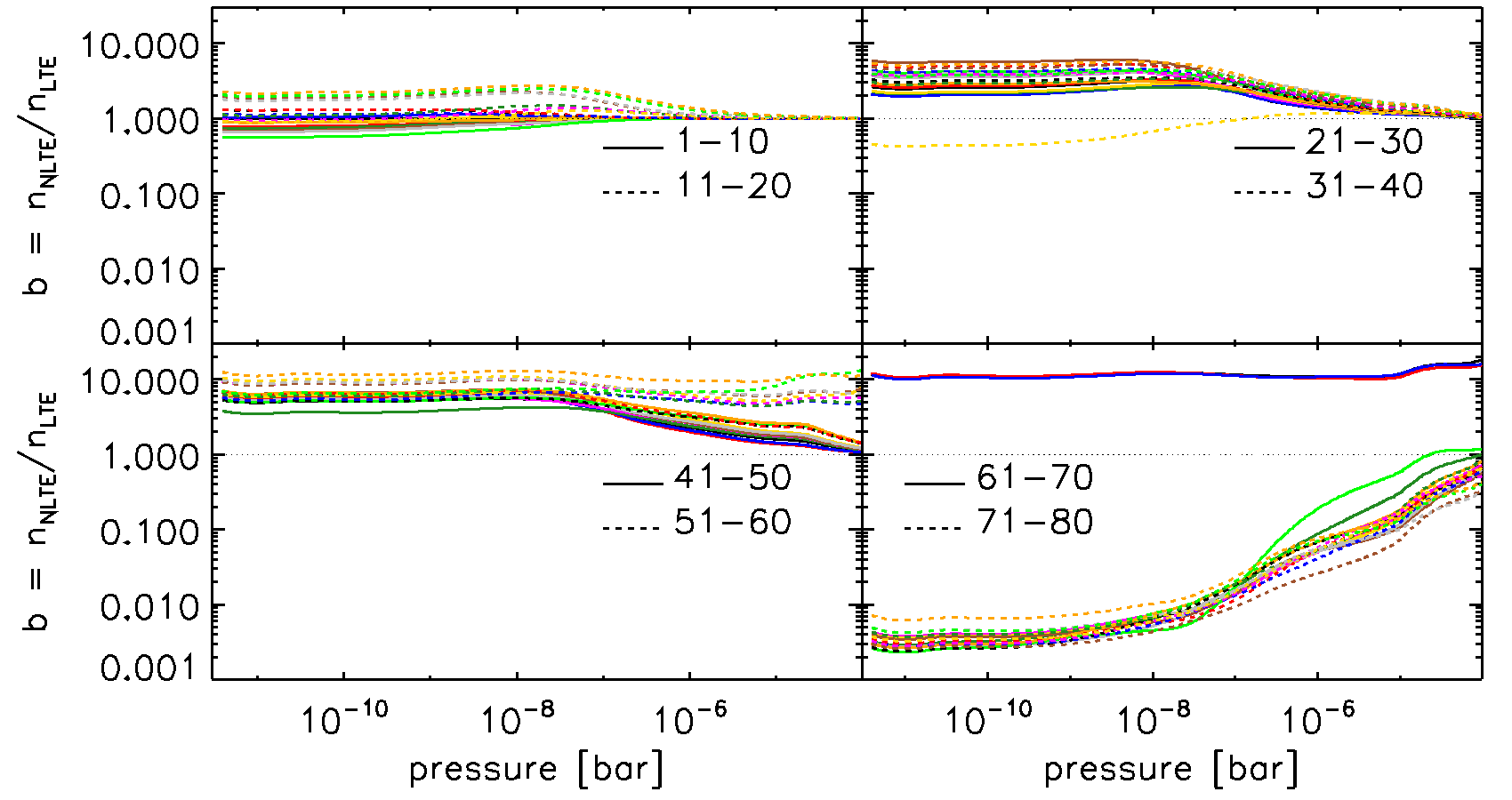}
\caption{Same as Figure~\ref{fig:dep.coeff.Fe1}, but for Fe{\sc ii}.}
\label{fig:dep.coeff.Fe2}
\end{figure*}
\begin{figure*}[h!]
\includegraphics[width=18cm]{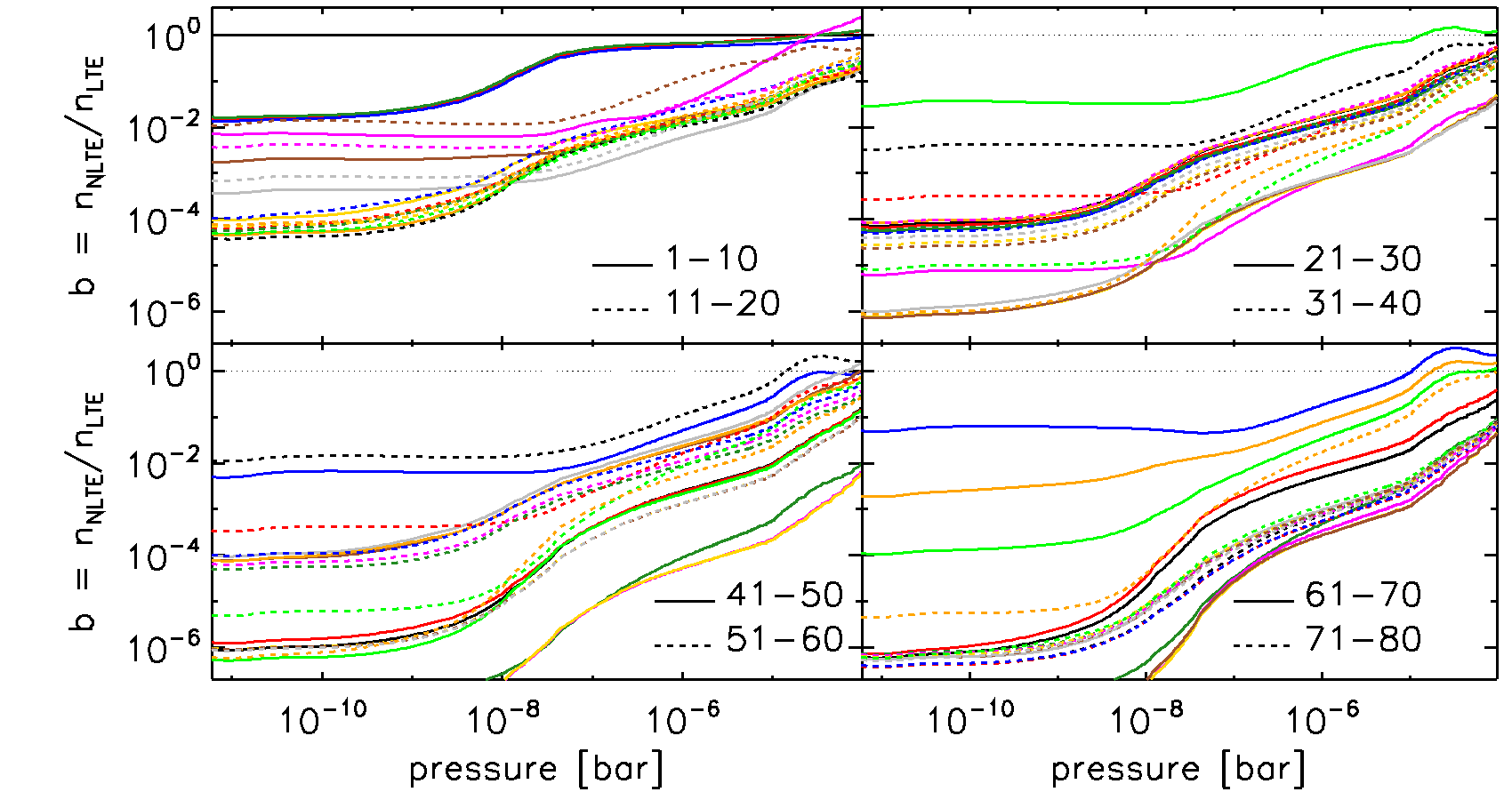}
\caption{Same as Figure~\ref{fig:dep.coeff.Fe1}, but for Mg{\sc i}.}
\label{fig:dep.coeff.Mg1}
\end{figure*}
\begin{figure}[h!]
\includegraphics[width=9cm]{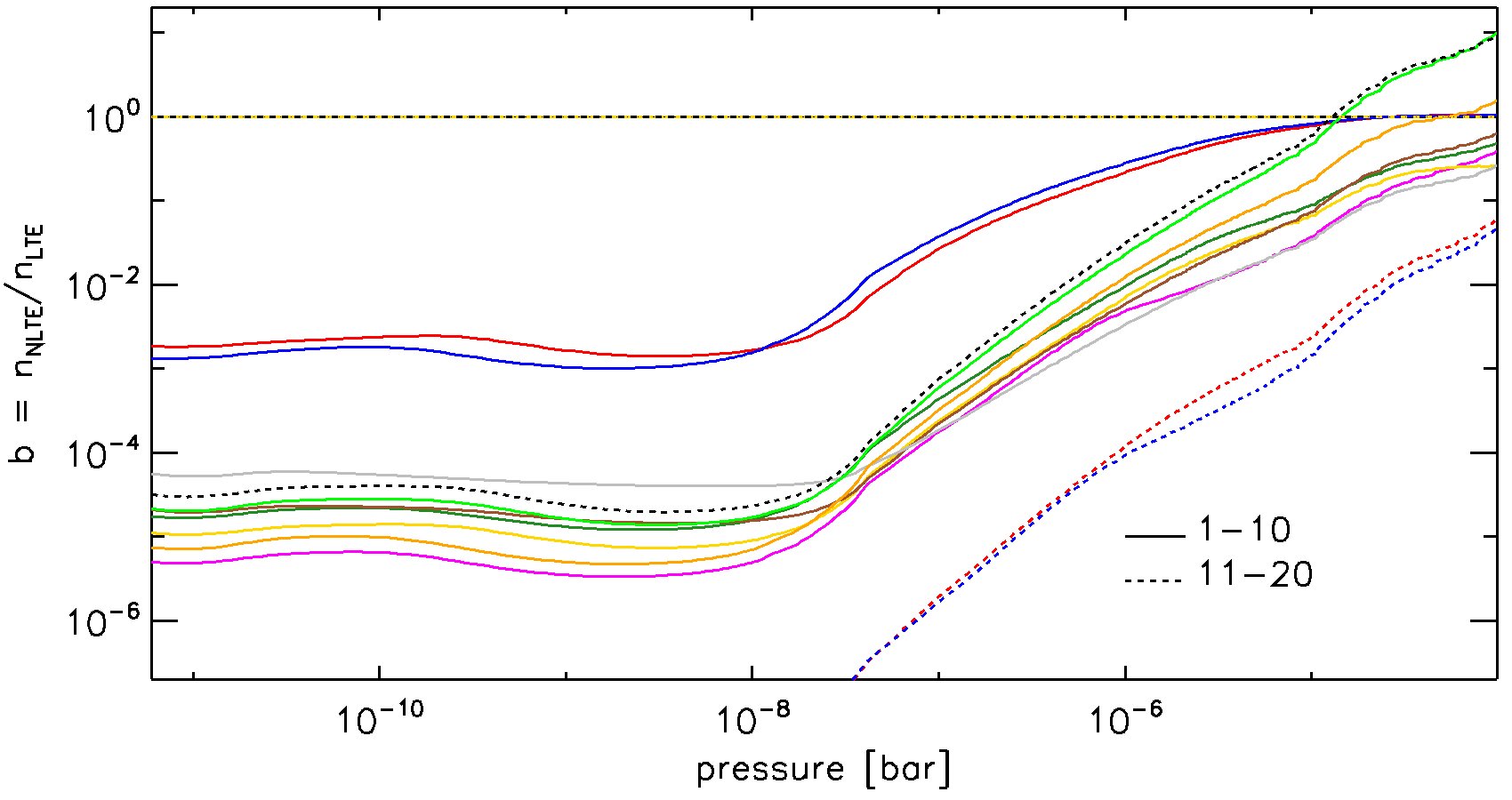}
\caption{Same as Figure~\ref{fig:dep.coeff.Fe1}, but for Mg{\sc ii} and up to the first 20 energy levels, which are those included in the 17.03 {\sc Cloudy} distribution.}
\label{fig:dep.coeff.Mg2}
\end{figure}
\begin{figure}[ht!]
		\centering
		\includegraphics[width=9cm]{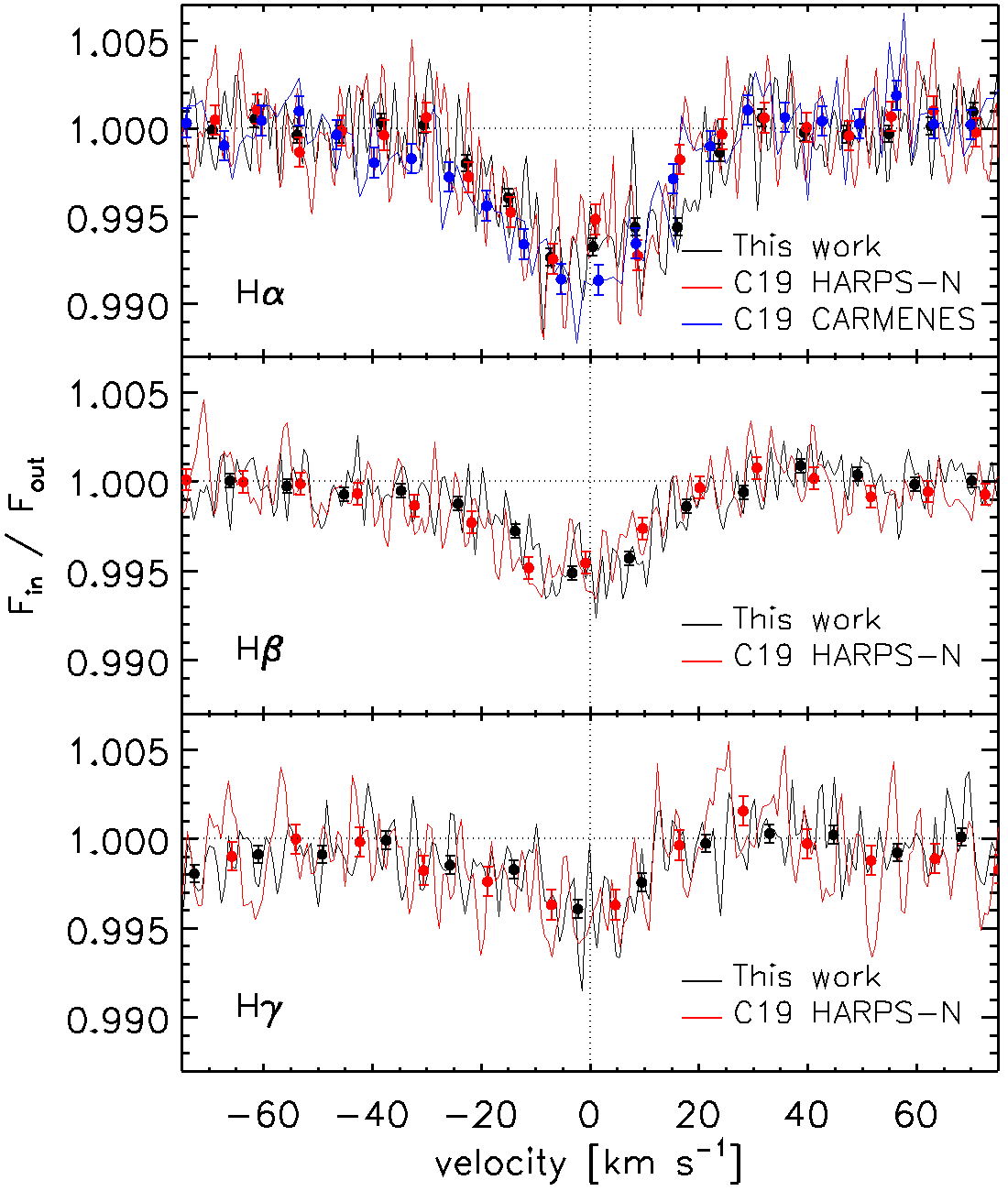}
		\caption{Comparison between our results obtained from combining six transits observed with HARPS-N (black) and what published by \citet{Casasayas_2019} employing HARPS-N (three transits; red) and CARMENES (one transit; blue). The top, middle and bottom panels show the comparison for the H$\alpha$, H$\beta$, and H$\gamma$ lines, respectively. The central wavelengths of the Balmer lines (in vacuum) used to convert the wavelengths into velocities are 6564.60\,\AA\ for H$\alpha$, 4862.71\,\AA\ for H$\beta$, and 4341.69\,\AA\ for H$\gamma$. Lines correspond to the original data, while to guide the eye dots are the data binned to about 7.5\,km\,s$^{-1}$. The vertical and horizontal dotted lines respectively at zero and one are for reference. The HARPS-N transmission spectra of \citet{Casasayas_2019} are about 20\% noisier than those presented here.}
  		\label{fig:comparisonWcasasayas}
\end{figure}
\begin{figure}[ht!]
		\centering
		\includegraphics[width=9cm]{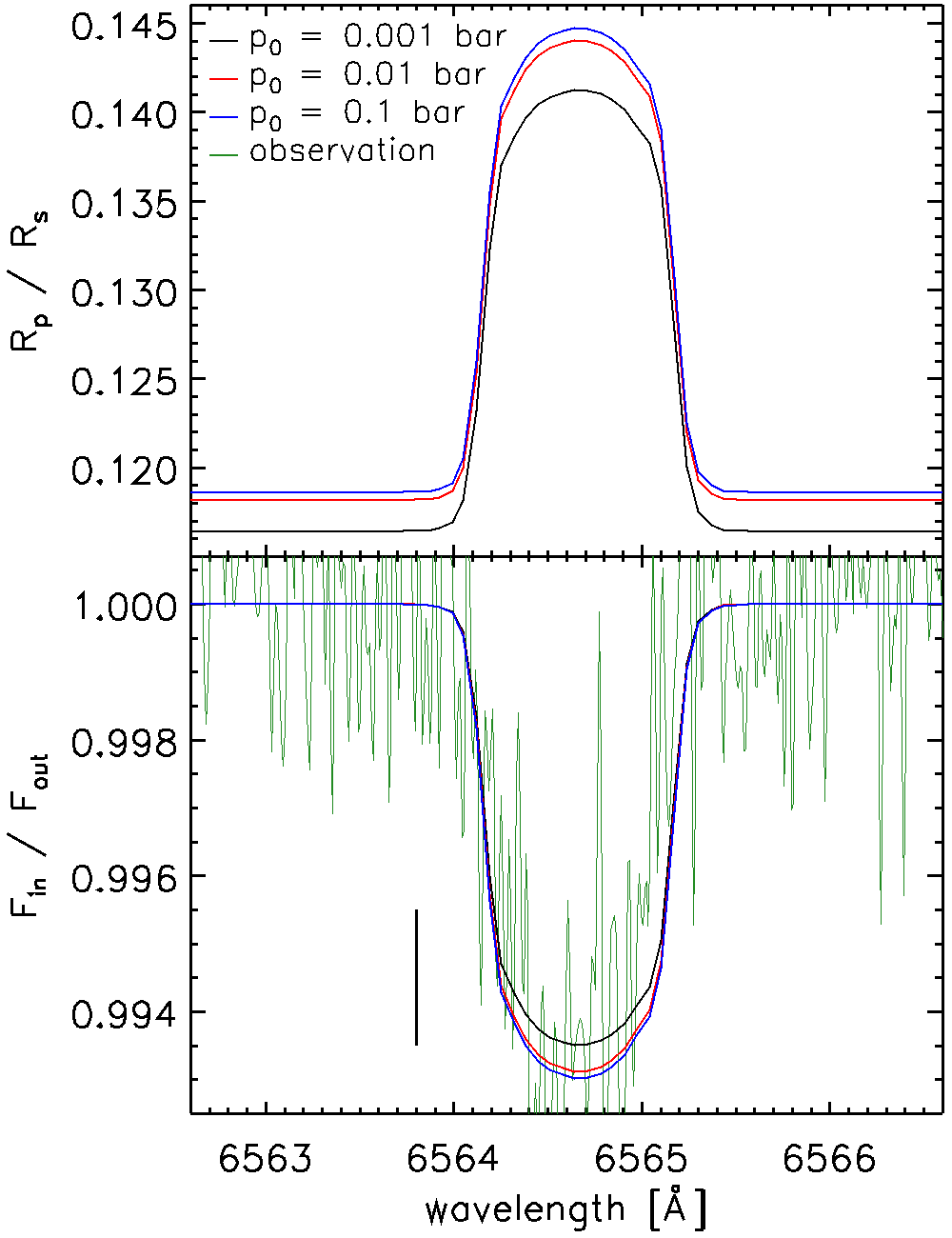}
		\caption{Comparison between {\sc Cloudy} synthetic transmission spectra of the H$\alpha$ line before (top) and after (bottom) normalisation computed considering different reference pressure levels at 0.001\,bar (black), 0.01\,bar (red), and 0.1\,bar (blue). For reference, in the bottom panel the green line shows the observed H$\alpha$ transmission spectrum, while the black straight line shows the average 1$\sigma$ uncertainty obtained from the observations. Wavelengths are in vacuum.}
  		\label{fig:comparisonTSdifferentP0}
\end{figure}
\begin{figure}[h!]
\includegraphics[width=9cm]{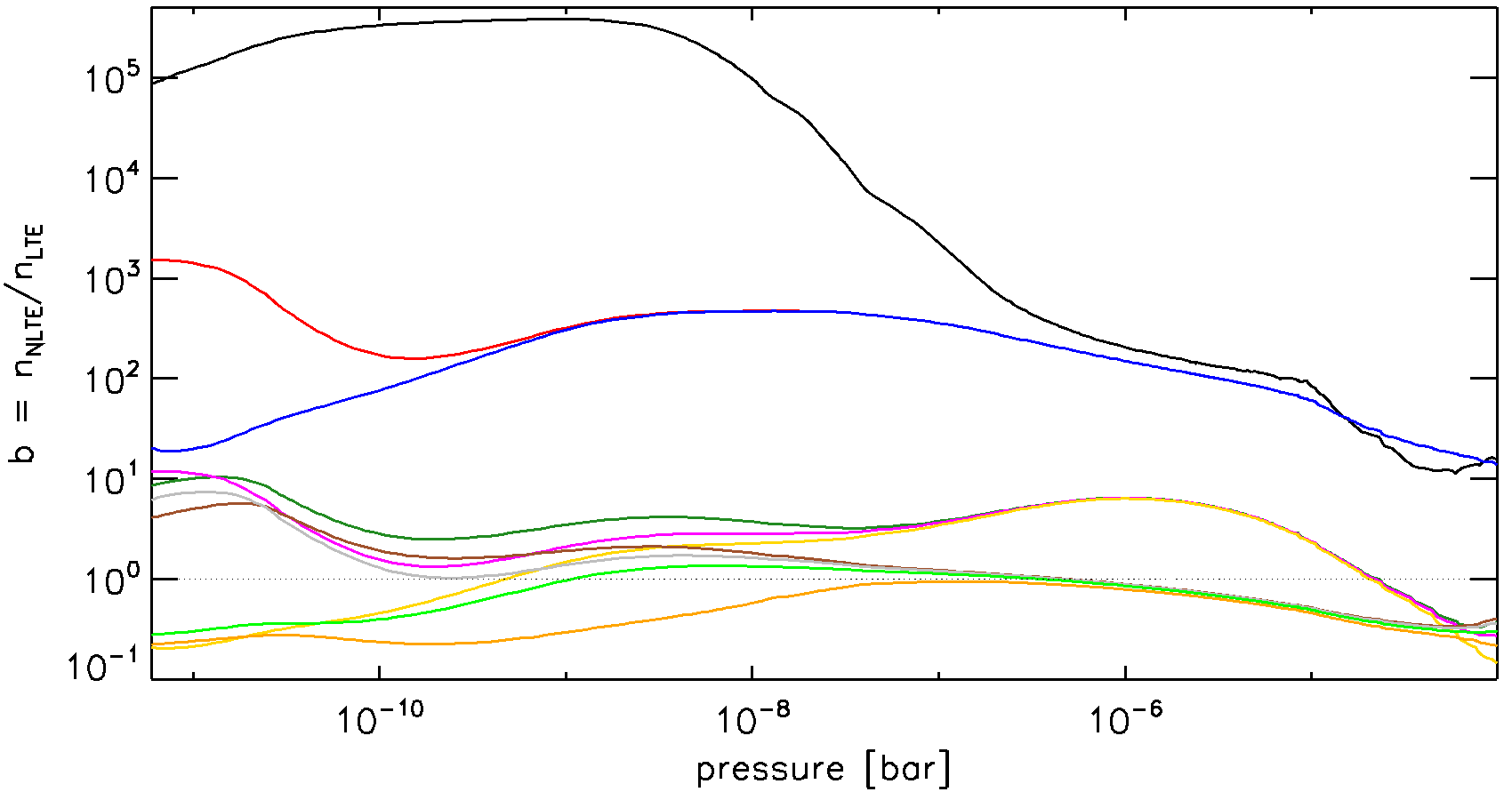}
\caption{Same as Figure~\ref{fig:dep.coeff.Fe1}, but for the first ten energy levels of H{\sc i}.}
\label{fig:dep.coeff.H1}
\end{figure}
\begin{figure*}[ht!]
		\centering
		\includegraphics[width=18cm]{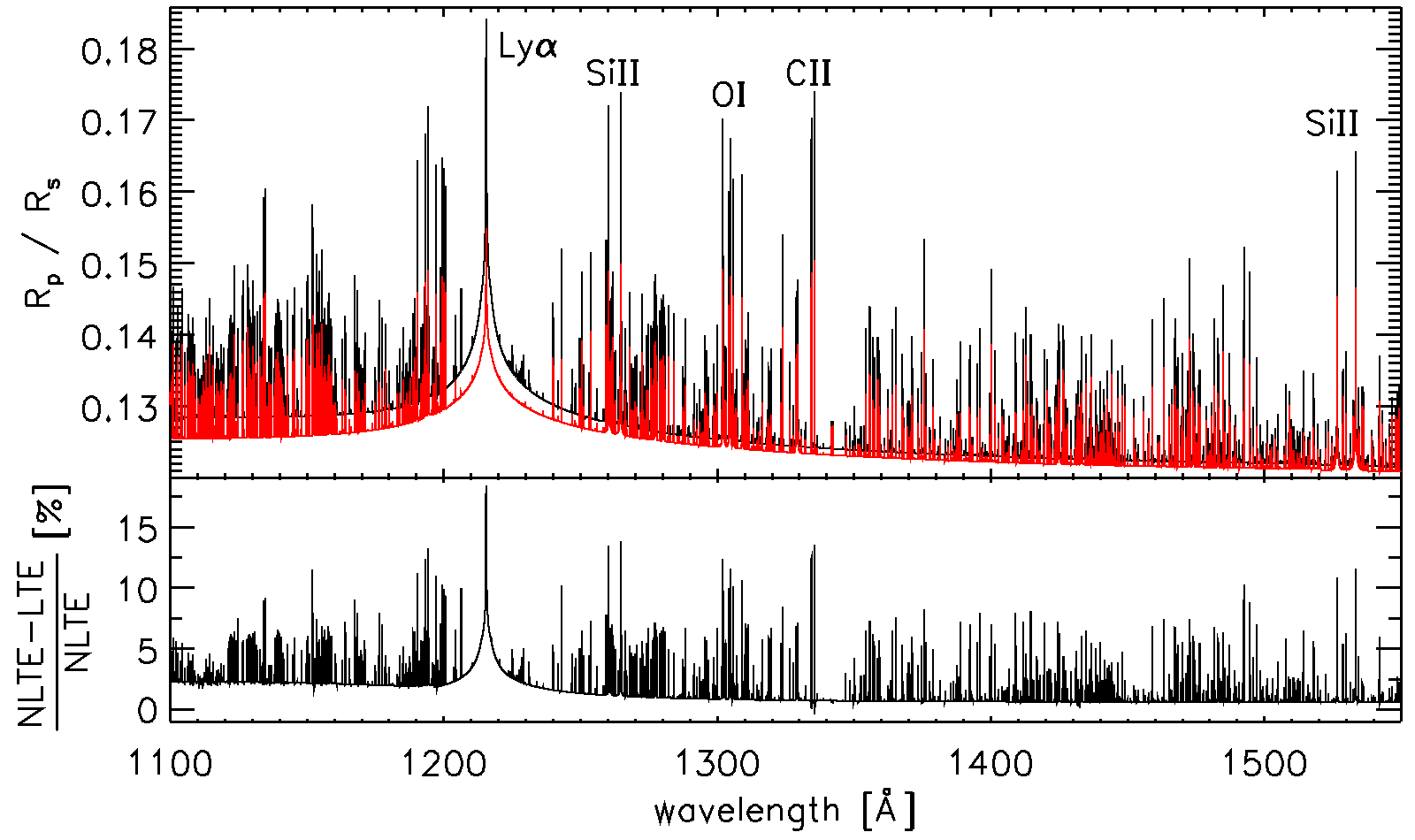}
		\caption{Same as Figure~\ref{fig:transmission_spectra}, but for the 1100--1550\AA\ band.} 
  		\label{fig:transmission_spectra_1100-1550}
\end{figure*}
\begin{figure*}[ht!]
		\centering
		\includegraphics[width=18cm]{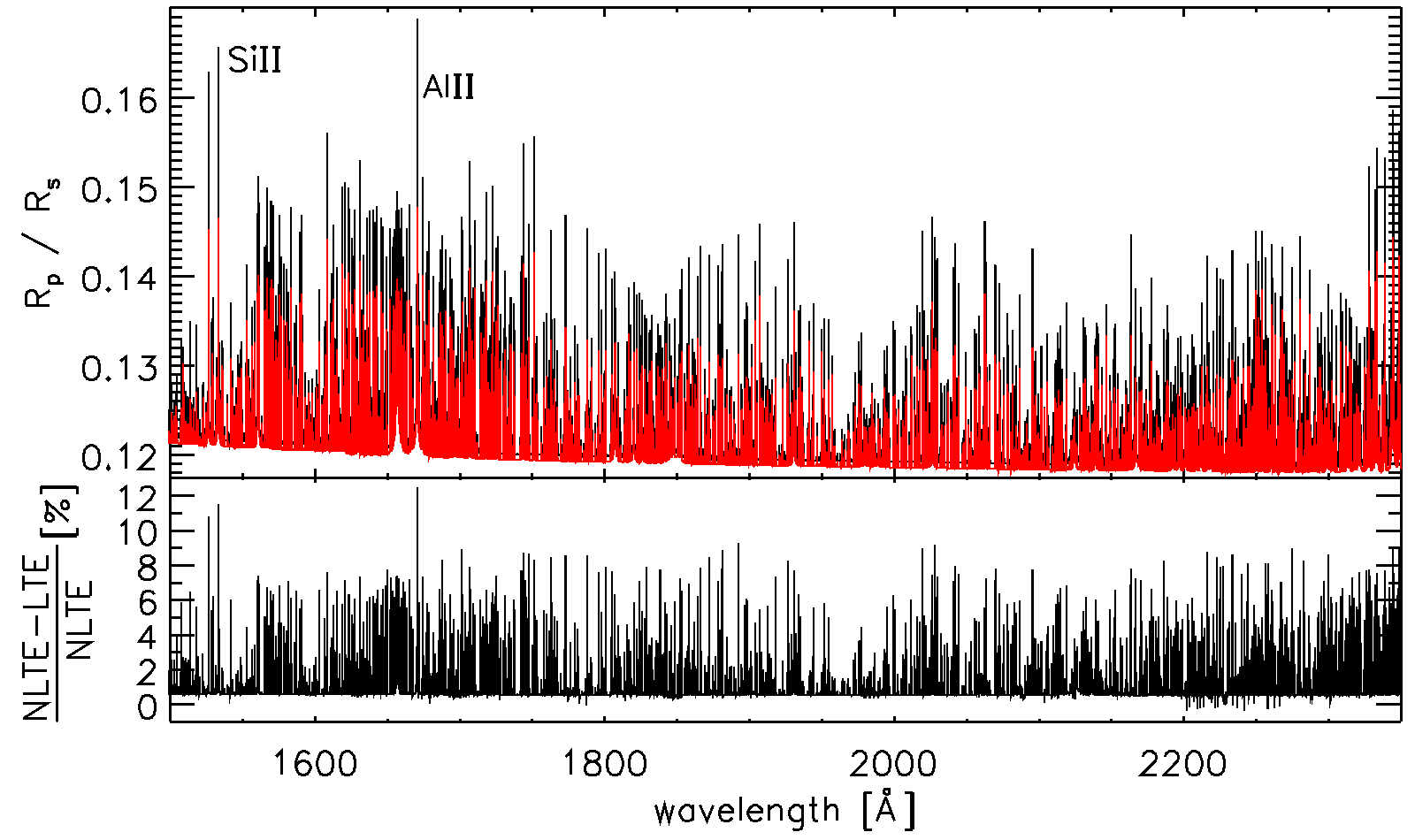}
		\caption{Same as Figure~\ref{fig:transmission_spectra}, but for the 1500--2350\AA\ band.} 
  		\label{fig:transmission_spectra_1500-2350}
\end{figure*}
\begin{figure*}[ht!]
		\centering
		\includegraphics[width=18cm]{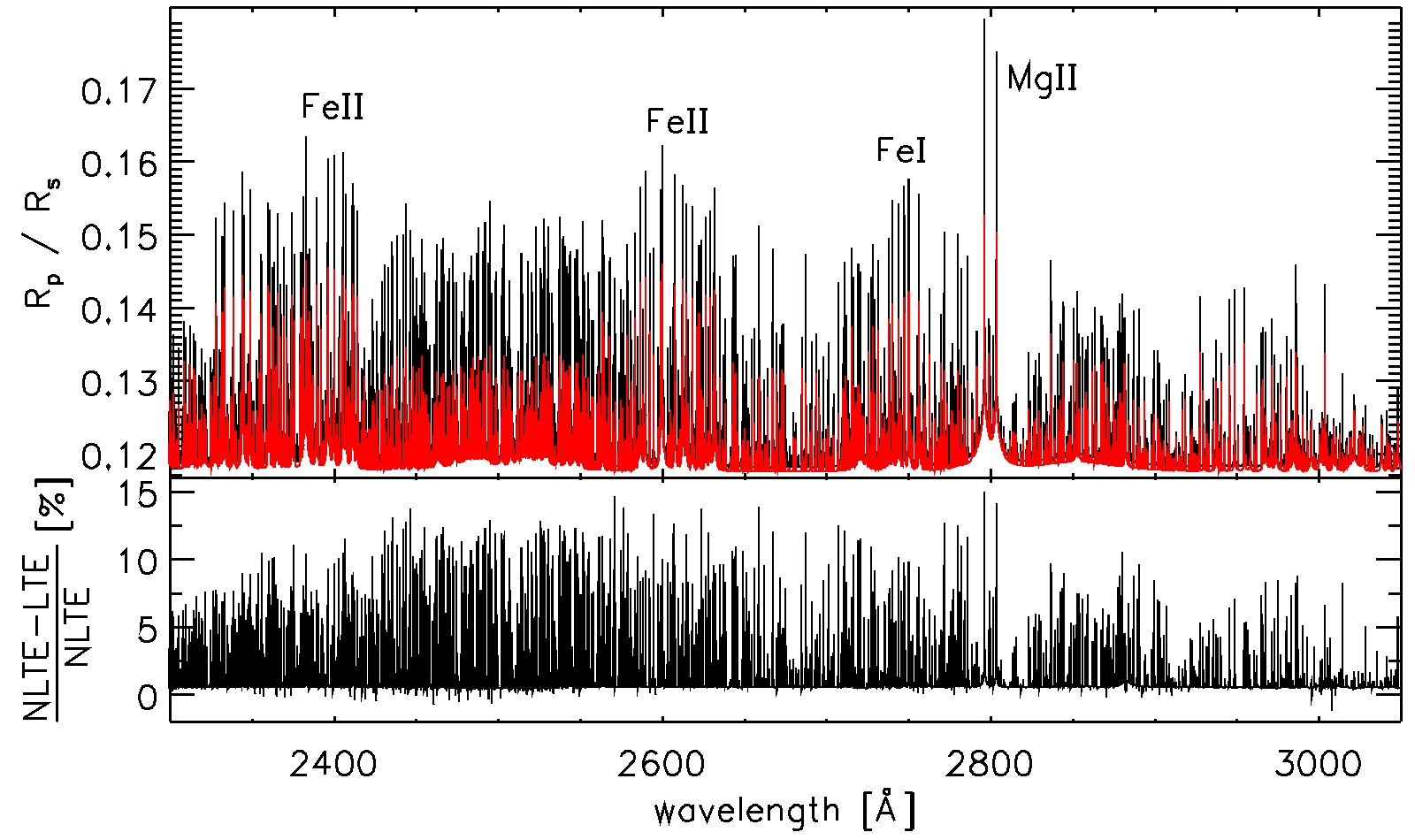}
		\caption{Same as Figure~\ref{fig:transmission_spectra}, but for the 2300--3050\AA\ band.} 
  		\label{fig:transmission_spectra_2300-3050}
\end{figure*}
\begin{figure*}[ht!]
		\centering
		\includegraphics[width=18cm]{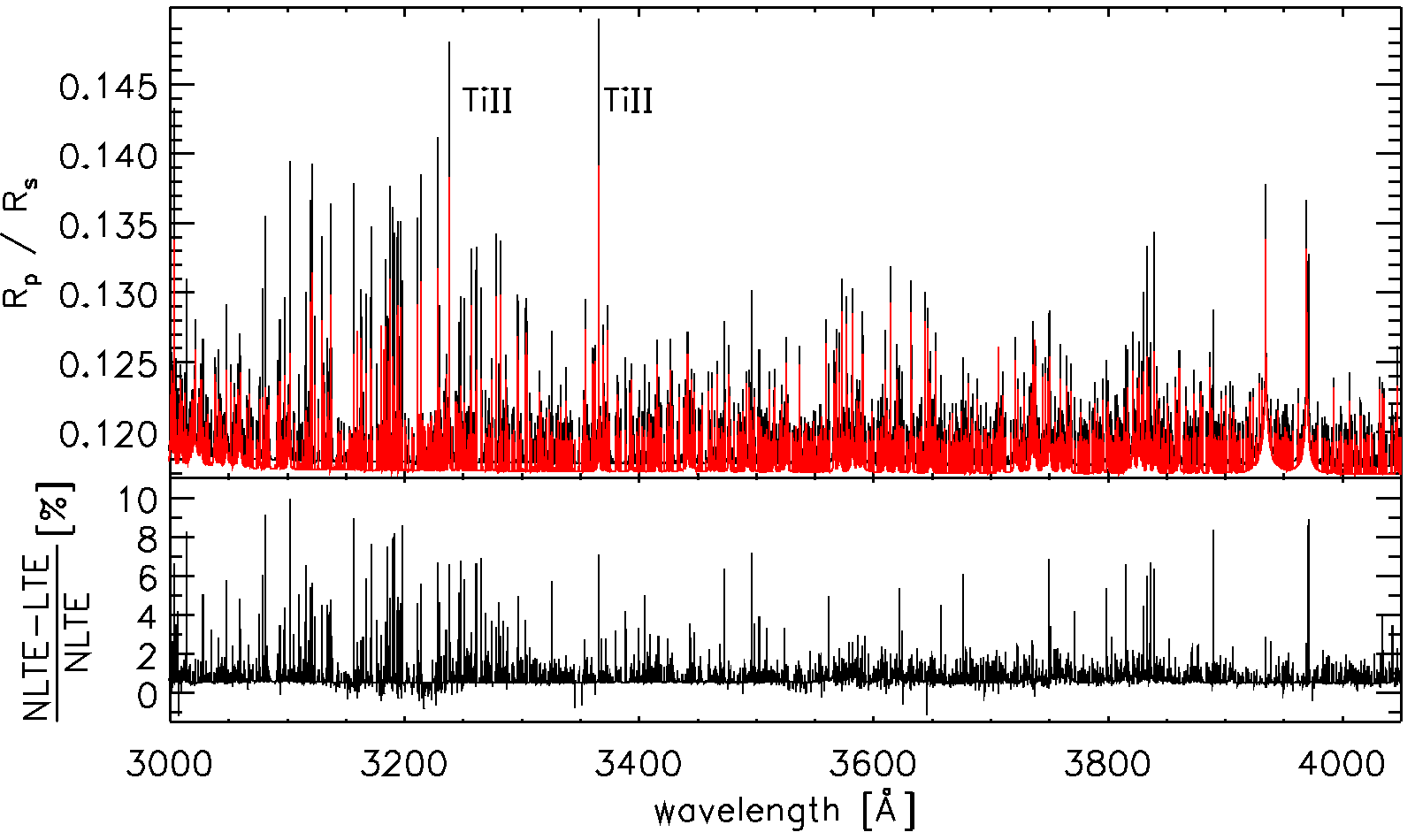}
		\caption{Same as Figure~\ref{fig:transmission_spectra}, but for the 3000--4050\AA\ band.} 
  		\label{fig:transmission_spectra_3000-4050}
\end{figure*}
\begin{figure*}[ht!]
		\centering
		\includegraphics[width=18cm]{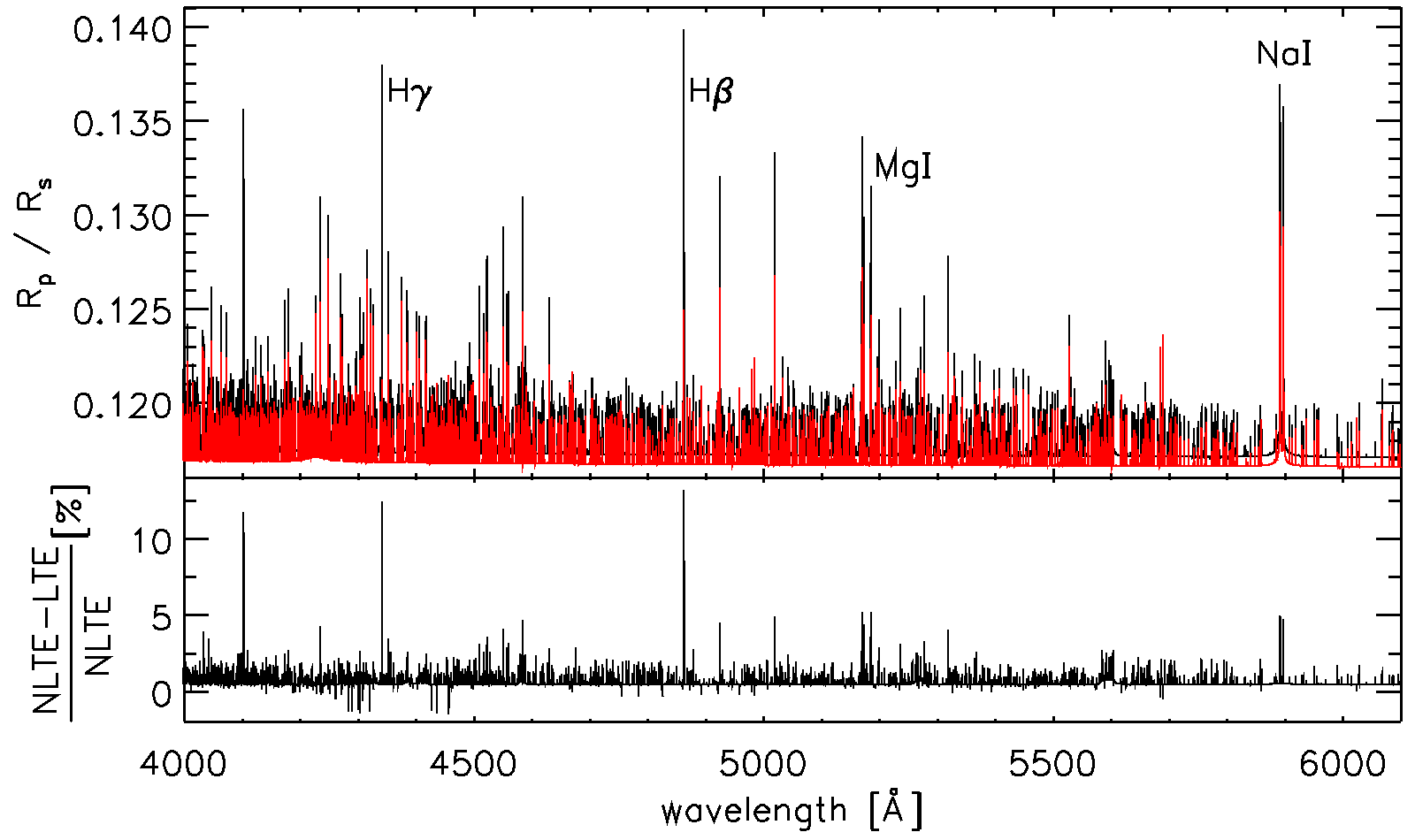}
		\caption{Same as Figure~\ref{fig:transmission_spectra}, but for the 4000--6100\AA\ band.} 
  		\label{fig:transmission_spectra_4000-6100}
\end{figure*}
\begin{figure*}[ht!]
		\centering
		\includegraphics[width=18cm]{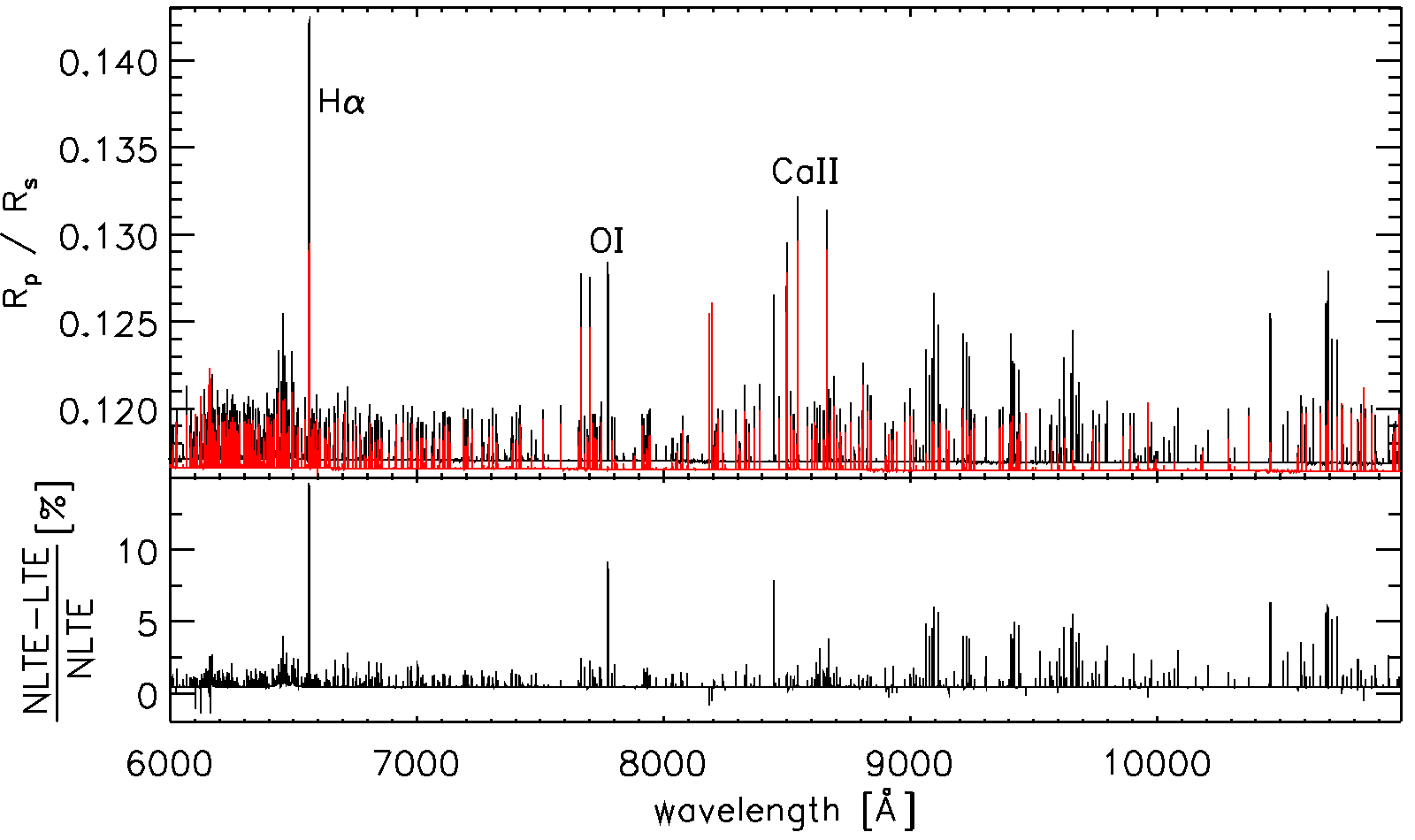}
		\caption{Same as Figure~\ref{fig:transmission_spectra}, but for the 6000--11000\AA\ band.} 
  		\label{fig:transmission_spectra_6000-11000}
\end{figure*}
\begin{figure*}[ht!]
		\centering
		\includegraphics[width=18cm]{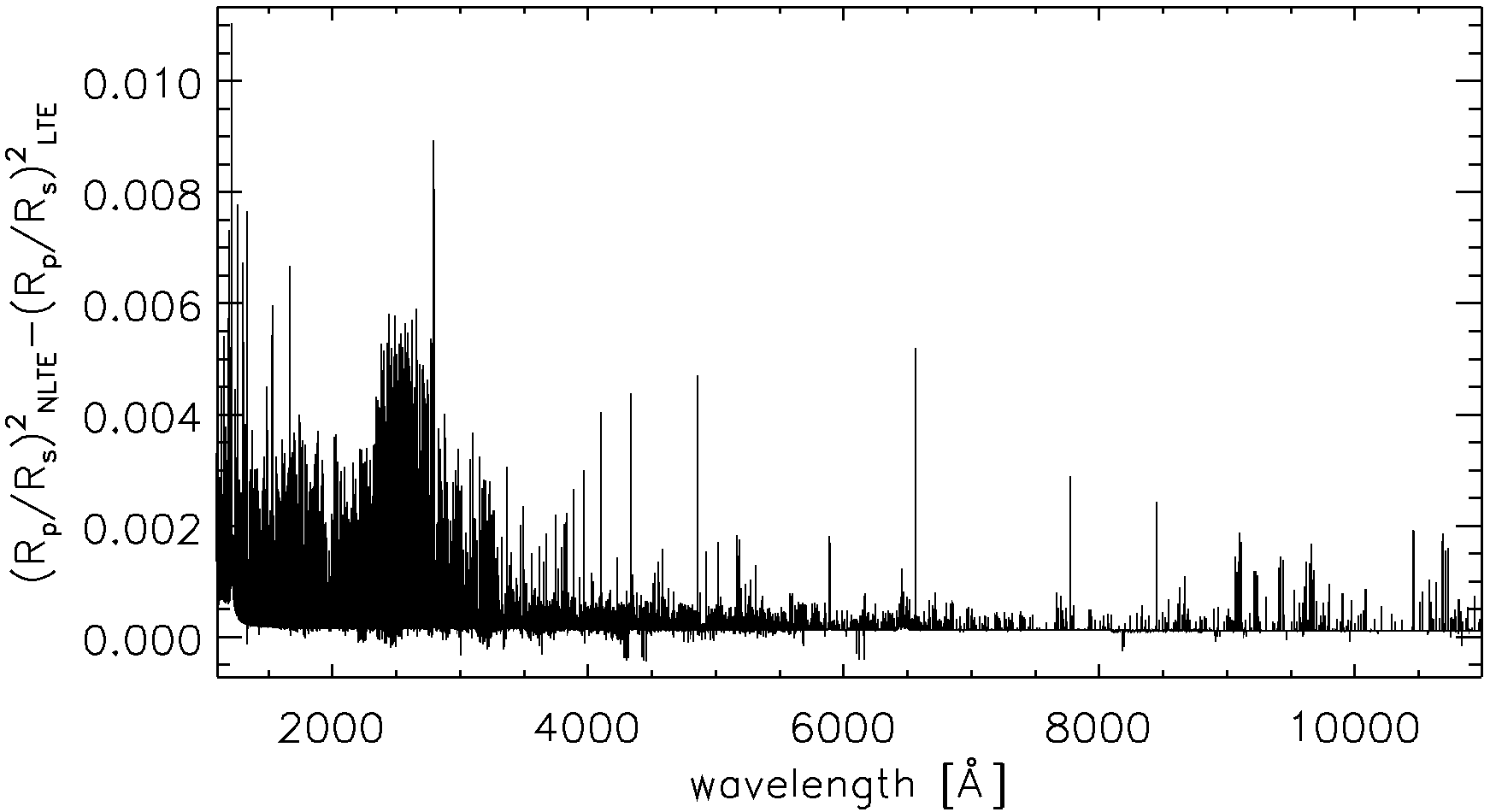}
		\includegraphics[width=18cm]{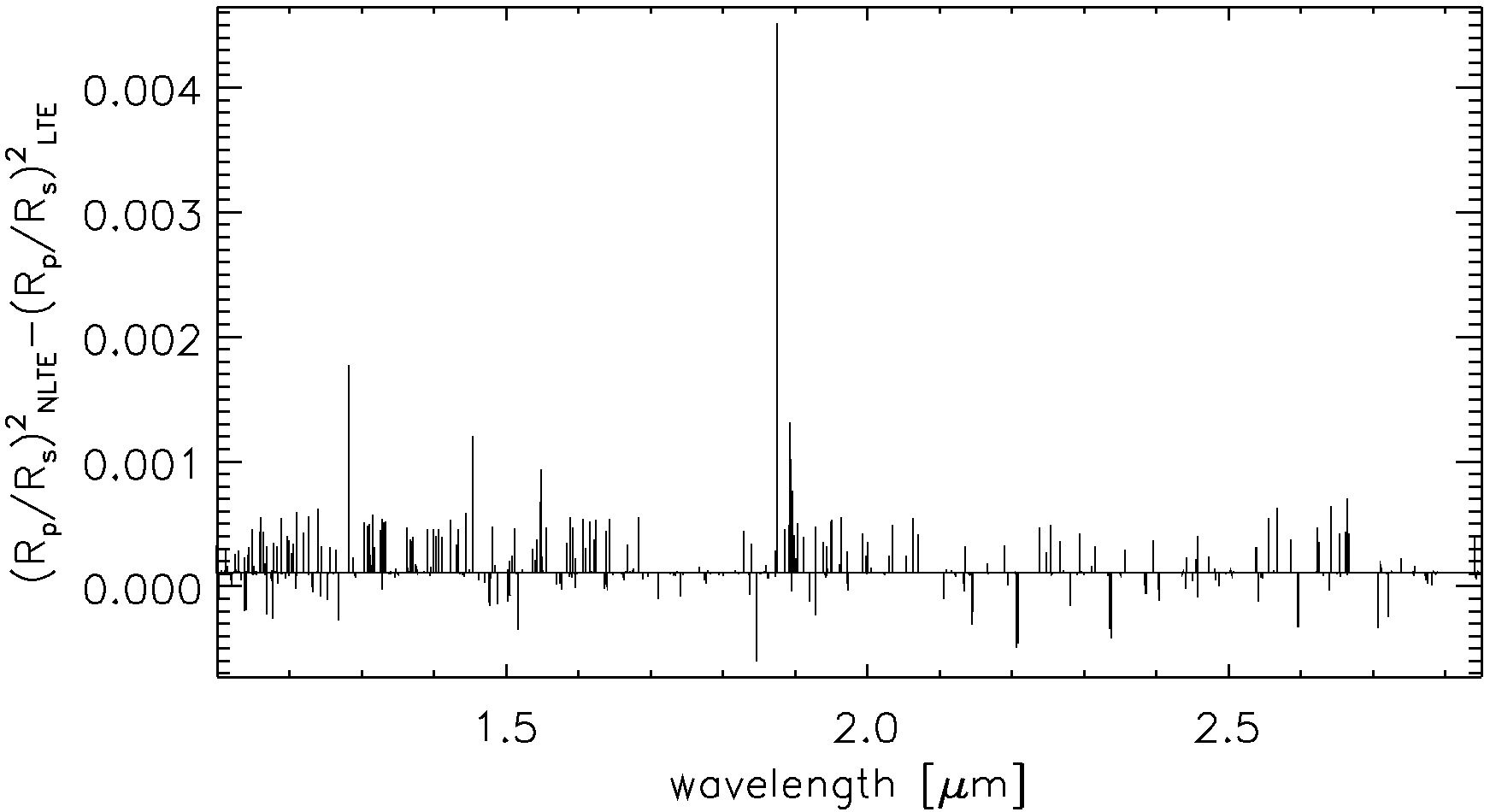}
		\caption{Transit depth difference between the NLTE and LTE transmission spectra shown in Figure~\ref{fig:transmission_spectra}. The top plot covers the UV and optical range, while the bottom plot covers the infrared band.} 
  		\label{fig:NLTEvsLTE_transit_depth_difference}
\end{figure*}
\begin{table*}[h]
\renewcommand{\arraystretch}{1.2}
\centering
\fontsize{8}{15}\selectfont
\caption{Gaussian fit parameters for H$\alpha$ and H$\beta$.}
\begin{tabular}{c|c|c|c|c|c|c|c|c}
\hline
\hline
& \multicolumn{4}{c|}{H$\alpha$} & \multicolumn{4}{c}{H$\beta$}\\
\hline
Night & Amplitude & CL & Center & FWHM & Amplitude & CL & Center & FWHM\\
& [\%] & $\sigma$ & [km\,s$^{-1}$] & [km\,s$^{-1}$] & [\%] & $\sigma$ & [km\,s$^{-1}$] & [km\,s$^{-1}$] \\
\hline
2017-08-16 & $0.771_{-0.118}^{+0.133}$ & $6.145$ & $1.590_{-3.198}^{+2.687}$ & $40.494_{-9.636}^{+31.781}$ & $0.313_{-0.101}^{+0.105}$ & $3.027$ & $-14.030_{-6.870}^{+6.073}$ & $44.779_{-13.457}^{+35.157}$ \\
2018-07-12 & $0.611_{-0.080}^{+0.075}$ & $7.879$ & $-2.462_{-2.270}^{+2.221}$ & $36.837_{-5.306}^{+5.850}$ & $0.625_{-0.073}^{+0.076}$ & $8.400$ & $-2.026_{-1.651}^{+1.802}$ & $28.330_{-3.522}^{+4.269}$ \\
2018-08-19 & $0.997_{-0.109}^{+0.114}$ & $8.955$ & $1.492_{-1.412}^{+1.492}$ & $25.635_{-3.083}^{+3.599}$ & $0.447_{-0.077}^{+0.087}$ & $5.437$ & $1.795_{-3.409}^{+3.229}$ & $43.184_{-11.205}^{+13.259}$ \\
2019-08-26 & $0.676_{-0.060}^{+0.064}$ & $10.944$ & $-1.226_{-1.849}^{+2.028}$ & $43.540_{-4.145}^{+5.167}$ & $0.593_{-0.064}^{+0.070}$ & $8.863$ & $1.464_{-1.935}^{+1.991}$ & $35.634_{-4.184}^{+4.638}$ \\
2019-09-02 & $0.925_{-0.070}^{+0.069}$ & $13.342$ & $-1.022_{-1.211}^{+1.175}$ & $33.998_{-2.590}^{+3.012}$ & $0.338_{-0.088}^{+0.088}$ & $3.830$ & $-4.307_{-3.045}^{+4.620}$ & $32.980_{-9.671}^{+55.881}$ \\
2022-07-31 & $0.845_{-0.116}^{+0.110}$ & $7.469$ & $-0.665_{-1.921}^{+2.007}$ & $30.108_{-4.221}^{+5.837}$ & $0.820_{-0.132}^{+0.130}$ & $6.250$ & $2.292_{-1.484}^{+1.354}$ & $18.252_{-2.554}^{+3.027}$ \\
Weighted mean & $0.789_{-0.035}^{+0.033}$ & $22.935$ & $-0.654_{-0.725}^{+0.747}$ & $34.678_{-1.707}^{+1.791}$ & $0.517_{-0.034}^{+0.034}$ & $15.049$ & $-0.907_{-0.948}^{+0.913}$ & $28.910_{-2.254}^{+2.228}$ \\
\hline
\end{tabular}
\label{tab:Ha_Hb_variability}
\end{table*}
\end{appendix}	
\end{document}